\documentclass{fac}

\usepackage{stmaryrd}
\usepackage{amsmath, amssymb}
\usepackage{amstext}
\usepackage{txfonts}
\usepackage{pxfonts}
\usepackage{graphicx}
\usepackage{wasysym}
\usepackage{listingsVDM}        
\usepackage{url}
\usepackage{relsize}
\usepackage{xspace}             
\usepackage[T1]{fontenc}        
\usepackage{color}              
\usepackage{textcomp}           
\usepackage{subfig}
\usepackage{enumitem}           
\usepackage[titletoc, title]{appendix}

\usepackage{chngcntr}
\AtBeginDocument{\counterwithin{lstlisting}{section}}

\newlist{formalismreqs}{enumerate}{1}
\setlist[formalismreqs]{align=left, label=F\arabic*, style=standard, leftmargin=!, labelwidth=2em, labelsep*=1em, labelindent=0em}

\newlist{reconfigreqs}{enumerate}{1}
\setlist[reconfigreqs]{align=left, label=R\arabic*, style=standard, leftmargin=!, labelwidth=2em, labelsep*=1em, labelindent=0em}

\newlist{workflow1reqs}{enumerate}{1}
\setlist[workflow1reqs]{label=C1.\arabic*, style=standard, leftmargin=*, labelwidth=2.5em, labelsep*=1em, labelindent=0em}

\newlist{workflow2reqs}{enumerate}{1}
\setlist[workflow2reqs]{label=C2.\arabic*, style=standard, leftmargin=*, labelwidth=2.5em, labelsep*=1em, labelindent=0em}

\newlist{designlist}{enumerate}{1}
\setlist[designlist]{label=Design \arabic*, style=standard, leftmargin=*, labelwidth=2.5em, labelsep*=1em, labelindent=0em}

\newlist{vdmtodo}{itemize}{1}
\setlist[vdmtodo]{align=left, label=\textbf{\emph{VDM:}}, style=standard, leftmargin=!, labelwidth=2em, labelsep*=1em, labelindent=0em}

\newcommand{\cpp}{C{\protect\raisebox{0.25ex}{\relsize{-1}$++$}}\xspace}
\newcommand{\vpp}{VDM{\protect\raisebox{0.25ex}{\relsize{-1}$++$}}\xspace}

\usepackage[english]{babel}

\newcommand{\COMMUNITY} {C\textsc{omm}U\textsc{nity}}

\newcommand{\trans}[1]{\mbox{$\stackrel{#1}{\longrightarrow}$}}

\newcommand{\mcIsb}[2]{\mbox{$\mathcal{#1}\sb{\!\!#2}$}}

\def \mid {\; | \;}

\catcode`@=11
\chardef\mathlig@atcode\count255

\def\actively#1#2{\begingroup\uccode`\~=`#2\relax\uppercase{\endgroup#1~}}
\def\mathlig@gobble{\afterassignment\mathlig@next@cmd\let\mathlig@next= }
\def\mathlig@delim{\mathlig@delim}
\def\mathlig@defcs#1{\expandafter\def\csname#1\endcsname}
\def\mathlig@let@cs#1#2{\expandafter\let\expandafter#1\csname#2\endcsname}
\def\mathlig@appendcs#1#2{\expandafter\edef\csname#1\endcsname{\csname#1\endcsname#2}}

\def\mathlig#1#2{\mathlig@checklig#1\mathlig@end\mathlig@defcs{mathlig@back@#1}{#2}\ignorespaces}

\def\mathlig@checklig#1#2\mathlig@end{%
 \expandafter\ifx\csname mathlig@forw@#1\endcsname\relax
 \expandafter\mathchardef\csname mathlig@back@#1\endcsname=\mathcode`#1%
 \mathcode`#1"8000\actively\def#1{\csname mathlig@look@#1\endcsname}%
 \mathlig@dolig#1\mathlig@delim
\fi
\mathlig@checksuffix#1#2\mathlig@end
}

\def\mathlig@checksuffix#1#2\mathlig@end{%
\ifx\mathlig@delim#2\mathlig@delim\relax\else\mathlig@checksuffix@{#1}#2\mathlig@end\fi
}
\def\mathlig@checksuffix@#1#2#3\mathlig@end{%
\expandafter\ifx\csname mathlig@forw@#1#2\endcsname\relax\mathlig@dosuffix{#1}{#2}\fi
\mathlig@checksuffix{#1#2}#3\mathlig@end
}

\def\mathlig@dosuffix#1#2{%
\mathlig@appendcs{mathlig@toks@#1}{#2}%
\mathlig@dolig{#1}{#2}\mathlig@delim
}

\def\mathlig@dolig#1#2\mathlig@delim{%
 \mathlig@defcs{mathlig@look@#1#2}{%
 \mathlig@let@cs\mathlig@next{mathlig@forw@#1#2}\futurelet\mathlig@next@tok\mathlig@next}%
 \mathlig@defcs{mathlig@forw@#1#2}{%
  \mathlig@let@cs\mathlig@next{mathlig@back@#1#2}%
  \mathlig@let@cs\checker{mathlig@chck@#1#2}%
  \mathlig@let@cs\mathligtoks{mathlig@toks@#1#2}%
  \expandafter\ifx\expandafter\mathlig@delim\mathligtoks\mathlig@delim\relax\else
  \expandafter\checker\mathligtoks\mathlig@delim\fi
  \mathlig@next
 }%
 \mathlig@defcs{mathlig@toks@#1#2}{}%
 \mathlig@defcs{mathlig@chck@#1#2}##1##2\mathlig@delim{%
  \ifx\mathlig@next@tok##1%
   \mathlig@let@cs\mathlig@next@cmd{mathlig@look@#1#2##1}\let\mathlig@next\mathlig@gobble
  \fi 
  \ifx\mathlig@delim##2\mathlig@delim\relax\else
   \csname mathlig@chck@#1#2\endcsname##2\mathlig@delim
  \fi
 }%
 \ifx\mathlig@delim#2\mathlig@delim\else
  \mathlig@defcs{mathlig@back@#1#2}{\csname mathlig@back@#1\endcsname #2}%
 \fi
}%

\mathchardef\mathlig@paren\mathcode`(

\begingroup \catcode`\"=12
\gdef\resetMathstrut@{%
  \setbox\z@\hbox{%
    \mathchardef\@tempa\mathlig@paren\relax
    \def\@tempb##1"##2##3{\the\textfont"##3\char"}%
    \expandafter\@tempb\meaning\@tempa \relax
  }%
  \ht\Mathstrutbox@\ht\z@ \dp\Mathstrutbox@\dp\z@
}
\endgroup

\catcode`@\mathlig@atcode

\mathchardef\ordinarypar\mathcode`\|

\mathchardef\ordinarycolon\mathcode`\:

\newtheorem{propcount}{dummy}

\newtheorem{defncount}{dummy}
\newtheorem{prop}[propcount]{Proposition}
\newtheorem{defn}[defncount]{Definition}

\newcommand{\wt}[1]{\widetilde{#1}} 

\newcommand{\webpii}{{\tt web}$\pi_{\infty}$}

\newcommand{\trs}[3]{\mbox{$\langle \hspace{-2pt} |$} #1 \; ; \; #2 \mbox{$| \hspace{-2pt} \rangle$}_{#3}}

\title{An Empirical Comparison of Formalisms for Modelling and Analysis of
Dynamic Reconfiguration of Dependable Systems}

\author[A. Bhattacharyya and A. Mokhov and K. Pierce]
    {Anirban Bhattacharyya$^1$ and Andrey Mokhov$^2$ and Ken Pierce$^1$\\
     $^1$School of Computing Science, Claremont Tower, Newcastle University, Newcastle upon Tyne, NE1 7RU, UK\\
     $^2$School of Electrical and Electronic Engineering, Merz Court, Newcastle University, Newcastle upon Tyne, NE1 7RU, UK}

\correspond{Anirban Bhattacharyya,
            School of Computing Science, Claremont Tower, Newcastle University, Claremont Road, Newcastle upon Tyne, NE1 7RU, UK.
            E-mail: A.Bhattacharyya@ncl.ac.uk}

\begin{document}

\makecorrespond

\maketitle

\parskip 0.1in
\parindent 0in

\begin{abstract}
\noindent This paper uses a case study to evaluate empirically three formalisms of different kinds
for their suitability for the modelling and analysis of dynamic reconfiguration of dependable systems.
The requirements on an ideal formalism for dynamic software reconfiguration are defined.
The reconfiguration of an office workflow for order processing is described,
and the requirements on the reconfiguration of the workflow are defined.
The workflow is modelled using the Vienna Development Method (VDM), conditional partial order graphs (CPOGs),
and the basic Calculus of Communicating Systems for dynamic process reconfiguration (basic~$\mathrm{CCS^{dp}}$),
and verification of the reconfiguration requirements is attempted using the models.
The formalisms are evaluated according to their ability to model the reconfiguration of the workflow,
to verify the requirements on the workflow's reconfiguration, and to meet the requirements on an ideal formalism.
\end{abstract}


\let\clearpage\relax

\vspace{-1cm}

\section{Introduction}\label{sec:intro}


The next generation of dependable systems is expected to have significant evolution requirements \cite{kn:CoyEtAl10}.
Moreover, it is impossible to foresee all the requirements that a system will have to meet in future
when the system is being designed \cite{kn:MenMagRum10}.
Therefore, it is highly likely that the system will have to be redesigned (i.e. reconfigured) during its lifetime,
in order to meet new requirements.
Furthermore, certain classes of dependable systems, such as control systems, must be dynamically reconfigured
\cite{kn:KarMasOstSch10}, because it is unsafe or impractical or too expensive to do otherwise.
The dynamic reconfiguration of a system is defined as the change at runtime of the structure of the system
-- consisting of its components and their communication links --
or the hardware location of its software components \cite{kn:Bha13} or their communication links.
For example, the dynamic upgrade of a software module in a telecommunications satellite during the execution of the old version of the module,
the removal of a faulty controller of an aero engine during flight,
re-establishing the exchange-to-mobile communication link during a conversation as the mobile crosses the boundary of communication zones,
and moving an executing software object between servers to achieve load balancing.
This paper focuses on dynamic software reconfiguration, because software is much more mutable than hardware.

Existing research in dynamic software reconfiguration can be grouped into three cases
from the viewpoint of interference between application and reconfiguration tasks, which is embedded in time (see Figure 1).
Interference is defined as the effect of the concurrent execution of a task on the execution of another task.
For example, an incorrect result of a computation performed by the affected task,
or a delay in the response time of the computation, or the delayed termination or replacement of the task.

\begin{figure}
\begin{center}
\includegraphics[width=0.6\textwidth]{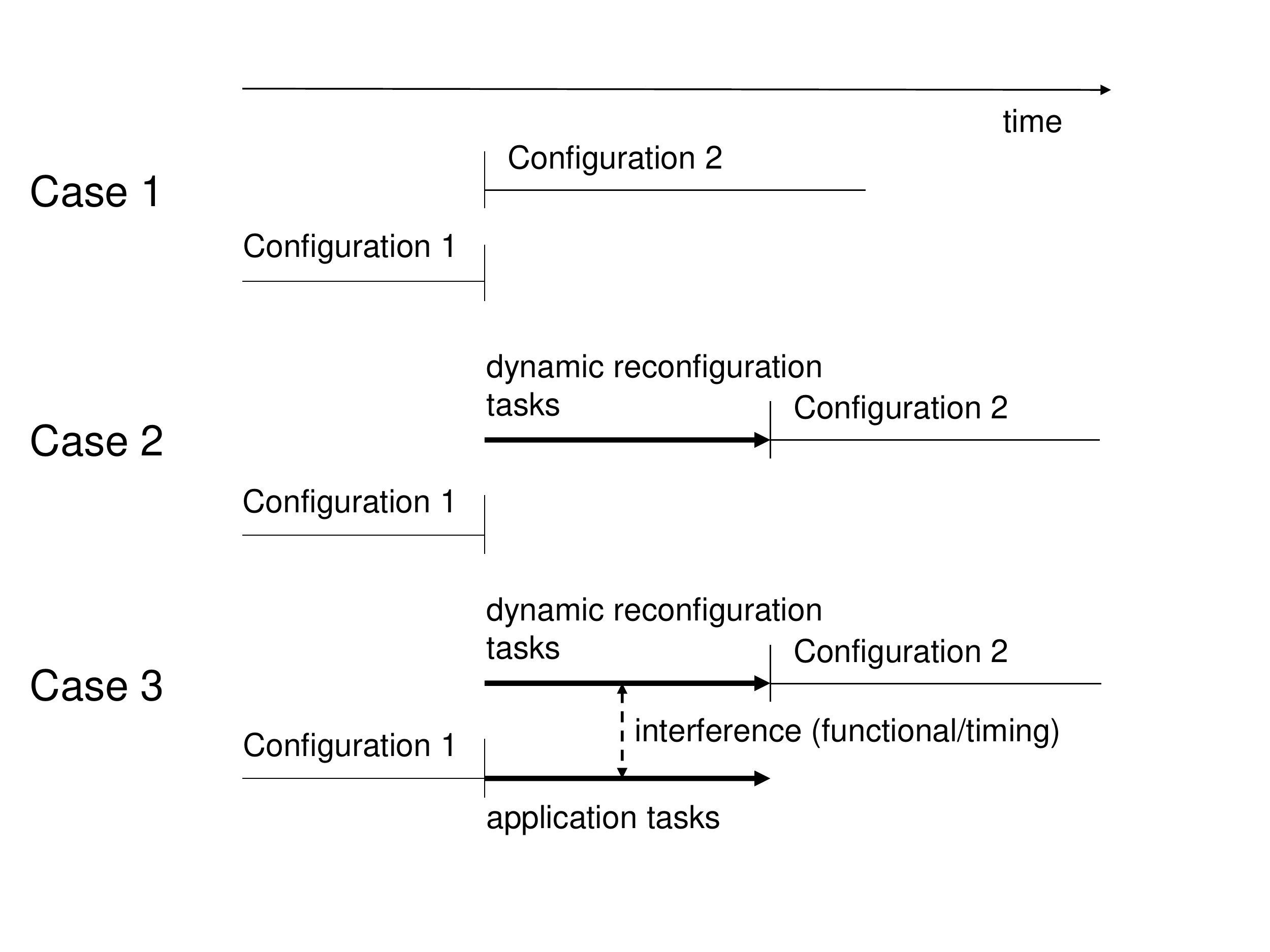} \vspace{-0.6cm}
\caption{Dynamic reconfiguration cases} \label{ReconfigCases}
\end{center}
\end{figure} 

Case 1 is the near-instantaneous reconfiguration of a system,
in which the duration of the reconfiguration interval is negligible in comparison to the durations of application tasks.
Any executing task in Configuration 1 that is not in Configuration 2 is aborted,
which can leave data in a corrupted or inconsistent state.
Alternatively, the reconfiguration is done at the end of the hyperperiod of Configuration 1
(i.e. the lowest common multiple of the periods of the periodic tasks in Configuration 1),
which can result in a significant delay in handling the reconfiguration-triggering event.
This is the traditional method of software reconfiguration,
and is applicable to small, simple systems running on a uniprocessor. 

Case 2 is the reconfiguration of a system in which the duration of the reconfiguration interval is significant
in comparison to the durations of application tasks,
and any executing task in Configuration 1 that can interfere with a reconfiguration task
is either aborted or suspended until the reconfiguration is complete.
This is the most common method of software reconfiguration
(see \cite{kn:SteVolKho97}, \cite{kn:AlmEtAl01}, \cite{kn:BloDay93}, and \cite{kn:KraMag90}),
and is applicable to some large, complex, distributed systems.
If the duration of the reconfiguration is bounded and the controlled environment can wait for the entire reconfiguration to complete,
then the method can be used for hard real-time systems;
otherwise, the environment can become irrecoverably unstable and suffer catastrophic failure if a time-critical service is delayed beyond its deadline.

Case 3 is the reconfiguration of a system in which the duration of the reconfiguration interval is significant
in comparison to the durations of application tasks,
and tasks in Configuration 1 execute concurrently with reconfiguration tasks.
This method avoids aborting tasks and reduces the delay on the application due to reconfiguration,
but it introduces the possibility of functional and timing interference between application and reconfiguration tasks.
If the interference can be controlled, then this method is the most suitable for large, complex, distributed systems,
including hard real-time systems, but it is also the least researched method of software reconfiguration.
Existing research in Case 3 has focused on timing interference between application and reconfiguration tasks,
and on achieving schedulability guarantees
(for example, see \cite{kn:ShaRajLehRam89}, \cite{kn:TinBurWel92}, \cite{kn:Ped99}, \cite{kn:Mon04}, \cite{kn:FisWin05}, and \cite{kn:Far06}).
There is little research on formal verification of functional correctness in the presence of
functional interference between application and reconfiguration tasks (see \cite{kn:MedTay00} and \cite{kn:BraEtAl04}).

Therefore, there is a requirement for formal representations of dynamic software reconfiguration that can express
functional interference between application and reconfiguration tasks, and can be analyzed to verify functional correctness.
There is also a significant requirement for modelling unplanned reconfiguration, that is,
reconfiguration that is not incorporated in the design of a system (see \cite{kn:CoyEtAl10}, \cite{kn:MenMagRum10}, and \cite{kn:KarMasOstSch10}).
This paper makes a contribution towards meeting these two key requirements.
Research shows that no single existing formalism is ideal for representing dynamic software reconfiguration \cite{kn:Wer99}.
Therefore, we use three formalisms of different kinds:
the Vienna Development Method (VDM, based on the state-based approach),
conditional partial order graphs (CPOGs, in which graph families are used for verification of workflow and reconfiguration requirements),
and the basic Calculus of Communicating Systems for dynamic process reconfiguration (basic $\mathrm{CCS^{dp}}$, based on the behavioural approach),
to produce representations of a case study, and evaluate how well the different representations meet the requirements.
The diversity of the three formalisms is manifested in the diversity of their semantics.
Thus, VDM has a denotational semantics in order to describe data structures and algorithms, which helps to refine a specification to an implementation;
CPOGs have an axiomatic semantics based on equations, which helps to make reasoning about graphs simple and computationally efficient;
and basic $\mathrm{CCS^{dp}}$ has a labelled transition system (LTS) semantics, which helps to describe process behaviour and reconfiguration simply.
In order to facilitate comparison of the formalisms, we have defined an LTS semantics for CPOGs and for the VDM model;
an LTS semantics for VDM is beyond the scope of this paper.

The rest of the paper is organized as follows:
Section \ref{sec:reqts} defines requirements on an ideal formalism for the modelling and analysis of dynamic software reconfiguration.
Section \ref{sec:casestudy} describes the case study, in which a simple office workflow for order processing is dynamically reconfigured,
and defines the requirements on the initial configuration, the final configuration, and on the reconfiguration of the workflow.
The case study is a modification of the case study used in \cite{kn:Bha13} to evaluate basic $\mathrm{CCS^{dp}}$.
The reconfiguration of the workflow is modelled and analyzed using
VDM (in Section \ref{sec:vdm}), CPOGs (in Section \ref{sec:cpog}), and basic $\mathrm{CCS^{dp}}$ (in Section \ref{sec:bccsdp}).
We have deliberately not used workflow-specific formalisms
(such as \cite{kn:YuLin05}, \cite{kn:ArdPer07}, and \cite{kn:HilEtAl11}) for two reasons.
First, because of our lack of fluency in them;
and second, because we believe the models should be produced using general purpose formalisms (if possible).
VDM, CPOGs, and basic $\mathrm{CCS^{dp}}$ are compared and evaluated in Section \ref{sec:comp}
using the results of the modelling and analysis exercise and the requirements on an ideal formalism defined in Section \ref{sec:reqts}.
Related work is reviewed in Section \ref{sec:relwork}.

This paper contains considerable material from the first author's doctoral thesis \cite{kn:Bha13}.
The core requirements on an ideal formalism F1--F11 in the following section are from the thesis,
requirement F12 is new and was suggested by one of the anonymous reviewers.
The case study is a modification of the thesis case study: Configuration 1 has been simplified by making it purely linear,
Configuration 2 has been made more complex by adding concurrently executing tasks,
and the reverse reconfiguration from Configuration 2 to Configuration 1 is now considered in the modelling.
The descriptions of the syntax and semantics of basic $\mathrm{CCS^{dp}}$ are from the thesis,
but the modelling now refers to all the designs of Configuration 1 (rather than to only Design 3)
and the analysis based on weak observational bisimulation is new.
The sections on VDM, CPOGs, and the comparison of formalisms are (of course) new.

\section{Requirements on an Ideal Formalism for Dynamic Software Reconfiguration}\label{sec:reqts}


No single existing formalism is ideal for the modelling and analysis of dynamic software reconfiguration \cite{kn:Wer99}.
However, it is possible to identify core requirements that an ideal formalism must meet
and to evaluate candidate formalisms against these requirements, such as those in this paper.
In this section, we identify and briefly justify a set of core requirements, labelled F1--F12,
and use them to evaluate the three formalisms in Section~\ref{sec:comp}.
We summarize our findings at the end of this section.

\subsection{Requirements on Formalisms}\label{sec:frmreqts}

We divide the requirements into two groups, the first relating specifically to modelling and verifying dynamic reconfiguration,
and the second relating more generally to `usable' formalisms.

\subsubsection*{Dynamic Reconfiguration Requirements}\label{sec:drreqts}

\begin{formalismreqs}
   \item \emph{It should be possible to model, and to identify instances of, software components and tasks, and their communication links.}
   A software component can be a program or a class (as in Smalltalk or C++) or a module (as in C), a task is a process (as in UNIX),
   and a communication link is a channel of communication (e.g. a socket-to-socket link over TCP/IP between communicating UNIX processes).
   Multiple tasks can be based on the same software component in order to process different transactions concurrently,
   and multiple software components can be used to provide fault tolerance.
   The dynamic reconfiguration of a software component or of a task involves the selective reconfiguration of its instances,
   which is facilitated by the use of instance identifiers.

   \item \emph{It should be possible to model the creation, deletion, and replacement of software components and tasks,
   and the creation and deletion of their communication links.}
   These are the fundamental operations used to change the software structure of a system.

   \item \emph{It should be possible to model the relocation of software components and tasks on physical nodes.}
   Software relocation helps to implement load balancing, which is used to improve performance and reliability in cloud computing.
   Thus, software relocation helps to increase the dependability of cloud computing.

   \item \emph{It should be possible to model state transfer between software components and between tasks.}
   In dependable systems with state, state transfer helps to implement Case 2 of dynamic reconfiguration (see Figure \ref{ReconfigCases})
   and to implement software relocation.

   \item \emph{It should be possible to model both planned and unplanned reconfiguration.}
   Planned reconfiguration is reconfiguration that is incorporated in the design of a system.
   Unplanned reconfiguration is reconfiguration that is \textbf{not} incorporated in the design of a system,
   which is relevant for legacy systems and for the evolution of systems.

   \item \emph{It should be possible to model the functional interference between application tasks and reconfiguration tasks.}
   This is the main modelling requirement in Case 3 of dynamic reconfiguration (see Figure \ref{ReconfigCases}),
   and is an outstanding research issue.

   \item \emph{It should be possible to express and to verify the functional correctness requirements of application tasks and reconfiguration tasks.}
   This is the main verification requirement of dynamic reconfiguration, and is an outstanding research issue in Case 3.
\end{formalismreqs}

\subsubsection*{General Requirements}\label{sec:genreqts}

\begin{formalismreqs}[start=8]

   \item \emph{It should be possible to model the concurrent execution of tasks.}
   Concurrency can cause functional interference between tasks, and thereby affect the functional correctness of a task,
   and it is a feature of many dependable systems. Therefore, it should be modelled.

   \item \emph{It should be possible to model state transitions of software components and tasks.}
   State affects the functionality of a task, and thereby affects the functional correctness of the task,
   and it is a feature of most dependable systems. Therefore, it should be modelled.

   \item \emph{The formalism should be as terse as possible.}
   Terseness supports abstraction, which is essential in removing unnecessary detail from a model,
   and thereby renders the model easier to understand. Thus, terseness facilitates the use of a model.

   \item \emph{The formalism should be supported by tools.}
   Otherwise, the formalism will not be used by software engineers.

   \item \emph{The formalism should be easy to learn and to use.}
   Otherwise, the rate of adoption of the formalism by users will be low.
\end{formalismreqs}

\subsection{Summary of Results}\label{sec:sumresults}

The evaluations of the three formalisms in Section~\ref{sec:comp} show that none of them is ideal,
since none of them meets all the requirements on an ideal formalism for dynamic software reconfiguration defined above.
However, the formalisms meet the requirements collectively, and (therefore) are complementary (albeit with extensions).

The main strength of basic $\mathrm{CCS^{dp}}$ is in modelling.
It can model:
abstractly and tersely the composition and concurrent execution of application and reconfiguration tasks using concurrent processes,
their functional interference using interleaved transitions,
their planned and unplanned reconfiguration using fraction processes,
cyclic processes using recursion,
and reconfiguration of fraction processes using other fractions.
Its main weaknesses are:
inability to control non-deterministic transitions,
inability to reconfigure selectively specific process instances,
computational complexity of process matching based on bisimulation,
computational complexity and restrictiveness of process congruence,
and lack of tools.

In contrast, the main strength of CPOGs is in verification.
A Boolean SAT solver can compare a model and its requirement in canonical form,
and efficient model checking is supported by predicates on actions and on action dependencies and the acyclic topology of CPOGs.
Correctness of a reconfiguration between configurations can be proved
using consistent histories of actions of the two configurations and by restricting interference through forbidden actions.
Functional interference between tasks can be modelled using either interleaved actions or simultaneous actions.
Its main weaknesses are:
inability to model composition and structure of software components and tasks,
low level of abstraction for modelling,
inability to model cyclic processes,
and lack of available tools.

In contrast to both basic $\mathrm{CCS^{dp}}$ and CPOGs, the main strength of VDM is in formal software development.
It can
model workflows, software components, and tasks as data types, which facilitates their refinement to an implementation,
its tools for development, simulation, and testing are mature and available,
and it is easy to use by system designers.
The main weaknesses of VDM-SL are:
lack of constructs for modelling concurrency and interference,
and lack of formal verification tools.

\section{Case Study: Dynamic Reconfiguration of an Office Workflow for Order Processing}\label{sec:casestudy}

\label{sec:wfactivs}

The case study described in this section involves the dynamic reconfiguration of a simple office workflow for order processing,
which is a simplified version of real workflows commonly found in large and medium-sized organisations \cite{kn:EllKedRoz95}.
Preliminary versions of the case study are in \cite{kn:MazEtAl11} and \cite{kn:Bha13}.
The case study was chosen for three reasons.
First, workflows are ubiquitous, which suggests that our workflow reconfiguration will be of interest to a large community.
Second, the case study is based on published work by other researchers, see \cite{kn:EllKedRoz95}.
Third, it is both simple to understand and complex enough to exercise all three formalisms.

The workflow consists of a network of several communicating tasks,
and the configuration of the workflow is the structure of the network.
The workflow does not contain any loop, because loops tend to reduce the scope of reconfiguration considerably.
A loop can have an invariant that is not an invariant of the reconfiguration,
and thereby can prevent reconfiguration of tasks constituting the loop during an execution of the loop.
The workflow contains the following tasks:

\begin{itemize}
   \item \texttt{Order Receipt}: an order for a product is received from an existing customer.
      The order identifier includes the customer identifier and the product identifier.
      An evaluation of the order is initiated to determine whether or not the order is viable.
   \item \texttt{Evaluation}: the product identity is used to check the availability of the product;
	   the customer identity is used to check the credit of the customer.
      If either check fails, the output is negative and the order is rejected; otherwise, the output is positive and the order is accepted.
   \item \texttt{Rejection}: if the order is rejected, a notification of rejection is sent to the customer and the workflow terminates.
   \item If the order is accepted, the following tasks are performed:
      \begin{itemize}
         \item \texttt{Shipping}: the product is shipped to the customer.
         \item \texttt{Billing}: the customer is billed for the cost of the product ordered plus shipping costs.
         \item \texttt{Archiving}: the order is archived for future reference.
         \item \texttt{Confirmation}: a notification of successful completion of the order is sent to the customer.
      \end{itemize}
\end{itemize}

\subsection{Configurations of the Workflow}
\label{sec:wfreqts2}
\label{sec:wfreqts1}

\begin{figure}
\begin{minipage}{.5\linewidth}
   \includegraphics[width=\linewidth]{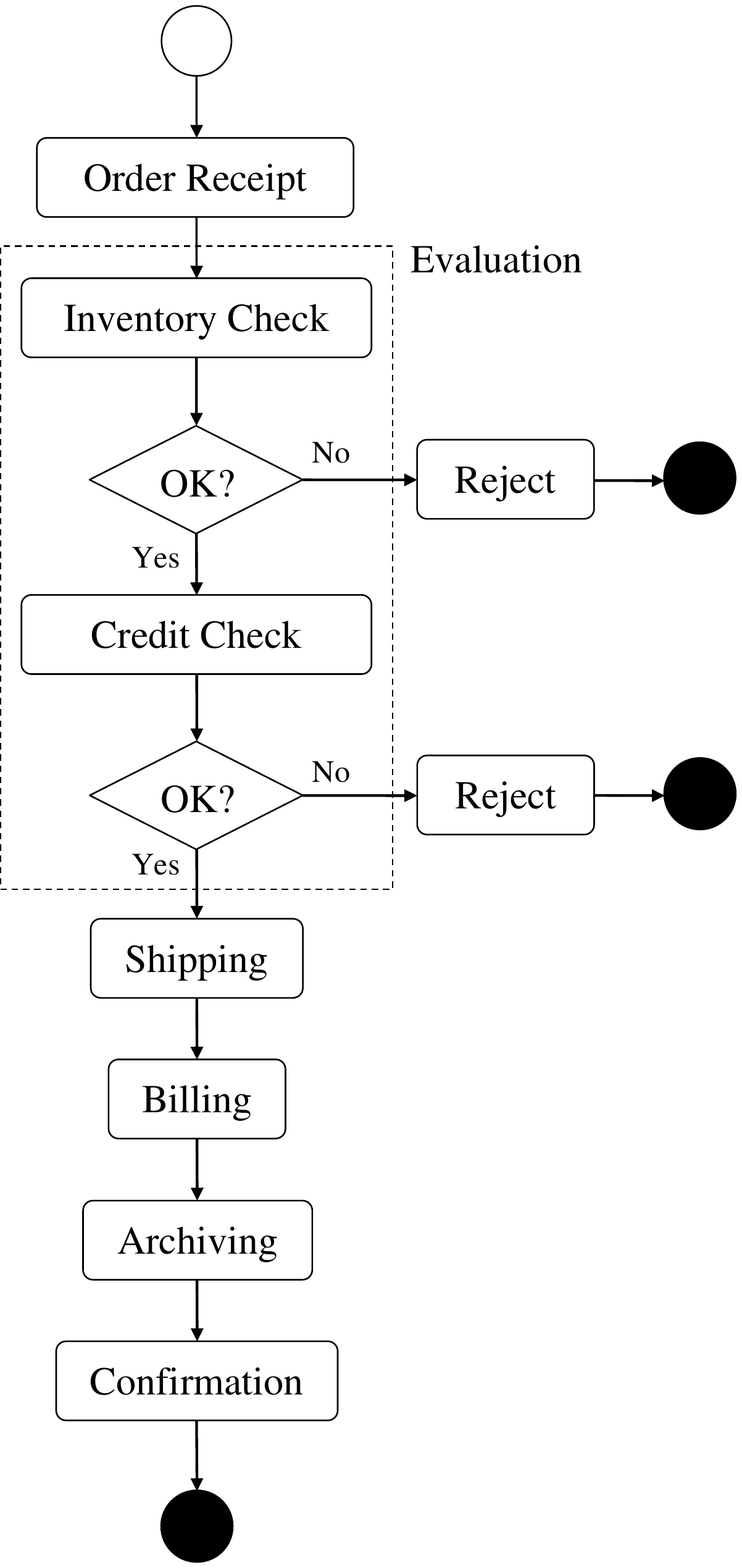}
   \vspace{-1.5cm}
   \captionof{figure}{Flow chart of the requirements on \\ Configuration 1}
   \label{OfficeWorkflowConfig1}
\end{minipage}
\quad
\begin{minipage}{0.5\linewidth}
   \vspace{0.10cm}
   \includegraphics[width=\linewidth]{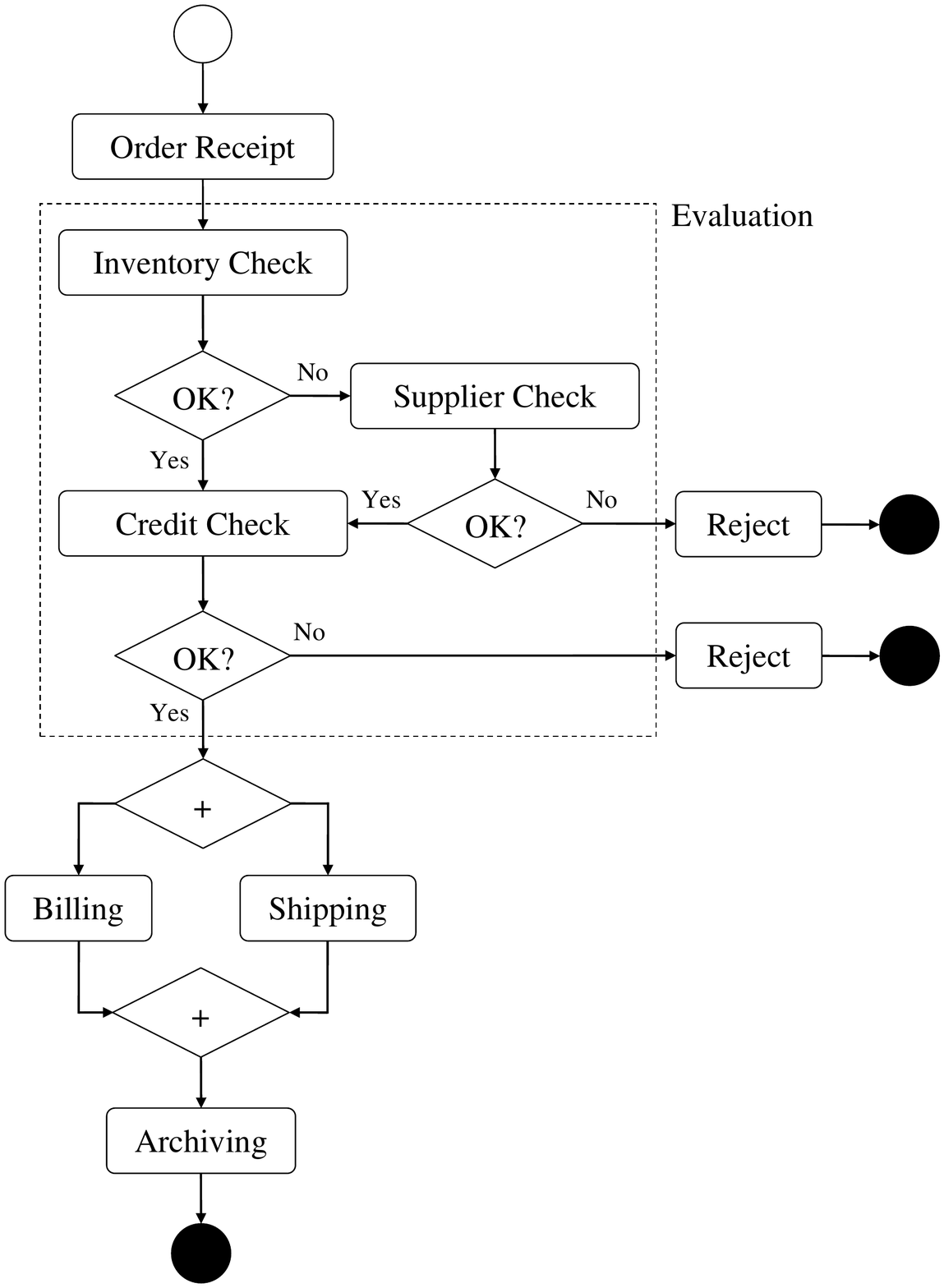}
   \vspace{-0.60cm}
   \captionof{figure}{Flow chart of the requirements on \\ Configuration 2}
   \label{OfficeWorkflowConfig2}
\end{minipage}
\end{figure}

There are two configurations of the workflow: Configuration~1 and Configuration~2.
Initially the workflow executes Configuration~1.
Subsequently, the workflow must be reconfigured through a process to Configuration~2.
Requirements on the two configurations are shown in Figures~\ref{OfficeWorkflowConfig1} and~\ref{OfficeWorkflowConfig2} and are explained below.
We then identify requirements on the reconfiguration from Configuration~1 to Configuration~2, and identify potential designs for this system.

The initial configuration of the workflow is Configuration~1 and must meet the following requirements for each order
(see Figure \ref{OfficeWorkflowConfig1}):

\begin{workflow1reqs}
   \item \texttt{Order Receipt} must be performed first. That is, it must begin before any other task.
   \item \texttt{Evaluation}: in evaluating the order,
      the product identity is used to perform an inventory check on the stock of the product,
      and the customer identity is used to perform a credit check on the customer.
      If either check fails, the output of \texttt{Evaluation} is negative; otherwise, the output is positive.
   \item \texttt{Evaluation} must be performed second.
   \item If the output of \texttt{Evaluation} is negative, \texttt{Rejection} must be the third and final task of the workflow.
   \item If the output of \texttt{Evaluation} is positive, the following conditions must be satisfied:
      \begin{enumerate}
         \item \texttt{Shipping} must be performed after \texttt{Evaluation}.
         \item \texttt{Billing} must be performed after \texttt{Shipping}.
         \item \texttt{Archiving} must be performed after \texttt{Billing}.
         \item \texttt{Confirmation} must be performed after \texttt{Archiving} and must be the final task to be performed.
         \item The customer must not receive more than one shipment of an order (safety requirement).
      \end{enumerate}
   \item Each task must be performed at most once.
   \item The order must be either rejected or satisfied (liveness requirement).
\end{workflow1reqs} 

\newpage
After some time, the management of the organization using the workflow decides to change it in order to increase sales and provide a faster service.
The new configuration of the workflow is Configuration~2 and must meet the following requirements for each order
(see Figure~\ref{OfficeWorkflowConfig2}):

\begin{workflow2reqs}
   \item \texttt{Order Receipt} must be performed first.
   \item \texttt{Evaluation}: in evaluating the order,
      the product identity is used to perform an inventory check on the stock of the product.
      If the inventory check fails, an external inventory check is made on the suppliers of the product.
	   The customer identity is used to perform a credit check on the customer.
      If either the inventory check or the supplier check is positive, and the credit check is positive, the order is accepted;
      otherwise, the order is rejected.
   \item \texttt{Evaluation} must be performed second.
   \item If the output of \texttt{Evaluation} is negative, \texttt{Rejection} must be the third and final task of the workflow.
   \item If the output of \texttt{Evaluation} is positive, the following conditions must be satisfied:
      \begin{enumerate}
         \item \texttt{Billing} and \texttt{Shipping} must be performed after \texttt{Evaluation}.
         \item \texttt{Billing} and \texttt{Shipping} must be performed concurrently.
         \item \texttt{Archiving} must be performed after \texttt{Billing} and \texttt{Shipping} and must be the final task to be performed.
         \item The customer must not receive more than one shipment of an order (safety requirement).
      \end{enumerate}
   \item Each task must be performed at most once.
   \item The order must be either rejected or satisfied (liveness requirement).
\end{workflow2reqs}

%

\subsection{Requirements on Reconfiguration of the Workflow}
\label{sec:wfreqts12}

In order to achieve a smooth transition from Configuration~1 to Configuration~2 of the workflow,
the process of reconfiguration must meet the following requirements:

\begin{reconfigreqs}
   \item Reconfiguration of a workflow should not necessarily result in the rejection of an order.

   In some systems, executing tasks of Configuration 1 are aborted during its reconfiguration to
   Configuration 2 (see Case 2 in Figure \ref{ReconfigCases}).
   The purpose of this requirement is to avoid the occurrence of Case 2 and ensure the occurrence of Case 3.
   \item Any order being processed that was received \textbf{before} the start of the reconfiguration must satisfy
         all the requirements on Configuration 2 (if possible);
         otherwise, all the requirements on Configuration 1 must be satisfied.
   \item Any order received \textbf{after} the start of the reconfiguration must satisfy all the requirements on Configuration 2.
\end{reconfigreqs}

\subsection{Designs of the Workflow and of its Reconfiguration}
\label{sec:recondes}
\label{sec:designs}

The two configurations of the workflow described above are stated as requirements because there are a number of ways in which an implementation,
and therefore a model, can realise the workflow. We identify four such possible designs:

\begin{designlist}
   \item There is at most one workflow, and the workflow handles a single order at a time.

   The workflow is sequential:
   after an order is received, the thread performs a sequence of tasks, with two choices at \texttt{Evaluation}.
   After the order has been processed, the thread is ready to receive a new order.
   This design corresponds to a cyclic executive.

   \item There is at most one workflow, and the workflow can handle multiple orders at a time.

   The workflow is mainly concurrent:
   after an order is received, the thread forks internally into concurrent threads,
   such that different threads perform the different tasks of the workflow --
   although the requirements on the configurations severely restrict the degree of internal concurrency of the workflow --
   and each thread performs the same task for different orders.

   \item There can be multiple workflows, and each workflow handles a single order.

   After an order is received, the thread forks into two threads:
   one thread processes the order sequentially -- as in Design 1 -- but terminates after the order has been processed;
   the other thread waits to receive a new order.

   \item There can be multiple workflows, and each workflow can handle multiple orders at a time.

   This design is a complex version of Design 2 with multiple workflows.

\end{designlist}

In the next three sections, we model the case study workflow and its reconfiguration using three formalisms of different kinds:
VDM, CPOGs, and basic $\mathrm{CCS^{dp}}$.
For each formalism, we use its 'idiom' to identify which of the above four designs of the configurations is the most suitable for the formalism,
which (in turn) affects how the reconfiguration of the workflow is performed.
Thus, we identify which of the three cases of reconfiguration outlined in Section~\ref{sec:intro}
(instantaneous, sequential, or concurrent) is the most suitable for the 'idiom' of the formalism.


%
%

\newpage
\section{VDM}\label{sec:vdm}


The Vienna Development Method (VDM) is a state-based formal method that was originally designed in the 1970s to give semantics to programming languages~\cite{Jones03i}. Since then it has been used widely both in academia and industry to define specifications of software systems. It is well-suited to formalise requirements and natural language specifications and to find defects. For example, the FeliCa contactless card technology, which is widely used in Japan, was developed using VDM~\cite{Fitzgerald&08a}. A specification was constructed in VDM that revealed a large number of defects (278) in the existing natural-language requirements and specifications. The VDM specification was used to generate test cases and as a reference when writing the \cpp code that was eventually deployed to millions of devices.

The VDM Specification Language (VDM-SL) was standardised as ISO/IEC 13817-1 in 1996~\cite{ISOVDM96a}. Developments beginning in the 1990s extended the language to cover object-orientation (\vpp~\cite{Fitzgerald&05}) and later to include abstractions for modelling real-time embedded software (VDM-RT~\cite{Verhoef&06b}). All three dialects are supported by two robust tools, the commercial VDMTools~\cite{Larsen01} and the open-source Overture\footnote{\url{http://www.overturetool.org/}} tool~\cite{Larsen&10a}. These tools offer type checking for VDM models, a number of analysis tools such as combinatorial testing~\cite{Larsen&10c}, and interpretation of an executable subset of VDM that allows models to be simulated. By connecting a graphical interface to an executable model, it is also possible to animate specifications~\cite{Fitzgerald&08a}, thereby allowing non-specialists to gain an understanding of the system described by the specification through interaction and interrogation of the model.

The models of our case study were developed in VDM-SL (using the Overture tool) and the remainder of this section uses that dialect. As part of the standardisation process, a full denotational semantics has been defined for VDM-SL~\cite{Larsen&95c}, as well as a proof theory and a comprehensive set of proof rules~\cite{Bicarregui&94}. Proofs in VDM typically verify internal consistency or are proofs of refinement~\cite{Jones90a}.

We proceed as follows: the VDM formalism is described in more detail in Section~\ref{sec:vdm-sl}, the modelling of the case study is described in Section~\ref{sec:vdm_approach},
the analysis of the model is described in Section~\ref{sec:vdm_analysis},
an LTS semantics of the model is defined in Section~\ref{sec:vdm_semantics}, and possible extensions to the model are described in Section~\ref{sec:vdm_extensions}. An evaluation of the model and formalism for describing reconfiguration is given in Section~\ref{sec:comp}.


\subsection{Formalism}
\label{sec:vdm-sl}

Specifications in VDM-SL are divided into modules, where each module contains a set of definitions. Modules can export definitions to, and import definitions from, other modules in the model. The definitions in a module are divided into distinct sections and preceded by a keyword. Definitions can include types, values, functions, state, and operations. Listing~\ref{lst:module} shows an empty VDM-SL module specification, divided into sections. 
We give an overview of the definitions later in this section, and further detail is introduced as required during explanation of the model.

\begin{lstlisting}[style=styleVDM,caption={Blank VDM-SL module definition},label={lst:module},float]
module MyModule

exports ...
imports ...

definitions

types
...

functions
...

state ... of
...
end

operations
...

values
...

end MyModule
\end{lstlisting}

A key part of the VDM-SL language is the powerful type system. VDM-SL contains a number of built-in scalar types including \textbf{\texttt{boolean}}s, numeric types, and characters. Non-scalar types include \textbf{\texttt{set}}s, \textbf{\texttt{seq}}uences, and \textbf{\texttt{map}}s. Custom types can be defined in the \textbf{\texttt{types}} section, based on built-in types and including type unions and record types with named fields. Custom types can be restricted by invariants and violations of these \textbf{\texttt{inv}}ariants can be flagged during interpretation of the models.

\emph{Functions} are pure and have no side effect. They can be defined implicitly (by pre- and post-conditions) or explicitly. Explicit functions can also be protected with a pre-condition. Only explicit functions can be interpreted by the tools. \emph{State} allows one or more global variables to be defined for the module. An invariant can be defined over the state to restrict its values. Again, invariant violations can be flagged during interpretation by the tool. \emph{Operations} are functions that are additionally able to read and write state variables. Therefore, operations can have side effects. Like functions, operations can be defined implicitly or explicitly. 
\emph{Values} define constants that can be used in functions and operations.


\subsection{Modelling}
\label{sec:vdm_approach}

In creating a specification that meets the workflow requirements given in Section~\ref{sec:wfactivs}, two approaches are possible in the VDM `idiom'.
The first is to build a data model that captures an order and its status,
with operations and pre-conditions ensuring that only valid transitions between statuses are possible
(e.g.\ order receipt to inventory check). Such a model could also include details of customers and suppliers.
The second approach is to model the entire workflow as sequences of actions and build an `interpreter' within the model to execute and reconfigure the workflow.

These approaches are not orthogonal; a data model would complement a workflow model.
The resulting specification would to be closer to any code to be written in implementing the order system than either approach separately.
In this paper we follow the second approach as this more closely matches the style of the requirements on workflows described in Section~\ref{sec:wfactivs}.
We return to the data model approach in Section~\ref{sec:vdm_extensions}, giving an example of how the model could be extended to incorporate a data model for orders.






As a general purpose modelling language, VDM-SL does not have built-in notions of concurrency or threading, nor does it have dedicated abstractions for modelling processes.
Following a standard VDM-SL paradigm to keep the model small, a single workflow was modelled,
with concurrency of the parallel actions modelled as non-deterministic interleaving; this is Design 1 from Section~\ref{sec:wfactivs}.
Reconfiguration is achieved by an operation that swaps a workflow during execution of the interpeter;
this is reconfiguration Case 1 as described in Section~\ref{sec:intro}.
In Section~\ref{sec:vdm_extensions} we describe extensions of the model that would facilitate exploration of Designs 2--4,
including features of the other VDM dialects (\vpp and VDM-RT).

Analysis of the model is done through testing. This is a common way of using VDM in industry~\cite{Agerholm:1998:FSV:298595.298861,Fitzgerald&08a},
using the formal model to record and test assumptions, then using the resulting model as a specification when writing code.
Tests are defined for manually checking all valid configurations, and for testing valid and invalid reconfigurations.
Section~\ref{sec:vdm_extensions} explains extensions that could facilitate greater automation in testing workflows.

The following subsections explain the modelling process in detail.
The model is split into three modules: \texttt{Configurations}, containing workflow definitions;
\texttt{Interpreter}, containing operations to execute and reconfigure workflows; and \texttt{Test}, which defines test cases for the model.

The contents of the modules are described as follows:
first, a set of types is defined that can capture the two workflows in the \texttt{Configurations} module.
Next, an interpreter is defined in the \texttt{Interpreter} module that can `execute' a workflow\footnote{Notice that the Overture tool
\emph{interprets} a VDM-SL specification in the sense that it does not perform a compilation beforehand.
We use the term \emph{execute} to describe what the interpreter in our model does to a workflow.} and be reconfigured during execution.
In order to test all possible paths through a configuration, a method for setting the outcome of external choices
(inventory check, credit check, and supplier check) is included.
The reconfiguration operation is presented in two forms:
the first allows reconfiguration to any arbitrary configuration; the second extends this with a pre-condition to permit only safe reconfiguration.
Finally, the \texttt{Test} module is described that includes operations to test all possible paths through the two configurations (in Section~\ref{sec:vdm_analysis}).


\subsection*{Workflows and Traces}
\label{sec:vdm_workflows}


The \texttt{Configurations} module defines types that are used to represent the workflows from Figures~\ref{OfficeWorkflowConfig1} and~\ref{OfficeWorkflowConfig2}
and are used by the interpreter (shown later). The module also defines two constants that instantiate these workflows,
a type to represent a trace of a workflow execution, and some useful auxiliary functions.
The types, functions, and values of this module are made available to both the other modules using the \texttt{\textbf{exports all}} declaration:

\begin{lstlisting}[style=styleVDM,caption={\texttt{Configurations} module definition},label={lst:configurations}]
module Configurations

exports all

definitions

...

end Configurations
\end{lstlisting}

The types for capturing workflows and their traces are built around a core type called \texttt{Action},
which enumerates all possible actions in a workflow (and therefore also in a trace).
Any action must be exactly one of the nine listed values. The \emph{type union} is defined using the pipe (\texttt{|}) operator,
and the individual values are \emph{quote} types (basic values that can only be compared for equality):

\begin{lstlisting}[style=styleVDM,caption={\texttt{Action} type definition},label={lst:action}]
types

-- actions in a workflow
Action = <OrderReceipt> | <InventoryCheck> | <Reject>
       | <CreditCheck> | <SupplierCheck> | <Shipping>
       | <Billing> | <Archiving> | <Confirmation>;
\end{lstlisting}


Based on this type, we define a trace as a sequence of events recording either the occurrence of an action,
or a special \texttt{<TERMINATE>} event indicating successful completion of a workflow.
The invariant on \texttt{Trace} states that if a termination event occurs, it must occur at the end
(i.e.\ if it appears in a trace of $n$ elements and is not the only element, then it does not appear in the first $n - 1$ elements):

\begin{lstlisting}[style=styleVDM,caption={\texttt{Event} and \texttt{Trace} type definitions},label={lst:event}]
types

-- record of an action or termination
Event = Action | <TERMINATE>;

-- trace of events
Trace = seq of Event
inv t == (<TERMINATE> in set elems t and len t > 1) =>
  <TERMINATE> not in set elems t(1, ..., len t - 1);
\end{lstlisting}

To define a workflow type, it is necessary to allow an order for actions to be specified;
this could be done with a sequence. However, in this study a recursive definition is used based around a type called \texttt{Workflow}:

\begin{lstlisting}[style=styleVDM,caption={\texttt{Workflow} type definition},label={lst:element}]
types

-- workflow with invariant
Workflow = Element
inv w == forall tr in set tracesof(w) & card elems tr = len tr;

-- elements that make up a workflow
Element =  [Simple | Branch | Par];
\end{lstlisting}

This definition states that an \texttt{Element} can be one of three other types (expanded below):
\texttt{Simple} represents a single transition such as order receipt to inventory check;
\texttt{Branch} represents an \emph{OK?} choice, such as the credit check;
and \texttt{Par} represents parallel composition.
The square brackets make the type \emph{optional}, meaning that it can take a fourth special value (\textbf{\texttt{nil}})
that represents termination (the black circles in Figures~\ref{OfficeWorkflowConfig1} and~\ref{OfficeWorkflowConfig2}).

The \texttt{Element} type can be used `as-is' to represent workflow configurations, but it is not restricted in any way.
For example, it can contain repeated actions (i.e.\ billing or shipping twice).
Therefore, we introduce a \texttt{Workflow} type\footnote{Separating the \texttt{Element} and \texttt{Workflow} definitions is necessary
in order to allow the Overture tool to check the invariant at runtime.} with an invariant that prevents duplicates
(by checking that for all possible traces of the workflow, the cardinality of the set of events in the trace is the same as the length of the trace).

The three types of element are defined as follows:

\begin{lstlisting}[style=styleVDM,caption={\texttt{Simple}, \texttt{Branch}, and \texttt{Par} type definitions},label={lst:elements}]
types

-- a simple element
Simple :: a : Action
          w : Workflow;

-- a conditional element
Branch ::  a : Action
           t : Workflow
           f : Workflow;

-- parallel elements
Par :: b1 : Action
       b2 : Action
        w : Workflow;
\end{lstlisting}

The above three definitions are \emph{record} types, that is, compound types with named elements.
Each type contains one or more actions to be executed, and one or more elements that follow this action.
Therefore, the definitions are recursive, and the recursion is terminated by a \textbf{\texttt{nil}} value at each leaf.
A simple workflow (called \texttt{T}) that rejects all orders could be defined as:

\begin{lstlisting}[style=styleVDM,caption={Example workflow definition},label={lst:rejectall}]
values

T = mk_Simple(<OrderReceipt>, mk_Simple(<Reject>, nil))
\end{lstlisting}

Notice that the \texttt{mk\textunderscore} keyword is a \emph{constructor} used to instantiate values of record types.
They are essentially (automatically defined) functions that construct a record with the parameters being assigned, in order, to named elements.

%

The \texttt{Configurations} module also defines two auxiliary functions that are useful for invariants and pre-conditions seen later.
The first (\texttt{prefixof}) determines if one trace is a prefix of another and the second (\texttt{tracesof}) recursively computes all traces of an element:

\begin{lstlisting}[style=styleVDM,caption={Headers of the auxiliary functions \texttt{prefixof} and \texttt{tracesof}},label={lst:prefixof}]
functions

-- true if a is a prefix of b, false otherwise
prefixof: Trace * Trace -> bool
prefixof(a, b) == ...

-- compute all traces of an element
tracesof: Element -> set of Trace
tracesof(el) == ...
\end{lstlisting}

%

%
%

%

Finally, the module defines values (constants) that describe the two configurations from Section~\ref{sec:wfactivs},
called \texttt{Configuration1} and \texttt{Configuration2} respectively:

\begin{lstlisting}[style=styleVDM,showlines=true,caption={Definitions of \texttt{Configuration1} and \texttt{Configuration2}},label={lst:config1and2}]
values

-- first configuration                         -- second configuration
Configuration1: Workflow =                     Configuration2: Workflow =
  mk_Simple(<OrderReceipt>,                      mk_Simple(<OrderReceipt>,
    mk_Branch(<InventoryCheck>,                    mk_Branch(<InventoryCheck>,
      mk_Branch(<CreditCheck>,                       mk_Branch(<CreditCheck>,
        mk_Simple(<Shipping>,                          mk_Par(<Billing>, <Shipping>,
          mk_Simple(<Billing>,                           mk_Simple(<Archiving>, nil)),
            mk_Simple(<Archiving>,                     mk_Simple(<Reject>, nil)
              mk_Simple(<Confirmation>, nil)         ),
            )                                        mk_Branch(<SupplierCheck>,
          )                                            mk_Branch(<CreditCheck>,
        ),                                               mk_Par(<Billing>, <Shipping>,
        mk_Simple(<Reject>, nil)                           mk_Simple(<Archiving>, nil)),
      ),                                                 mk_Simple(<Reject>, nil)
      mk_Simple(<Reject>, nil)                         ),
    )                                                  mk_Simple(<Reject>, nil)
  );                                                 )
                                                   )
                                                 );

\end{lstlisting}

%
%
%

\subsection*{Interpreter}
\label{sec:vdm_interpreter}

The \texttt{Interpreter} module allows a workflow to be executed.
The module exports its definitions so that they can be accessed by the \texttt{Test} module,
and it imports required type definitions from the \texttt{Configurations} module,
including the definitions of actions, the \texttt{prefixof} and \texttt{tracesof} auxiliary functions,
and the values of \texttt{Configuration1} and \texttt{Configuration2}.

\begin{lstlisting}[style=styleVDM,caption={\texttt{Interpreter} module definition},label={lst:interpreter}]
module Interpreter

exports all

imports from Configurations types Workflow, Trace, ...
\end{lstlisting}


In the VDM idiom, models typically have persistent state, and operations that alter this state.
The state of the \texttt{Interpreter} module records the trace of the execution so far (\texttt{trace}) and the remaining workflow to be executed  (\texttt{workflow}).
A state is similar to a record type and is defined in a similar manner (the state definition acts as its own section type):

\begin{lstlisting}[style=styleVDM,caption={\texttt{Interpreter} state definition},label={lst:stateS}]
-- interpreter state
state S of
     trace : Trace
  workflow : Workflow
init s == s = mk_S([], nil)
end;
\end{lstlisting}

The above state definition contains an \texttt{init} clause that gives initial values to both components of the state (they are both `empty').
An invariant can also be defined with an \texttt{inv} clause.
The module provides operations to set (and reset) the state of the interpreter,
to step through execution of a workflow or execute it in a single step, and to access the current value of the trace.
Operations for reconfiguration are also included and are described below (see \textit{Reconfiguration}).

An operation called \texttt{Init} is used to prime (set and reset) the interpreter with a workflow passed as a parameter and an empty trace:

\begin{lstlisting}[style=styleVDM,caption={\texttt{Init} operation definition},label={lst:init}]
operations

Init: Workflow ==> ()
Init(w) == (
    trace := [];
    workflow := w
);
\end{lstlisting}

The basic operation of the interpreter is to move an action from the head of \texttt{workflow} and append it to the end of the trace.
Once the workflow is empty, the \texttt{<TERMINATE>} element is added to the trace and the execution ends.
Since there is no data model underlying the workflow, no additional work is done when moving an action from the workflow to the trace.
However, an extension is considered towards this in Section~\ref{sec:vdm_extensions}.


The absence of a data model also means that the external choices in the workflow (the inventory check, credit check, and supplier check) must be made in some other manner.
The main analysis method for this model is testing (described in Section~\ref{sec:vdm_analysis}).
Therefore, it is desirable to be able to control these external choices to ensure test coverage.
In order to do this, a \texttt{Choices} type is introduced, which is a mapping from (choice) actions to Boolean.
The invariant ensures that the domain of the map is exactly the set of actions that represent external choices:

\begin{lstlisting}[style=styleVDM,caption={\texttt{Choices} type definition},label={lst:choices}]
types

-- collapse probabilities
Choices = map Action to bool
inv c == dom c = {<InventoryCheck>, <CreditCheck>, <SupplierCheck>}
\end{lstlisting}

For example, a run of the workflow where there is sufficient inventory and sufficient credit can be achieved using the \texttt{Choices} given below.
Notice that in this case a supplier check will not be needed because there is sufficient stock:


\begin{lstlisting}[style=styleVDM,caption={\texttt{NoProblems} value definition},label={lst:noproblems}]
values

-- all branches true
NoProblems = {<InventoryCheck> |-> true,
               <SupplierCheck> |-> true,
                 <CreditCheck> |-> true};
\end{lstlisting}


The \texttt{Interpreter} module defines two operations that execute a workflow, \texttt{Step} and \texttt{Execute}, with the following signatures:

\begin{lstlisting}[style=styleVDM,caption={\texttt{Step} and \texttt{Execute} operation headers},label={lst:stephead}]
operations

-- perform a single step of the interpreter
Step: Choices ==> Event

-- execute workflow in one go
Execute: Choices ==> ()
\end{lstlisting}

The \texttt{Step} operation performs a single step of execution, updating the trace and moving to the next step of the workflow.
This operation selects the outcome of \texttt{Branch} elements based on the \texttt{Choices} passed as a parameter,
and the order of execution of actions in \texttt{Par} elements are selected randomly leading to interleaving of the actions.
The \texttt{Step} operation returns the last event that occurred, which is used for reconfiguration (described below).
The \texttt{Execute} operation uses \texttt{Step} operation to execute a workflow and produce a full trace.
The \texttt{\textbf{let}} expression is used to ignore the value returned by \texttt{Step}, since it is not needed for a simple execution run:


\begin{lstlisting}[style=styleVDM,caption={\texttt{Execute} operation definition},label={lst:interpret}]
operations

-- execute workflow in one go
Execute: Choices ==> ()
Execute(c) == (
    while workflow <> nil do
    	let - = Step(c) in skip;
    trace := trace ^ [<TERMINATE>]
);
\end{lstlisting}

\subsection*{Reconfiguration}
\label{sec:vdm_reconfiguration}


Reconfiguration is enabled by an operation in the \texttt{Interpreter} module.
This operation replaces the current workflow in the state by another workflow during execution.
We consider the case of a single thread of execution moving from some point in Configuration1 to an appropriate point in Configuration2, with extensions discussed later.

The following operation is defined in the \texttt{Interpreter} module that replaces the workflow in the state by the workflow passed as a parameter to the operation.
The point at which this operation is called, and the workflow passed to the operation, are left to the caller.
In this way, the model is able to capture unplanned reconfiguration.

\begin{lstlisting}[style=styleVDM,caption={Unprotected \texttt{Reconfigure} operation definition},label={lst:reconfigure}]
operations

-- reconfigure, replacing current workflow
Reconfigure: Workflow ==> ()
Reconfigure(w) ==
	workflow := w;
\end{lstlisting}

In this unprotected form, the calling thread is able to make arbitrary changes to the workflow, resulting in traces that do not meet the requirements described earlier.
For example, double billing a customer by reconfiguring to a workflow with a \texttt{<Billing>} element after billing had already occurred.
This is avoided by adding an invariant to the state that disallows configurations that could generate illegal traces.
Additionally, a pre-condition is added to the \texttt{Reconfigure} operation to protect the invariant, ensuring that the operation only processes valid reconfigurations.
Ideally, invariants should be protected by pre-conditions on operations in this fashion, so that invariants form a `last line of defence'.

It is a requirement that traces produced by the interpreter must be traces of Configuration 1 or Configuration 2.
Therefore, the invariant states that, given the current trace (which can be empty),
the remaining workflow can only produce traces valid under Configuration 1 or Configuration 2.
Similarly, the pre-condition checks that the new workflow can only produce valid traces of Configuration 2
(since we currently only consider reconfigurations from Configuration 1 to Configuration 2).
With the pre-condition added, the \texttt{Reconfigure} operation is defined as follows:


\begin{lstlisting}[style=styleVDM,caption={\texttt{Reconfigure} operation with pre-condition},label={lst:reconfigurepre}]
-- reconfigure, replacing current workflow
Reconfigure: Workflow ==> ()
Reconfigure(w) ==
	workflow := w
pre w <> nil and
    branch_check(trace, workflow, w) and
    forall t in set {trace ^ ftr | ftr in set tracesof(w)} &
	  (exists x in set tracesof(Configuration2) & t = x);
\end{lstlisting}

The pre-condition requires first that the new workflow is not empty.
Second, the \texttt{branch\_check} auxiliary function (described below) is used to check that the reconfiguration takes account of the outcome of branching actions.
Finally, the \texttt{tracesof} auxiliary function is used to check that
for all traces in the set of traces produced by appending the possible traces of the new configuration to the current trace,
that the resulting trace is a valid trace of Configuration 2.
The \texttt{\&} denotes `such that' in these quantifications.
Therefore, all traces that could occur after reconfiguration are valid under Configuration 2.

The \texttt{branch\_check} auxiliary function is defined as follows:

\begin{lstlisting}[style=styleVDM,caption={\texttt{branch\_check} auxiliary function},label={lst:branch_check}]
branch_check: Trace * Workflow * Workflow -> bool
branch_check(tr, w, w') ==
  (last(tr) = <InventoryCheck> and first(w) = {<Reject>}
    => {<SupplierCheck>} subset first(w')) and
  (last(tr) = <InventoryCheck> and first(w) = {<CreditCheck>}
    => {<CreditCheck>} subset first(w')) and
  (last(tr) = <CreditCheck> and first(w) = {<Reject>} 	
    => {<Reject>} subset first(w')) and
  (last(tr) = <CreditCheck> and first(w) = {<Shipping>}
    => first(w') subset {<Billing>, <Shipping>});
\end{lstlisting}

The above check is necessary in order to take account of the outcome of a branching action.
For example, if the inventory check passes, then the external supplier check should not be performed.
Since actions are not parameterised in this model, the only way to tell which branch (true or false) was taken by the action
is to examine the first element of the remaining workflow,
and to check that the first element of the new workflow is a valid replacement.
The \texttt{branch\_check} function explicitly encodes this checking for the inventory check and credit check.
A more general solution is to pass parameters to workflow actions (e.g.\ to indicate the outcome of an evaluation action)
and to pass a verification condition parameter to \texttt{Reconfigure}. Extensions are discussed in Section~\ref{sec:vdm_extensions}.

An invariant is added to the state that is defined similarly to the pre-condition of \texttt{Reconfigure}.
However, the invariant requires that the traces must belong to either \texttt{Configuration1} or \texttt{Configuration2},
and \texttt{prefixof} is used rather than equality since the trace is incomplete in the intermediate state.

\begin{lstlisting}[style=styleVDM,caption={\texttt{Interpreter} state definition with invariant},label={lst:stateSinv}]
-- interpreter state
state S of
     trace : Trace
  workflow : Workflow
init s == s = mk_S([], nil)
inv mk_S(trace, workflow) ==
  workflow <> nil =>
    (forall t in set {trace ^ tr | tr in set tracesof(workflow)} &
      (exists x in set tracesof(Configuration1) union tracesof(Configuration2) &
        prefixof(t, x)))
end;
\end{lstlisting}


\subsection{Analysis}
\label{sec:vdm_analysis}

The \texttt{Test} module defines operations that test both configurations with the five combinations of external choices.
These operations call the operations of the \texttt{Interpreter} module (using the back tick operator: \texttt{`}) and output the trace,
which can then be printed to the console in Overture.
For example, the operation that tests \texttt{Configuration1} with the \texttt{NoProblems} choices is defined below.
This operation initialises the interpreter with \texttt{Configuration1}, executes the interpreter and returns the completed trace:


\begin{lstlisting}[style=styleVDM,caption={\texttt{Config1NoProblems} operation definition},label={lst:config1noproblems}]
module Test

operations

-- Test Configuration 1 / NoProblems
Config1NoProblems: () ==> Trace
Config1NoProblems() == (
	Interpreter`Init(Configuration1);
	Interpreter`Execute(NoProblems);
	return Interpreter`GetTrace()
);
\end{lstlisting}

When printed to the console, the output of the \texttt{Config1NoProblems} operation shows the following trace:


\begin{lstlisting}[style=styleVDM,caption={Output of \texttt{Config1NoProblems} definition},label={lst:config1noproblemsout}]
Test`Config1NoProblems() =
  [<OrderReceipt>, <InventoryCheck>, <CreditCheck>,
   <Shipping>, <Billing>, <Archiving>, <Confirmation>,
   <TERMINATE>]
\end{lstlisting}

Tests are also defined for the other combinations of choices, and named accordingly.
For example, the test where the \texttt{Choices} map yields false for the credit check is called \texttt{Config1NoCredit} (see Appendix~\ref{sec:appendix:vdm}).

In order to test the reconfiguration operation, and demonstrate the outcome of a valid and invalid reconfiguration request,
the test module defines an operation called \texttt{TestReconfig}, that takes the \texttt{Choices} required for execution (\texttt{c}),
an action (\texttt{rp}) and a workflow (\texttt{w}) as parameters:

\begin{lstlisting}[style=styleVDM,caption={\texttt{TestReconfig} operation signature},label={lst:testreconfig}]
operations

TestReconfig: Choices * Action * Workflow ==> ()
TestReconfig(c, rp, w) == ...
\end{lstlisting}

The operation initialises the interpreter, then steps through the execution until the action \texttt{rp} is seen,
then attempts to reconfigure the interpreter to the workflow \texttt{w}.
If the reconfiguration is valid under the requirements, the final trace will be printed.
Otherwise, a message is printed stating that the reconfiguration is invalid (and that the pre-condition would fail if execution continued).
In the model as presented, the \texttt{w} passed to the operation is constructed manually as a suffix of \texttt{Configuration2}.
This is a weakness of the model. Discussion of automating such testing appears in Section~\ref{sec:vdm_extensions} below.

Using \texttt{TestReconfig}, two operations are defined that demonstrate a valid and invalid reconfiguration respectively.
The first, \texttt{TestReconfigSuccess}, reconfigures from \texttt{Configuration1} to \texttt{Configuration2} after the inventory check
(where there is no inventory in stock, so a supplier check is performed). This is a valid reconfiguration, and the console output is as follows:

\begin{lstlisting}[style=styleVDM,caption={Output of \texttt{TestReconfigSuccess} operation},label={lst:testreconfigsuccess}]
[<OrderReceipt>, <InventoryCheck>]
Reconfiguring Configuration1 <@to@> Configuration2...
[<OrderReceipt>, <InventoryCheck>, <SupplierCheck>,
 <CreditCheck>, <Billing>, <Shipping>, <Archiving>,
 <TERMINATE>]
\end{lstlisting}

The second operation, \texttt{TestReconfigFail},
attempts to reconfigure from \texttt{Configuration1} to the parallel composition of shipping and billing in \texttt{Configuration2} after shipping has already occurred.
This is an invalid reconfiguration, since shipping will occur twice. The output on the console is as below:

\begin{lstlisting}[style=styleVDM,caption={Output of \texttt{TestReconfigFail} operation},label={lst:testreconfigfail}]
[<OrderReceipt>, <InventoryCheck>, <CreditCheck>,
 <Shipping>]
Reconfiguring Configuration1 <@to@> Configuration2...
These potential <@traces@> are <@not@> valid under Configuration2:
* [<OrderReceipt>, <InventoryCheck>, <CreditCheck>,
   <Shipping>, <Billing>, <Shipping>, <Archiving>]
* [<OrderReceipt>, <InventoryCheck>, <CreditCheck>,
   <Shipping>, <Shipping>, <Billing>, <Archiving>]
\end{lstlisting}

Notice that the Overture tool terminates with an error due to pre-condition failure:

\begin{lstlisting}[style=styleVDM,caption={Pre-condition failure reported by the Overture tool},label={lst:prefail}]
Reconfiguration could generate invalid <@traces@>; <@pre@>-
condition will fail.

<@\color{red}Error 4071: Precondition failure: pre\_Reconfigure in@>
<@\color{red}`Interpreter'@>
\end{lstlisting}


\subsection{LTS Semantics}
\label{sec:vdm_semantics}

As part of the ISO standard, a full denotational semantics has been defined for VDM-SL~\cite{Larsen&95c,ISOVDM96a}.
Structured operational semantics (SOS) have also been defined for some of the new language features introduced in newer dialects~\cite{Lausdahl&13a}.
However, in order to facilitate comparison with the other two formalisms, we develop a labelled transition system (LTS) semantics for the VDM model.
Such an LTS would be unwieldy for the whole semantics defined in the ISO standard, so we restrict this LTS to this model of the case study.
This means that if the model were to be changed, the LTS would have to be updated.
In the general case, this is unwieldy and is not typically part of a VDM development, but it is instructive in this instance.

The LTS rules are given in Table~\ref{tab:vdmlts}.
The rules are relations that define transitions between tuples of type $(Trace \times Workflow)$, corresponding to the state (\texttt{S}) of the interpreter.
The labels on the transitions correspond to the actions that the workflow performs (which are appended to the trace),
or $\tau$ for unobservable steps (that do not append items to the trace).
The LTS rules are defined using the \texttt{choices} and \texttt{pre\_Reconfigure} functions defined in the VDM model
in order to render the rule definitions concise and human-readable.

The Init rule states that the interpreter can be given a workflow when it does not currently have one and its trace is empty.
This is the initial state of S as defined above: \texttt{\textbf{init} s == s = \textbf{mk\_S}([], \textbf{nil})}.
The Terminate rule allows the interpreter to terminate when it has no workflow left.
The Reset rule allows a terminated interpreter to be reset with another workflow.

The Simple, Branch-T, Branch-F, Par-1 and Par-2 rules encode the logic of the \texttt{Step} operation introduced  above (though not given in full).
They correspond to the three elements that form the \texttt{Workflow} type.
The action of a \texttt{Simple} element can always happen.
The action performed by a \texttt{Branch} action depends on the \texttt{choices} parameter that is passed to the \texttt{Execute} operation.
For the \texttt{Par} element, the two rules have no hypotheses and therefore it is non-deterministic choice of which action is performed first.
These rules make it clear that this model has an interleaving semantics.
The Reconfigure rule states that the workflow that is still to be executed can be replaced in one atomic step,
assuming that the new workflow will not violate the pre-condition of the \texttt{Reconfigure} operation.

We give three examples of application of the LTS rules in Figure~\ref{fig:vdmlts:all}.
The first (top) demonstrates a complete run of the workflow and is equivalent to the \texttt{Config1NoProblems} test.
The second (middle) demonstrates a run in which the credit check yields false and is equivalent to the \texttt{Config1NoCredit} test.
The third (bottom) demonstrates a reconfiguration from Configuration 1 to Configuration 2 after the inventory check
and is equivalent to the \texttt{TestReconfigSuccess} test.
Model checking of the reconfiguration is facilitated by the trace of actions and the LTS rule applications, and is shown in Appendix \ref{sec:appendix:vdmmc}.

\begin{table}\renewcommand{\arraystretch}{2.5}
\[
\begin{array}{|llll|}
\hline
\;\; \mbox{Init} &
\frac{w\;\in\;Workflow}
{([],\;\textbf{nil})\;\trans{\tau}\;([],\;w)} & & \\
\;\; \mbox{Terminate} &
\frac{<TERMINATE>\;\notin\;\textbf{elems}\;tr}
{(tr,\;\textbf{nil})\;\trans{\tau}\;(tr^{\;\wedge\;}[<TERMINATE>],\;\textbf{nil})} & & \\
\;\; \mbox{Reset} &
\frac{<TERMINATE>\;\in\;\textbf{elems}\;tr}
{(tr,\;\textbf{nil})\;\trans{\tau}\;([],\;\textbf{nil})} & & \\
\;\; \mbox{Simple} &
\frac{}
{(tr,\;mk\_Simple(a,\;e))\;\trans{a}\;(tr^{\;\wedge\;}[a],\;e)} & & \\
\;\; \mbox{Branch-T} &
\frac{choices(a)}
{(tr,\;mk\_Branch(a,\;t,\;f))\;\trans{a}\;(tr^{\;\wedge\;}[a],\;t)} &
\;\;\;\;\;\;\;\; \mbox{Branch-F} &
\frac{\neg choices(a)}
{(tr,\;mk\_Branch(a,\;t,\;f))\;\trans{a}\;(tr^{\;\wedge\;}[a],\;f)} \;\; \\
\;\; \mbox{Par-1} &
\frac{}
{(tr,\;mk\_Par(b1,\;b2,\;e))\;\trans{b1}\;(tr^{\;\wedge\;}[b1],\;mk\_Simple(b2,\;e))} &
\;\;\;\;\;\;\;\; \mbox{Par-2} &
\frac{}
{(tr,\;mk\_Par(b1,\;b2,\;e))\;\trans{b2}\;(tr^{\;\wedge\;}[b2],\;mk\_Simple(b1,\;e))} \;\; \\[2ex]
\;\; \mbox{Reconfigure} &
\frac{w'\;\in\;Workflow\;\wedge\;pre\_Reconfigure(tr,\;w')}
{(tr,\;w)\;\trans{\tau}\;(tr,\;w')} & & \\[2ex]
\hline
\end{array}
\]
\caption{Labelled transition system semantics of the basic VDM model}\label{tab:vdmlts}\renewcommand{\arraystretch}{1.0}
\end{table}

\begin{figure}
\setlength{\tabcolsep}{3pt}\scriptsize
\begin{tabular}{lll}
$([],\;$\texttt{\textbf{nil}}$)$ & & \\
$\trans{\tau}$                                 & $([], mk\_Simple(\langle{}OrderReceipt\rangle, ...))$
    & by Init \\
$\xrightarrow{\langle{}OrderReceipt\rangle}$   & $([\langle{}OrderReceipt\rangle], mk\_Branch(\langle{}InventoryCheck\rangle, ...))$
    & by Simple \\
$\xrightarrow{\langle{}InventoryCheck\rangle}$ & $([\langle{}OrderReceipt\rangle, \langle{}InventoryCheck\rangle], mk\_Branch(\langle{}CreditCheck\rangle, ...))$
    & by Branch-T \\
$\xrightarrow{\langle{}CreditCheck\rangle}$    & $([\langle{}OrderReceipt\rangle, \langle{}InventoryCheck\rangle, \langle{}CreditCheck\rangle], mk\_Simple(\langle{}Shipping\rangle, ...))$
    & by Branch-T \\
$\xrightarrow{\langle{}Shipping\rangle}$       & $([\langle{}OrderReceipt\rangle, \langle{}InventoryCheck\rangle, \langle{}CreditCheck\rangle, \langle{}Shipping\rangle], mk\_Simple(\langle{}Billing\rangle, ...))$
    & by Simple \\
$\xrightarrow{\langle{}Billing\rangle}$        & $([\langle{}OrderReceipt\rangle, \langle{}InventoryCheck\rangle, \langle{}CreditCheck\rangle, \langle{}Shipping\rangle, \langle{}Billing\rangle], mk\_Simple(\langle{}Archiving\rangle, ...))$
    & by Simple \\
$\xrightarrow{\langle{}Archiving\rangle}$      & $([\langle{}OrderReceipt\rangle, \langle{}InventoryCheck\rangle, \langle{}CreditCheck\rangle, \langle{}Shipping\rangle, \langle{}Billing\rangle, \langle{}Archiving\rangle], mk\_Simple(\langle{}Confirmation\rangle, \texttt{\textbf{nil}}))$
    & by Simple \\
$\xrightarrow{\langle{}Confirmation\rangle}$   & $([\langle{}OrderReceipt\rangle, \langle{}InventoryCheck\rangle, \langle{}CreditCheck\rangle, \langle{}Shipping\rangle, \langle{}Billing\rangle, \langle{}Archiving\rangle, \langle{}Confirmation\rangle], \texttt{\textbf{nil}})$
    & by Simple \\
$\trans{\tau}$                                 & $([\langle{}OrderReceipt\rangle, \langle{}InventoryCheck\rangle, \langle{}CreditCheck\rangle, \langle{}Shipping\rangle, \langle{}Billing\rangle, \langle{}Archiving\rangle, \langle{}Confirmation\rangle, \langle{}TERMINATE\rangle], \texttt{\textbf{nil}})$
    & by Terminate \\
$\trans{\tau}$                                 & $([],\;$\texttt{\textbf{nil}}$)$ & by Reset \\
\end{tabular}

\bigskip
\begin{tabular}{lll}
$([],\;$\texttt{\textbf{nil}}$)$ & & \\
$\trans{\tau}$                                    & $([], mk\_Simple(\langle{}OrderReceipt\rangle, ...))$
    & by Init \\
$\xrightarrow{\langle{}OrderReceipt\rangle}$      & $([\langle{}OrderReceipt\rangle], mk\_Branch(\langle{}InventoryCheck\rangle, ...))$
    & by Simple \\
$\xrightarrow{\langle{}InventoryCheck\rangle}$    & $([\langle{}OrderReceipt\rangle, \langle{}InventoryCheck\rangle], mk\_Branch(\langle{}CreditCheck\rangle, ...))$
    & by Branch-T \\
$\xrightarrow{\langle{}CreditCheck\rangle}$       & $([\langle{}OrderReceipt\rangle, \langle{}InventoryCheck\rangle, \langle{}CreditCheck\rangle], mk\_Simple(\langle{}Reject\rangle, \texttt{\textbf{nil}}))$
    & by Branch-F \\
$\xrightarrow{\langle{}Reject\rangle}$            & $([\langle{}OrderReceipt\rangle, \langle{}InventoryCheck\rangle, \langle{}CreditCheck\rangle, \langle{}Reject\rangle], \texttt{\textbf{nil}}))$
    & by Simple \\
$\trans{\tau}$                                    & $([\langle{}OrderReceipt\rangle, \langle{}InventoryCheck\rangle, \langle{}CreditCheck\rangle, \langle{}Reject\rangle, \langle{}TERMINATE\rangle], \texttt{\textbf{nil}})$
    & by Terminate \\
$\trans{\tau}$                                    & $([],\;$\texttt{\textbf{nil}}$)$
    & by Reset \\
\end{tabular}

\bigskip
\begin{tabular}{lll}
$([],\;$\texttt{\textbf{nil}}$)$ & & \\
$\trans{\tau}$                                    & $([], mk\_Simple(\langle{}OrderReceipt\rangle, ...))$
    & by Init \\
$\xrightarrow{\langle{}OrderReceipt\rangle}$      & $([\langle{}OrderReceipt\rangle], mk\_Branch(\langle{}InventoryCheck\rangle, ...))$
    & by Simple \\
$\xrightarrow{\langle{}InventoryCheck\rangle}$    & $([\langle{}OrderReceipt\rangle, \langle{}InventoryCheck\rangle], mk\_Simple(\langle{}Reject\rangle,\texttt{\textbf{nil}}))$
    & by Branch-F \\
$\trans{\tau}$                                    & $([\langle{}OrderReceipt\rangle, \langle{}InventoryCheck\rangle], mk\_Branch(\langle{}SupplierCheck\rangle, ...))$
    & by Reconfigure \\
$\xrightarrow{\langle{}SuplierCheck\rangle}$      & $([\langle{}OrderReceipt\rangle, \langle{}InventoryCheck\rangle, \langle{}SupplierCheck\rangle], mk\_Branch(\langle{}CreditCheck\rangle, ...))$
    & by Branch-T \\
$\xrightarrow{\langle{}CreditCheck\rangle}$       & $([\langle{}OrderReceipt\rangle, \langle{}InventoryCheck\rangle, \langle{}SupplierCheck\rangle, \langle{}CreditCheck\rangle], mk\_Par(\langle{}Billing\rangle, \langle{}Shipping\rangle, ...))$
    & by Branch-T \\
$\xrightarrow{\langle{}Billing\rangle}$           & $([\langle{}OrderReceipt\rangle, \langle{}InventoryCheck\rangle, \langle{}SupplierCheck\rangle, \langle{}CreditCheck\rangle, \langle{}Billing\rangle], mk\_Simple(\langle{}Shipping\rangle, ...))$
    & by Par-1 \\
$\xrightarrow{\langle{}Shipping\rangle}$          & $([\langle{}OrderReceipt\rangle, \langle{}InventoryCheck\rangle, \langle{}SupplierCheck\rangle, \langle{}CreditCheck\rangle, \langle{}Billing\rangle, \langle{}Shipping\rangle], mk\_Simple(\langle{}Archiving\rangle, \texttt{\textbf{nil}}))$
    & by Simple \\
$\xrightarrow{\langle{}Archiving\rangle}$         & $([\langle{}OrderReceipt\rangle, \langle{}InventoryCheck\rangle, \langle{}SupplierCheck\rangle, \langle{}CreditCheck\rangle, \langle{}Billing\rangle, \langle{}Shipping\rangle, \langle{}Archiving\rangle], \texttt{\textbf{nil}})$
    & by Simple \\
$\trans{\tau}$                                    & $([\langle{}OrderReceipt\rangle, \langle{}InventoryCheck\rangle, \langle{}SupplierCheck\rangle, \langle{}CreditCheck\rangle, \langle{}Billing\rangle, \langle{}Shipping\rangle, \langle{}Archiving\rangle, \langle{}TERMINATE\rangle], \texttt{\textbf{nil}})$
    & by Terminate \\
$\trans{\tau}$                                    & $([],\;$\texttt{\textbf{nil}}$)$
    & by Reset \\
\end{tabular}
\setlength{\tabcolsep}{6pt}
\caption{Examples of LTS rule applications for the VDM model}\label{fig:vdmlts:all}
\end{figure}

\subsection{Extensions}
\label{sec:vdm_extensions}

The model describes an interpreter with a single thread.
Therefore, interference is not considered beyond the non-deterministic execution of parallel compositions.
To allow multiple threads of control, the state of the interpreter must be extended to allow a set of workflows (and their associated traces) to be defined.
The reconfiguration operation must then be extended to reconfigure each thread in turn.
This would allow the model to exhibit concurrent application and reconfiguration actions (Case 3 in Figure~\ref{ReconfigCases}).
Actions are atomic in the current model, so there is no way to \emph{abort} actions.
To extend the model to allow this, a notion of beginning and completing actions would be required.
This could be achieved either by having `begin' and `end' forms of each action,
as used in~\cite{ColemanJones07} to investigate fine-grained concurrency in programming languages,
or by defining a `current action' in the state, which is placed there and removed at a later state.
If reconfiguration occurs between the beginning and end of an action, or if there is a current action present in the state, then an abort occurs.

As mentioned at the beginning of this section, there are two further dialects of VDM.
These use VDM-SL as their core specification language, but extend it with additional features.
\vpp adds object-orientation and concurrency through threads.
VDM-RT extends \vpp to add features for modelling real-time embedded systems.
These are a global `wall clock' that is advanced as expressions are evaluated (to predict real-world execution time),
and models of compute nodes connected by buses on which objects can be deployed.
The (simulated) time taken to evaluate language expressions depends on the speed of the compute nodes and buses,
and can be overridden to tune model where measurements of the actual speed of the target system can be made,
in order to make better predictions of behaviour of the final code.

These two dialects are often used as part of a development process~\cite{Larsen&09b} that begins with a sequential model of the system,
which can be constructed in VDM-SL or in \vpp. Conversion from VDM-SL to \vpp is straightforward as the core language is the same,
the main changes are to turn modules into classes and to turn the state definition into instance variables.
The sequential model is then extended to allow concurrent execution,
then a real-time version is made that captures the resources and configuration of the target hardware for which code will be written.
\vpp could be used to extend the model and capture true concurrency, allowing Designs 2, 3 and 4 (in Section~\ref{sec:recondes}) to be modelled.

In addition, extending the model to use \vpp allows additional features of the Overture tool to be used.
In the above model, tests were manual created and executed for all possible configurations.
This is obviously cumbersome and unscalable for more complicated models.
For \vpp models, Overture has \emph{combinatorial testing} features that allow tests to be created using regular expressions.
Overture also has true unit testing for \vpp models.

To extend the model to allow reconfiguration from Configuration 2 back to Configuration 1,
the pre-condition on \texttt{Reconfigure} needs to be relaxed to permit future traces from both configurations (not just Configuration 2 in the current model).
To extend the model to add further configurations, a few steps are necessary.
First, the configuration must be defined as a value in the \texttt{Configurations} module (for example, \texttt{Configuration3}).
Tests for this new configuration should be added to the \texttt{Test} module and executed.
Finally, the pre-condition must be extended to consider traces of the new configuration to be valid.
This is simple if it is acceptable to switch between any configuration at any time.
However, the definitions would be more complicated if there were restrictions on reconfiguration.
For example, if there are `points of no return' in between different configurations.

In Section \ref{sec:vdm_approach}, extending the current model with a model of data was suggested.
This extension represents an augmentation of the current workflow models with the data and operations necessary to allow customers to place orders.
This could include data types for representing orders, such as the following:

\begin{lstlisting}[style=styleVDM,caption={Example data types for modelling the order system},label={lst:extdata}]
CustId = token;
OrderId = token;

Order ::      custid : CustId
         inventoryOK : [bool]
            creditOK : [bool]
              accept : [bool];

state Office of
  orders : map OrderId to Order
end
\end{lstlisting}

The above defines identifiers for customers and orders using \texttt{\textbf{token}} types
(a countably infinite set of distinct values that can be compared for equality and inequality).
The \texttt{Order} type is a record that identifies a customer and the status of the order:
whether or not the checks have been passed, and whether the order is accepted. The state of the model stores all orders in the \texttt{orders} map.

To continue this model, operations should be defined to receive and evaluate orders, and to accept or reject them,
then to notify the customer, to bill and ship, and finally to archive.
Each should manipulate the data model and be protected by pre-conditions to ensure consistency and make explicit any assumption about the system.
The following implicit operation, defined only in terms of pre- and post-conditions, evaluates an order.
The pre-condition states that the order must be known (in the domain of the \texttt{orders} state variable) and be unprocessed.
The post-condition states that the \texttt{orders} map should contain an updated record for this order with the result of the inventory and credit check.
The \texttt{\textbf{mu}} expression allows record values to be overwritten, rather than constructing a new copy with the \texttt{\textbf{mk\_}} constructor.

\begin{lstlisting}[style=styleVDM,caption={Example implicit operation to \texttt{Evaluate} an order},label={lst:evalorder}]
EvaluateOrder(oid:OrderId)
ext wr orders
pre oid in set dom orders and
    orders(oid).inventoryOK = nil and
    orders(oid).creditOK = nil and
    orders(oid).accept = nil
post exists iOK, cOK : bool & orders = orders~ ++
       {oid |-> mu(orders(oid),
         inventoryOK |-> iOK, creditOK |-> cOK)};
\end{lstlisting}

This extended data model could then be connected to the existing interpreter,
such that when elements of the workflow are executed, calls are made to the operations that manipulate the data model.
The extended model would demonstrate whether it was possible to build an actual order system that met the requirements,
particularly when new configurations are introduced.
The data and operations could then be used as a specification during implementation in some programming language.


\section{Conditional Partial Order Graphs}\label{sec:cpog}

\newcommand{\rYes}{\overset{\!\textrm{Yes}}{\longrightarrow}}
\newcommand{\rNo}{\overset{\!\textrm{No}}{\longrightarrow}}

\begin{figure*}
\begin{centering}
\includegraphics[width=1\textwidth]{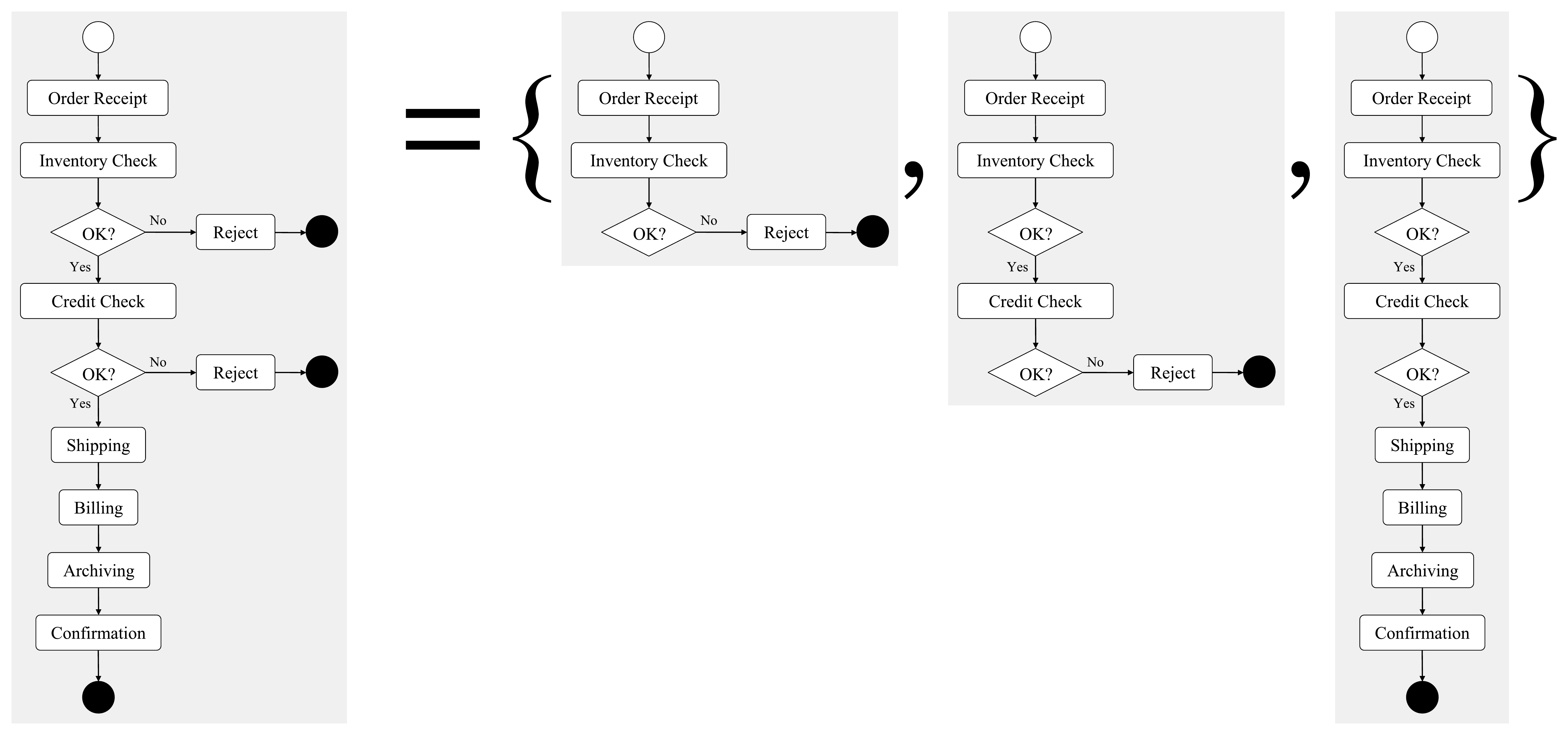}
\par\end{centering}
\caption{Expressing workflow requirements on Configuration 1
by a family of graphs\label{fig:Representing-workflow-requiremen} (see Figure~\ref{OfficeWorkflowConfig1})}
\end{figure*}

Conditional partial order graphs~(CPOGs)~\cite{2009_mokhov_phd} were
originally introduced for reasoning about properties of \emph{families of graphs}
in the context of asynchronous microcontroller design~\cite{2010_mokhov_ieee} and
processors with reconfigurable microarchitecture~\cite{2014_mokhov_jlpea}.
CPOGs are related to \emph{featured transition systems} (FTSs)
that are used to model \emph{feature families} of software product lines~\cite{2010_classen_fts}.
The key difference between FTSs and CPOGs is that FTSs use transition systems as the underlying formalism,
whereas CPOGs are built on top of partial orders, which enables the modelling of true concurrency.
In this paper, CPOGs are used to represent compactly graph families in which each graph expresses a workflow \emph{scenario}
(i.e. a particular collection of outcomes of branching actions of a workflow)
or a particular collection of requirements on workflow actions and their order.
For example, the requirements on Configuration~1 of the case study workflow
(shown in Figure~\ref{OfficeWorkflowConfig1}) can be expressed
using a family of three simple (i.e. branch-free) graphs (see Figure~\ref{fig:Representing-workflow-requiremen}).
We use the term \emph{family} instead of the more general term \emph{set} to emphasise the fact
that the graphs are annotated with branch decisions, in this case:

\begin{itemize}
\item Inventory check OK: \textsf{No}
\item Inventory check OK: \textsf{Yes}, credit check OK: \textsf{No}
\item Inventory check OK: \textsf{Yes}, credit check OK: \textsf{Yes}
\end{itemize}

This can be expressed equivalently using the following predicates:

\begin{itemize}
\item $\text{\textlnot}(\mathsf{InventoryCheck\ OK)}$
\item $(\mathsf{InventoryCheck\ OK)}\wedge\text{\textlnot}(\mathsf{CreditCheck\ OK)}$
\item $(\mathsf{InventoryCheck\ OK)}\wedge(\mathsf{CreditCheck\ OK)}$
\end{itemize}
where \textsf{InventoryCheck} and \textsf{CreditCheck} are the names of the
two branch actions.
Henceforth, we adopt the predicate notation.

The remainder of this section is organized as follows.
In Section~\ref{cpog-axioms} an axiomatic semantics of CPOGs is given and basic workflow modelling primitives are introduced,
which are used in Section~\ref{cpog-analysis} to model the case study workflow
and to demonstrate how CPOGs can be used for workflow verification by reduction to the Boolean Satisfiability~(SAT)~problem.
An LTS semantics of CPOGs is defined in Section~\ref{cpog-lts-semantics},
thereby establishing a formal link with the other two formalisms discussed in this paper.
Dynamic reconfiguration is discussed in Section~\ref{cpog-dynamic},
and we show that CPOGs use true concurrency semantics when modelling functional interference between workflow and reconfiguration actions.
We comment on our experience of automating the verification of the workflow requirements in Section~\ref{cpog-implementation},
and give an overall evaluation of the CPOG-based approach in Section~\ref{sec:comp}.

\subsection{Formalism\label{cpog-axioms}}

\begin{figure*}
\begin{centering}
\includegraphics[scale=0.4]{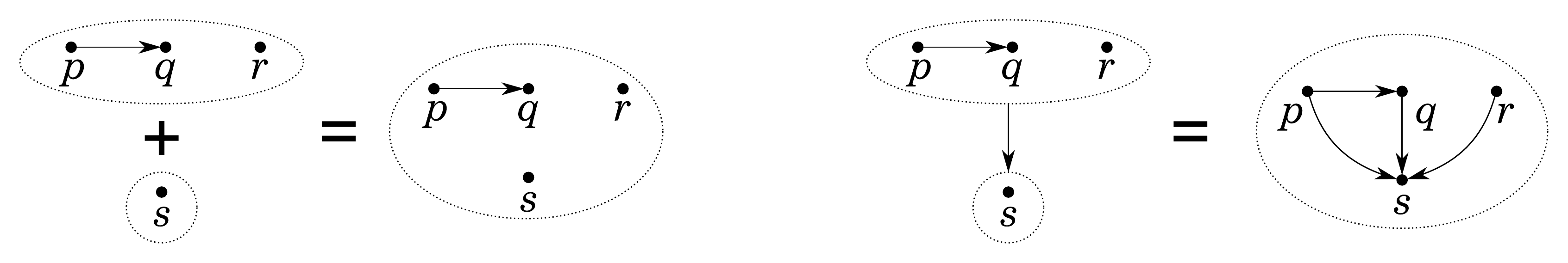}
\par\end{centering}
\caption{Parallel and sequential composition example (no common vertex)\label{fig:Overlay-and-sequence-no-common}}
\end{figure*}

\begin{figure*}
\begin{centering}
\includegraphics[scale=0.4]{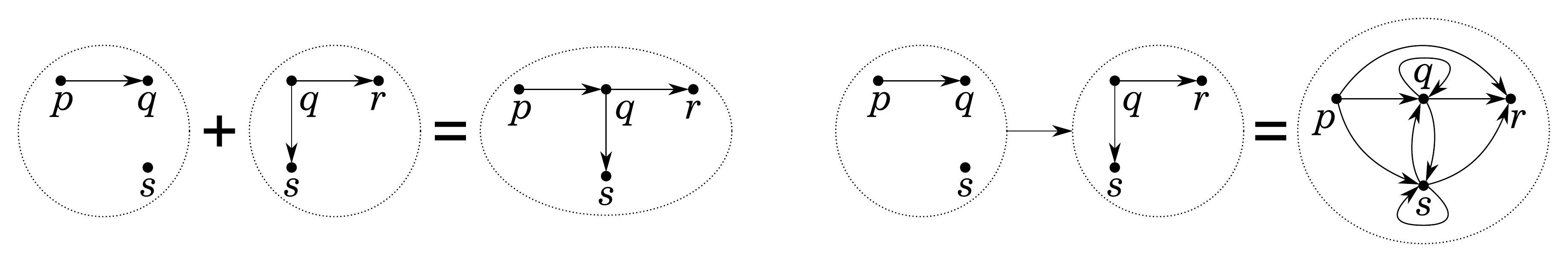}
\par\end{centering}
\caption{Parallel and sequential composition example (common vertices)\label{fig:Overlay-and-sequence}}
\end{figure*}

\begin{figure}
\begin{centering}
\subfloat[Distributivity]{\begin{centering}
\includegraphics[scale=0.4]{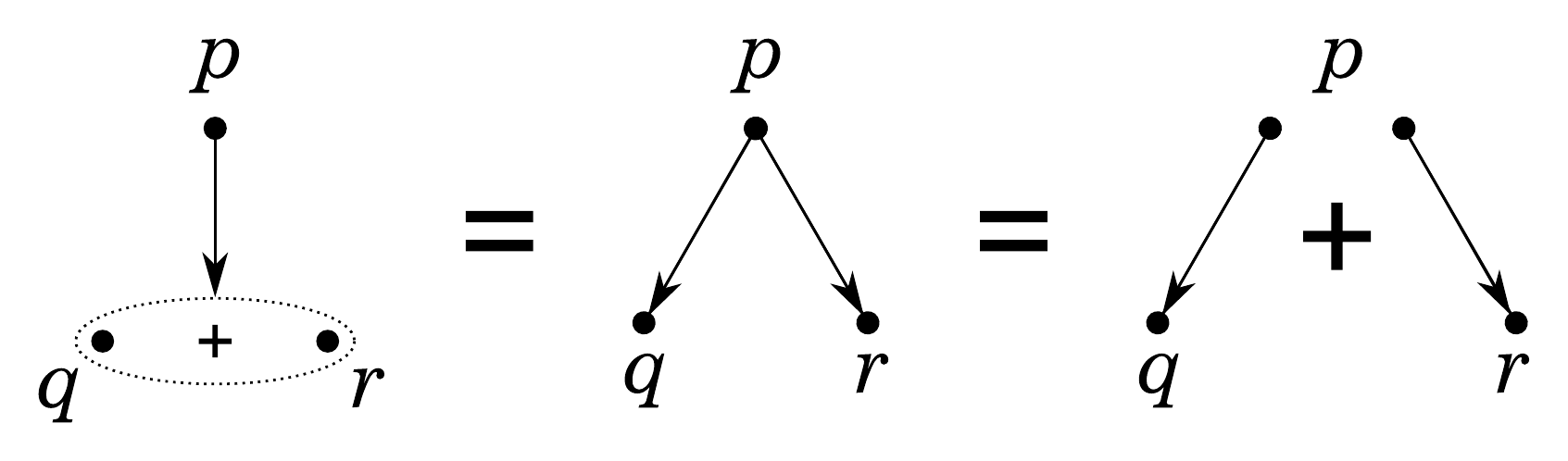}
\par\end{centering}

}
\par\end{centering}

\begin{centering}
\subfloat[Decomposition]{\begin{centering}
\includegraphics[scale=0.4]{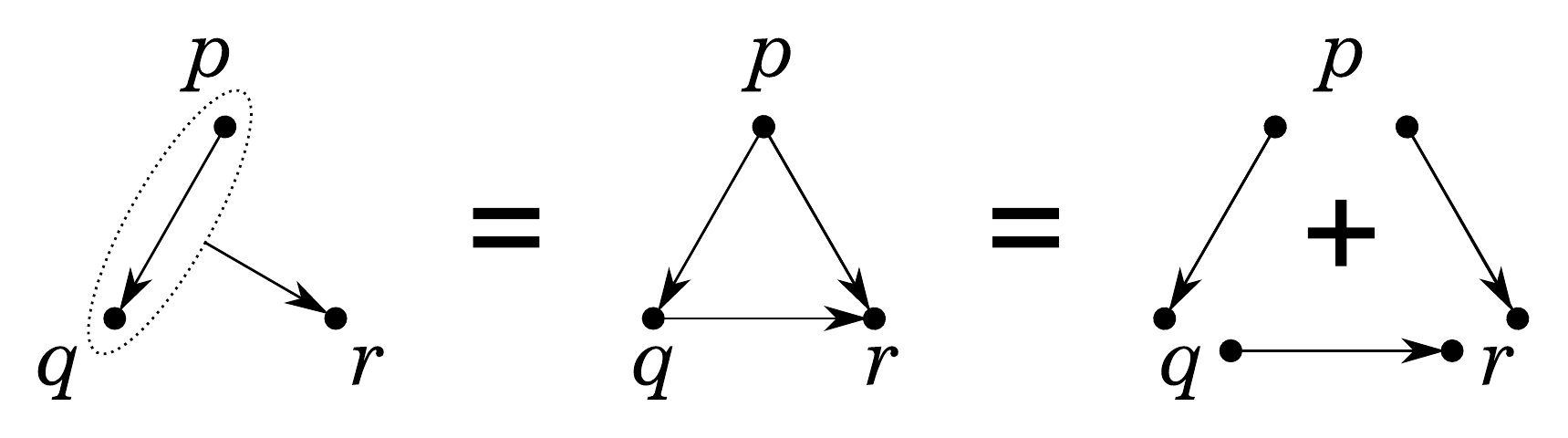}
\par\end{centering}

}
\par\end{centering}

\begin{centering}
\subfloat[Branching graph family]{\begin{centering}
\includegraphics[scale=0.4]{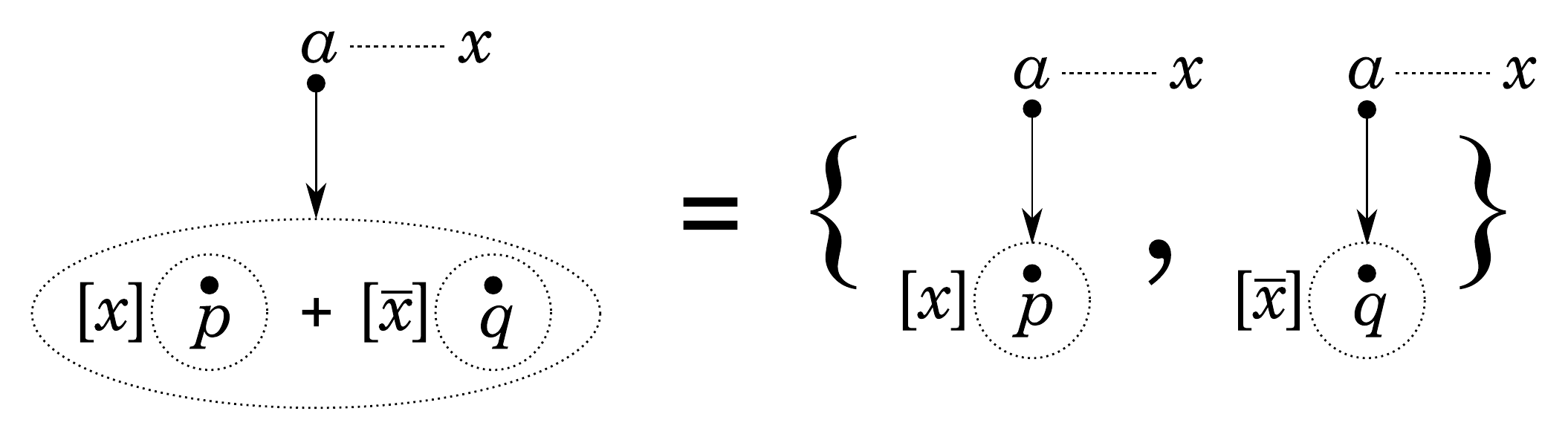}
\par\end{centering}

}
\par\end{centering}

\begin{centering}
\subfloat[Transitive reduction/closure]{\begin{centering}
\includegraphics[scale=0.4]{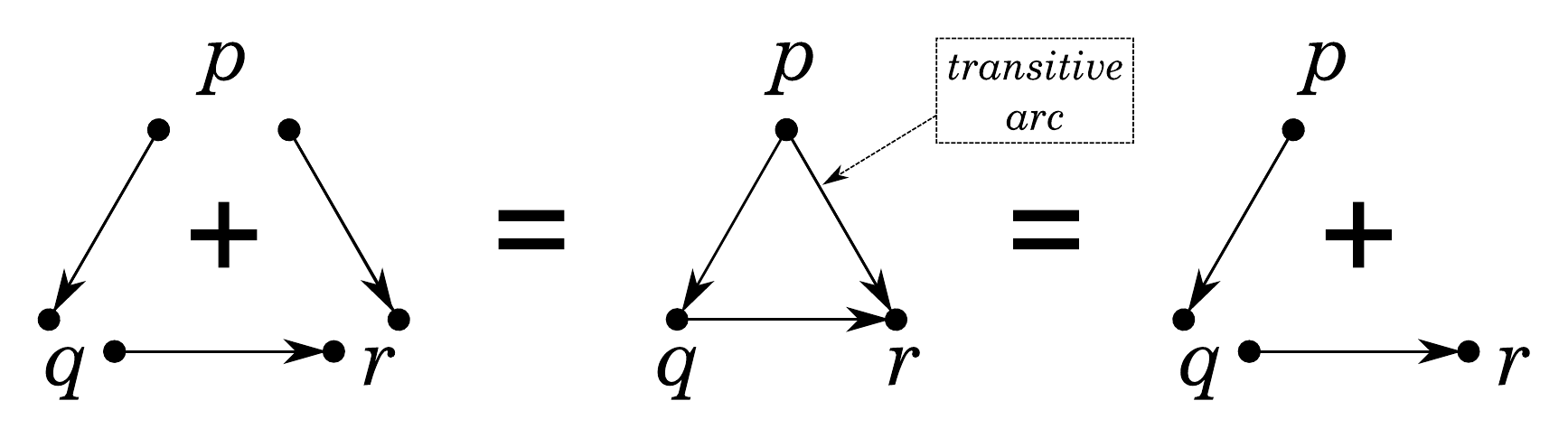}
\par\end{centering}

}
\par\end{centering}

\caption{Manipulating parameterised graphs\label{fig:Several-axioms-of}}
\end{figure}

The axiomatic semantics of CPOGs is outlined below, and was first
introduced in~\cite{2014_mokhov_pg} to provide efficient compositionality and abstraction facilities for graphs.
Each element of the algebra represents a family of graphs, and is defined as follows:

Let $\mathcal{A}$ be the alphabet of names of \emph{actions} (e.g. $\mathsf{OrderReceipt}$ and $\mathsf{InventoryCheck}$)
that represent tasks and subtasks. Actions can have \emph{order dependencies} between them.
$\mathcal{A}$ is used to construct models of workflows and to define their requirements using the following axioms:
\begin{itemize}
\item The empty workflow is denoted by $\varepsilon$, that is, the empty family of graphs.
\item A workflow consisting of a single action $a \!\in\! \mathcal{A}$ is denoted simply by $a$.
It corresponds to a family consisting of a single graph that contains a single vertex $a$.
\item The \emph{parallel composition} of workflows $p$ and $q$ is denoted by $p+q$~
(e.g. the concurrent execution of \texttt{Billing} and \texttt{Shipping} shown in Figure~\ref{OfficeWorkflowConfig2}
is denoted by $\mathsf{Billing}+\mathsf{Shipping}$).
The operator~$+$ is commutative, associative, and has $\varepsilon$ as the identity:

\begin{enumerate}
\item $p+q=q+p$
\item $p+(q+r)=(p+q)+r$
\item $p+\varepsilon=p$
\end{enumerate}
Intuitively, $p+q$ means `execute workflows $p$ and $q$ concurrently, synchronising on common actions'.
\item The \emph{sequential composition} of workflows $p$ and $q$ is denoted by $p\rightarrow q$~
(e.g. the sequential execution of \texttt{Shipping} followed by \texttt{Billing} shown in Figure~\ref{OfficeWorkflowConfig1}
is denoted by $\mathsf{\mathsf{Shipping}}\rightarrow\mathsf{\mathsf{Billing}}$).
The operator~$\rightarrow$ has a higher precedence than~$+$,
is associative, has the same identity ($\varepsilon$) as~$+$, distributes over~$+$,
and can be decomposed into pairwise sequences:

\begin{enumerate}
\item $p\rightarrow(q\rightarrow r)=(p\rightarrow q)\rightarrow r$
\item $p\rightarrow\varepsilon=p$ and $\varepsilon\rightarrow p=p$
\item $p\rightarrow(q+r)=p\rightarrow q+p\rightarrow r$ ~and~
$(p+q)\rightarrow r=p\rightarrow r+q\rightarrow r$
\item $p\rightarrow q\rightarrow r=p\rightarrow q+p\rightarrow r+q\rightarrow r$
\end{enumerate}
Intuitively, $p\rightarrow q$ means `execute workflow $p$, then execute workflow $q$'.
In other words, introduce order dependencies between all actions in $p$ and all actions in $q$.
Note that if $p$ and $q$ contain common actions they cannot be executed in sequence, as illustrated by an example below.\\

Figure~\ref{fig:Overlay-and-sequence-no-common} shows an example of
parallel and sequential composition of graphs. It can be seen that the
parallel composition does not introduce any new order dependency
between actions in different graphs; therefore, the actions can be executed concurrently.
Sequential composition, on the other hand, imposes order on the actions
by introducing new dependencies between actions $p$, $q$, and $r$ in the top graph
and action $s$ in the bottom graph. Hence, the resulting workflow behaviour
is interpreted as the behaviour specified by the top graph followed
by the behaviour specified by the bottom graph. Another example of
these operations is shown in Figure~\ref{fig:Overlay-and-sequence}.
Since the graphs have common vertices, their compositions are more
complicated, in particular, their sequential composition contains
the self-dependencies $(q,q)$ and $(s,s)$ which lead to a \emph{deadlock} in the resulting workflow.
Action $p$ can occur, but all the remaining actions are locked: neither $q$ nor $s$ can proceed due to the self-dependencies,
while $r$ cannot proceed without $q$ and $s$.
Figures~\ref{fig:Several-axioms-of}(a) and~\ref{fig:Several-axioms-of}(b)
illustrate the distributivity and decomposition axioms respectively.

\item A \emph{conditional workflow} is denoted by $[x]p$,
where $x$ is a predicate expressing a certain condition (e.g. `Inventory Check OK') and $p$ is a workflow.
Intuitively, $[x]p$ means `execute workflow $p$ only if condition $x$ holds'.
We postulate that $[1]p=p$ and $[0]p=\varepsilon$. This allows us to model branching in flow charts.
For example, the algebraic expression
\[
a\rightarrow([x]p+[\text{\textlnot}x]q)
\]
corresponds to a branching performed after action $a \!\in\! \mathcal{A}$,
which is followed by workflow $p$ if predicate $x$ holds, and by workflow $q$ if predicate $x$ does not hold.
Alternatively, we can say that the above expression corresponds to a family of graphs,
in which vertex $a$ is followed by actions from the graphs coming either from family
$p$ or from family $q$, as illustrated in Figure~\ref{fig:Several-axioms-of}(c),
where a dotted line between vertex $a$ and variable $x$ indicates that
the variable is assigned a value during the execution of action $a$. We will use
the following short-hand notation for a clearer correspondence with flow charts:
\[
\begin{cases}
a\rYes p\overset{\textrm{df}}{=}a\rightarrow[\textrm{A OK}]p\\
a\rNo p\overset{\textrm{df}}{=}a\rightarrow[\text{\textlnot}(\textrm{A OK)}]p
\end{cases}
\]
where predicate `A OK' corresponds to the successful completion
of a branching action $a$. For example, the expression
\[
\begin{array}{l}
\mathsf{InventoryCheck}\rYes \mathsf{CreditCheck}\  +\
\mathsf{InventoryCheck}\rNo \mathsf{Reject}\rightarrow\mathsf{End}
\end{array}
\]
corresponds to the first branching in Figure~\ref{OfficeWorkflowConfig1}: if the inventory check
is completed successfully the workflow continues with the credit check,
otherwise the order is rejected and the workflow ends (actions $\mathsf{Start}$
and $\mathsf{End}$ denote respectively the start and end circles used in the flow chart).
Notice that operators $\rYes$ and $\rNo$ bind less tightly
than $\rightarrow$ to reduce the number of parentheses.
Operator $[x]$ has the highest precedence and has the following useful properties (proved in \cite{2014_mokhov_pg}): \\

\begin{tabular}{ll}
1.~~$[x\wedge y]p=[x][y]p$ &~~~~~
2.~~$[x\vee y]p=[x]p+[y]p$ \vspace{2mm} \\
3.~~$[x](p+q)=[x]p+[x]q$ &~~~~~
4.~~$[x](p\rightarrow q)=[x]p\rightarrow[x]q$ \\
\end{tabular}
\end{itemize}

Notice that the condition operator can be used not only to create conditional workflows,
but also to create \emph{conditional workflow dependencies}.
For example, in the expression $p + q + [x](p \rightarrow q)$,
if $x=0$, the expression simplifies to the parallel composition $p + q$,
but if $x=1$, the expression simplifies to the sequential composition $p \rightarrow q$.
In other words, actions in $q$ conditionally depend on actions in $p$.

To summarise, the following operators are used to create and manipulate graph
families corresponding to workflows (in decreasing order of precedence): \\
$[x]p$, $p \rightarrow q$, $a \overset{\!\!\textrm{Yes}}{\longrightarrow} p$,
$a \overset{\!\textrm{No}}{\longrightarrow} p$, and $p + q$, where $a$ is an
action, $x$ is a predicate, and $p$ and $q$ are workflows.

The algebraic notation is used to translate flow charts into
mathematical descriptions amenable to automated verification,
as described in the next subsection.

\subsection{Modelling and Analysis\label{cpog-analysis}}

The granularity of concurrency in CPOGs is a single action,
and the granularity of reconfiguration is a single action and a single dependency between actions.
Multiple actions and dependencies can also be reconfigured by a single reconfiguration action
that sets Boolean variables in the predicates that guard the actions and their dependencies.
Thus, it is possible to model all three cases of dynamic reconfiguration identified in Section \ref{sec:intro}.
A CPOG can easily express concurrent actions, and (therefore) Case 3 of reconfiguration fits the CPOG 'idiom' best.
However, we have modelled Case 1 of reconfiguration for simplicity.
Notice that the different designs of the case study workflow configurations are represented as the same CPOG model,
because CPOGs cannot represent cyclic processes.

We now describe how to specify workflow requirements
and their reconfiguration as families of graphs,
how to reason about the correctness of such specifications,
and how to manipulate them using the operators of the algebra.

The algebraic notation can be used to translate
the flow chart in Figure~\ref{OfficeWorkflowConfig1} into the following expression:
\[
\begin{array}{ccl}
c_{1} & = & \mathsf{Start}\rightarrow\mathsf{OrderReceipt}\rightarrow ( \\
 &  & \ \ \mathsf{InventoryCheck}\rNo \mathsf{Reject}\rightarrow\mathsf{End}\ + \\
 &  & \ \ \mathsf{InventoryCheck}\rYes (  \\
 &  & \ \ \ \ \mathsf{CreditCheck}\rNo \mathsf{Reject}\rightarrow\mathsf{End}\ +\\
 &  & \ \ \ \ \mathsf{CreditCheck}\rYes \mathsf{Shipping}\rightarrow\mathsf{Billing} \rightarrow \mathsf{Archiving}\rightarrow\mathsf{Confirmation}\rightarrow\mathsf{End} \\
 &  & \ \ )\\
 &  & )
\end{array}
\]
The expression can be rewritten using the axioms in order to prove
that certain requirements hold. For example, any expression can be
rewritten into a so-called \emph{canonical form}~\cite{2014_mokhov_pg},
which is sufficient for checking most of the requirements listed in
Section~\ref{sec:wfreqts1}.

\begin{prop}\label{prop-canonical}Any workflow expression can be rewritten in
the following canonical form~\cite{2014_mokhov_pg}:
\begin{equation}
\left(\sum_{a\in V}[f_{a}]a\right)+\left(\sum_{a,b\in V}[f_{ab}](a\rightarrow b)\right)\label{eq:canonical-form}
\end{equation}
where:
\begin{itemize}
\item $V$ is a subset of actions that appear in the original expression;
\item for all $a \!\in\! V$, $f_{a}$ are canonical forms of Boolean expressions
and are distinct from 0;
\item for all $a, b \!\in\! V$, $f_{ab}$ are canonical forms of Boolean expressions
such that $f_{ab}\Rightarrow f_{a}\wedge f_{b}$
(that is, a dependency between actions $a$ and $b$ can exist only if both actions exist).
\end{itemize}
\end{prop}
In other words, the canonical form of an expression lists all constituent
actions and all pairwise dependencies between them (along with their
predicates). The canonical form can contain redundant transitive dependencies,
such as $p\rightarrow r$ in presence of both $p\rightarrow q$ and $q\rightarrow r$.
Such terms can be eliminated to simplify the resulting expression.
This corresponds to the well-known \emph{transitive reduction} procedure,
which can be formalised by adding the following axiom~\cite{2014_mokhov_pg}:

\(
\mbox{{if\ }}q\neq\varepsilon\mbox{{\ then\ }}p\rightarrow q+p\rightarrow r+q\rightarrow r=p\rightarrow q+q\rightarrow r
\)

The axiom can be used to add or to remove transitive dependencies as necessary,
see Figure~\ref{fig:Several-axioms-of}(d).

Rewriting a workflow expression manually by following the CPOG
axioms is tedious and error-prone. This motivated us to automate computation
of the canonical form and the transitive reduction, as will be discussed in
Section~\ref{cpog-implementation}. By applying these procedures to~$c_{1}$
we obtain its reduced canonical form shown below.
For brevity, we will denote predicates \textsf{InventoryCheck}~OK and
\textsf{CreditCheck}~OK simply by $x$ and $y$ respectively.
Actions are visually separated from dependencies by a horizontal line.

\(
\begin{array}{cclc}
c_{1} & = & [1]\mathsf{Start}+[1]\mathsf{OrderReceipt} & +\\
 &  & [1]\mathsf{InventoryCheck}+[\overline{x}\vee\overline{y}]\mathsf{Reject} & +\\
 &  & [1]\mathsf{End}+[x]\mathsf{\mathsf{CreditCheck}} & +\\
 &  & [x\wedge y]\mathsf{Billing}+[x\wedge y]\mathsf{Shipping} & +\\
 &  & [x\wedge y]\mathsf{Archiving}+[x\wedge y]\mathsf{Confirmation} & +\\
[\smallskipamount]\cline{3-4}
\noalign{\vskip 1mm}
 &  & [1](\mathsf{Start}\rightarrow\mathsf{OrderReceipt}) & +\\
 &  & [1](\mathsf{OrderReceipt}\rightarrow\mathsf{InventoryCheck}) & +\\
 &  & [x](\mathsf{InventoryCheck}\rightarrow\mathsf{\mathsf{CreditCheck}}) & +\\
 &  & [\overline{x}](\mathsf{InventoryCheck}\rightarrow\mathsf{Reject}) & +\\
 &  & [x\wedge\overline{y}](\mathsf{CreditCheck}\rightarrow\mathsf{Reject}) & +\\
 &  & [\overline{x}\vee\overline{y}](\mathsf{Reject}\rightarrow\mathsf{End}) & +\\
 &  & [x\wedge y](\mathsf{\mathsf{CreditCheck}}\rightarrow\mathsf{Shipping}) & +\\
 &  & [x\wedge y](\mathsf{Shipping}\rightarrow\mathsf{Billing}) & +\\
 &  & [x\wedge y](\mathsf{Billing}\rightarrow\mathsf{Archiving}) & +\\
 &  & [x\wedge y](\mathsf{Archiving}\rightarrow\mathsf{Confirmation}) & +\\
 &  & [x\wedge y](\mathsf{Confirmation}\rightarrow\mathsf{End})
\end{array}
\)

The resulting expression gives us plenty of valuable information that can be
used for analysis of the workflow. In particular,
the following correctness properties can be verified:
\begin{itemize}
\item The starting and ending actions are part of the workflow regardless
of possible outcomes of the branching actions, as indicated by `unconditional'
terms $[1]\mathsf{Start}$ and $[1]\mathsf{End}$.
\item The billing, shipping, archiving, and confirmation actions are performed
if and only if both the inventory and internal credit checks are successful;
in fact, all these actions are pre-conditioned with the predicate $x\wedge y$.
\item An order is rejected if either the inventory check or the internal
credit check fails, as confirmed by term $[\overline{x}\vee\overline{y}]\mathsf{Reject}$.
\item The first action after $\mathsf{Start}$ is always $\mathsf{OrderReceipt}$,
as confirmed by term $[1](\mathsf{Start}\rightarrow\mathsf{OrderReceipt})$
and the fact that all other actions transitively depend on
$\mathsf{OrderReceipt}$.
\item The shipping and billing actions are always performed sequentially:
whenever the actions occur \\
($[x\wedge y]\mathsf{Billing}$ and $[x\wedge y]\mathsf{Shipping}$) so does the dependency
$[x\wedge y](\mathsf{Shipping}\rightarrow\mathsf{Billing})$.
\item There are no cyclic dependencies and therefore each action is performed at most once.
\end{itemize}

We now translate the workflow requirements for Configuration~2 shown in
Figure~\ref{OfficeWorkflowConfig2} into our notation, verify several
relevant correctness properties, and highlight the differences between the two
configurations.
The flow chart in Figure~\ref{OfficeWorkflowConfig2} can be translated
into the following expression:

\(
\begin{array}{ccl}
c_{2} & = & \mathsf{Start}\rightarrow\mathsf{OrderReceipt}\rightarrow ( \\
 &  & \ \ \mathsf{InventoryCheck}\rYes cc\ +  \\
 &  & \ \ \mathsf{InventoryCheck}\rNo ( \\
 &  & \ \ \ \ \mathsf{SupplierCheck}\rYes cc \\
 &  & \ \ \ \ \mathsf{SupplierCheck}\rNo\mathsf{Reject}\rightarrow\mathsf{End}\\
 &  & \ \ )\\
 &  & )\\
\end{array}
\)

where expression $cc$ corresponds to the part of the workflow
starting with the $\mathsf{CreditCheck}$ action:

\(
\begin{array}{ccl}
cc & = & \mathsf{CreditCheck}\rNo \mathsf{Reject}\rightarrow\mathsf{End}\ +\\
 &  & \mathsf{CreditCheck}\rYes (\mathsf{Billing} + \mathsf{Shipping}) \rightarrow  \mathsf{Archiving}\rightarrow\mathsf{End}
\end{array}
\)

The ability to abstract and share/instantiate common behaviour (e.g. $cc$ in
the above workflow) is essential for specifying real-life reconfigurable
systems, where monolithic specifications are impractical. Our prototype
implementation supports abstraction and sharing (see
Section~\ref{cpog-implementation} for further details).

Expression $c_2$ can now be prepared for further analysis by converting it into
the canonical form and transitively reducing the result as described above.
For brevity, we denote predicates \textsf{InventoryCheck}~OK
and \textsf{SupplierCheck}~OK by $x_1$ and $x_2$ respectively,
thereby emphasising that roles of inventory and supplier checks are similar;
the predicate \textsf{CreditCheck}~OK is denoted by $y$ as before.
The resulting expression is shown below.

\(
\begin{array}{cclc}
c_{2} & = & [1]\mathsf{Start}+[1]\mathsf{OrderReceipt}+[1]\mathsf{End} & +\\
 &  & [1]\mathsf{InventoryCheck}+[\overline{x_1}]\mathsf{SupplierCheck} & +\\
 &  & [x_1\vee x_2]\mathsf{CreditCheck}+[\overline{x_1} \wedge\overline{x_2}\vee\overline{y}]\mathsf{Reject} & +\\
 &  & [(x_1 \vee x_2) \wedge y]\mathsf{Billing} & +\\
 &  & [(x_1 \vee x_2) \wedge y]\mathsf{Shipping} & +\\
 &  & [(x_1 \vee x_2) \wedge y]\mathsf{Archiving} & +\\
[\smallskipamount]\cline{3-4}
\noalign{\vskip 1mm}
 &  & [1](\mathsf{Start}\rightarrow\mathsf{OrderReceipt}) & +\\
 &  & [1](\mathsf{OrderReceipt}\rightarrow\mathsf{InventoryCheck}) & +\\
 &  & [\overline{x_1}](\mathsf{InventoryCheck}\rightarrow\mathsf{\mathsf{SupplierCheck}}) & +\\
 &  & [x_1](\mathsf{InventoryCheck}\rightarrow\mathsf{CreditCheck}) & +\\
 &  & [\overline{x_1}\wedge x_2](\mathsf{SupplierCheck}\rightarrow\mathsf{CreditCheck}) & +\\
 &  & [\overline{x_1}\wedge \overline{x_2}](\mathsf{SupplierCheck}\rightarrow\mathsf{Reject}) & +\\
 &  & [(x_1 \vee x_2) \wedge \overline{y}](\mathsf{CreditCheck}\rightarrow\mathsf{Reject}) & +\\
 &  & [\overline{x_1} \wedge\overline{x_2}\vee\overline{y}](\mathsf{Reject}\rightarrow\mathsf{End}) & +\\
 &  & [(x_1 \vee x_2) \wedge y](\mathsf{\mathsf{CreditCheck}}\rightarrow\mathsf{Billing}) & +\\
 &  & [(x_1 \vee x_2) \wedge y](\mathsf{\mathsf{CreditCheck}}\rightarrow\mathsf{Shipping}) & +\\
 &  & [(x_1 \vee x_2) \wedge y](\mathsf{\mathsf{Billing}}\rightarrow\mathsf{Archiving}) & +\\
 &  & [(x_1 \vee x_2) \wedge y](\mathsf{\mathsf{Shipping}}\rightarrow\mathsf{Archiving}) & +\\
 &  & [(x_1 \vee x_2) \wedge y](\mathsf{Archiving}\rightarrow\mathsf{End})
\end{array}
\)

Using the derived expression, the following properties of Configuration~2 can be verified:
\begin{itemize}
\item The billing and shipping actions are concurrent as indicated by the lack
of any dependency between them. This is different from Configuration~1 where
the actions could only occur in sequence: \\
$[x\wedge y](\mathsf{Shipping}\rightarrow\mathsf{Billing})$.
\item The credit check is conducted when either inventory or supplier check is
successful, as indicated by term $[x_1\vee x_2]\mathsf{CreditCheck}$. Again,
this is different from Configuration~1, where there was no way around the
inventory check.
\item Consequently, an order is rejected under the condition
$\overline{x_1} \wedge\overline{x_2}\vee\overline{y}$, that is, when either both
inventory and supplier checks have failed, or the credit check has failed. By
negating this condition we obtain $(x_1 \vee x_2) \wedge y$, which guards the
billing, shipping, and archiving actions, as well as dependencies between them.
\item The confirmation action is missing in $c_2$, as intended. We can also
highlight this by adding a redundant term $[0]\mathsf{Confirmation}$ to $c_2$.
\end{itemize}

As demonstrated above, Configurations~1 and~2 have several important differences.
Hence, if a system's configuration is dynamically changed
from one configuration to another, the system may end up in an impossible state
according to the new configuration. Such situations may lead to the system's failure,
and (therefore) must be prevented. In Section~\ref{cpog-dynamic} we discuss how the
CPOG-based modelling and verification approach can be used to describe formally such
situations, determine under which circumstances they can occur,
and derive practicable \emph{reconfiguration guidelines} to prevent their occurrence.

\subsection{LTS semantics\label{cpog-lts-semantics}}

In this section, we define an LTS semantics of CPOGs in order to facilitate comparison with the other two formalisms.
The definition is based on CPOG firing rules introduced in~\cite{2009_mokhov_phd}
and the canonical form construction~\cite{2014_mokhov_pg} (see Proposition~\ref{prop-canonical}).

According to~(\ref{eq:canonical-form}), any algebraic CPOG expression can be
uniquely represented by a directed graph $(V,E)$ whose vertices~$V$
and arcs~$E\subseteq V\times V$ are labelled with Boolean conditions $f_a$ and
$f_{ab}$ (respectively) that are defined on a set of Boolean variables~$X$.
A variable $x \!\in\! X$ is said to be \emph{controlled} by a vertex $v_x \!\in\! V$ if
the value of the variable $x$ is changed by the action associated with
the vertex $v_{x}$. Notice there is at most one such vertex for each variable.
Therefore, the value of each variable can be changed at most once. 

Given an assignment of variables $\psi:X\rightarrow\{0,1\}$,
a vertex $v \!\in\! V$ and an arc $(v_1, v_2) \!\in\! E$ are termed \emph{active} if their respective conditions evaluate
to 1 under the assignment, denoted by $f_v\vert_{\psi}=1$ and $f_{v_1v_2}\vert_{\psi}=1$ respectively;
otherwise, $v$ and $(v_1, v_2)$ are termed \emph{inactive}.
There are $2^{|X|}$ possible variable assignments.
Therefore, a CPOG can describe up to $2^{|X|}$ possible graphs comprised of active vertices and arcs.

The \emph{preset} of a vertex $v \!\in\! V$, denoted by $preset(v, \psi)$, is
defined to be the set of active (determined by $\psi$) vertices that are
connected to $v$ by active arcs $(u,v) \!\in\! E$:
\[
preset(v, \psi)=\{\ u\ \vert \ u\in V\wedge f_u\vert_{\psi}=1\wedge
f_{uv}\vert_{\psi}=1\ \}.
\]

A \emph{history} is a subset of active vertices $H \subseteq V$ whose
corresponding actions have already occurred. The number of possible histories is bounded by $2^{|V|}$.


A \emph{labelled transition system}~(LTS) is a triple $(S, A, \rightarrow)$,
where $S$ is the set of states, $A$ is the alphabet of \emph{actions} that label state transitions,
and $\rightarrow : S \times A \times S$ is the transition relation.
We will use the shorthand notation $s_{1}\xrightarrow{\ a\ }s_{2}$
to denote $(s_{1}, a, s_{2})$ belongs to the relation $\rightarrow$.

A CPOG defines an LTS with $2^{|V|}\cdot2^{|X|}$ states that correspond
to all possible histories and variable assignments.
Hence, every state $s \!\in\! S$ is a pair $(H, \psi)$. We fix the initial state
$s_{0} \!\in\! S$ to be equal to $(\emptyset,\mathbf{0})$
where $\mathbf{0}$ is the \emph{zero} variable assignment: $\mathbf{0}(x)=0$ for all $x \!\in\! X$.

The alphabet of actions of the LTS directly corresponds to the CPOG vertices. Therefore, $A = V$.

The transition relation can be captured by the following rule:
\[
\begin{array}{c}
preset(w, \psi) \subseteq H ~\wedge~ w \!\notin\! H ~\wedge~ f_w\vert_{\psi} = 1\wedge
\forall x\in\!X~(v_{x} \neq w \implies \psi'(x)=\psi(x))
\\
\hline (H, \psi) \xrightarrow{\ \ w\ \ }(H \cup \{w\}, \psi')
\end{array}
\]
In other words, the system can transition from state $s_1 = (H, \psi)$ to
state $s_2 = (H \cup \{w\}, \psi')$
by executing the action corresponding to an active vertex $w\in V$ if the vertex
preset belongs to the history $H$ and the action has not been performed previously.
The new variable assignment $\psi'$ must be identical to the old variable assignment $\psi$
on all variables that are not controlled by vertex $w$.

Notice that the above rule defines the execution of a single action,
and thereby defines an interleaving semantics for pseudo-concurrent actions.
In order to define the semantics of true concurrency,
the LTS syntax must be extended to enable a set of actions to be performed simultaneously,
which results in the following alternative rule for the transition relation:
\[
\begin{array}{c}
\forall w \!\in\! W ~(preset(w, \psi) \subseteq H ~\wedge~ w \!\notin\! H ~\wedge~ f_w\vert_{\psi}=1) \wedge
\forall x\in\!X~(v_{x} \notin W \implies \psi'(x)=\psi(x))\\
\hline (H, \psi) \xrightarrow{\ \ W\ \ }(H \cup W, \psi')
\end{array}
\]
Thus, the alternative rule allows multiple actions in a set $W \subseteq V$
to be performed simultaneously. Hence, every diamond
\[
\begin{cases}
s_{1}\xrightarrow{\ X\ }s'\,\xrightarrow{\ Y\ }s_{2}\\
s_{1}\xrightarrow{\ Y\ }s''\xrightarrow{\ X\ }s_{2}
\end{cases}
\]
is augmented with a diagonal transition $s_{1}\xrightarrow{\ X\cup Y\ }s_{2}$.

\begin{figure}
\begin{centering}
\subfloat[Early choice between two sequential behaviours]{\begin{centering}
\includegraphics[scale=0.3]{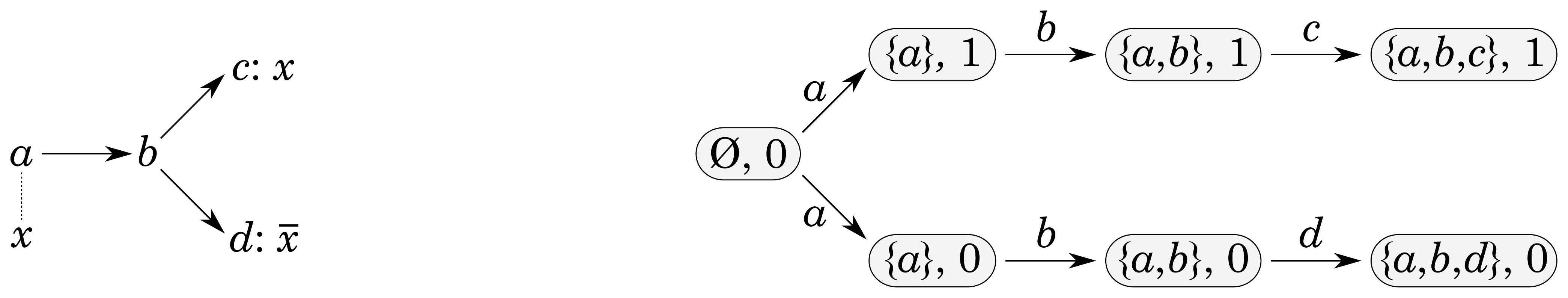}
\par\end{centering}

}
\par\end{centering}
\begin{centering}
\subfloat[Late choice between two sequential behaviours]{\begin{centering}
\includegraphics[scale=0.3]{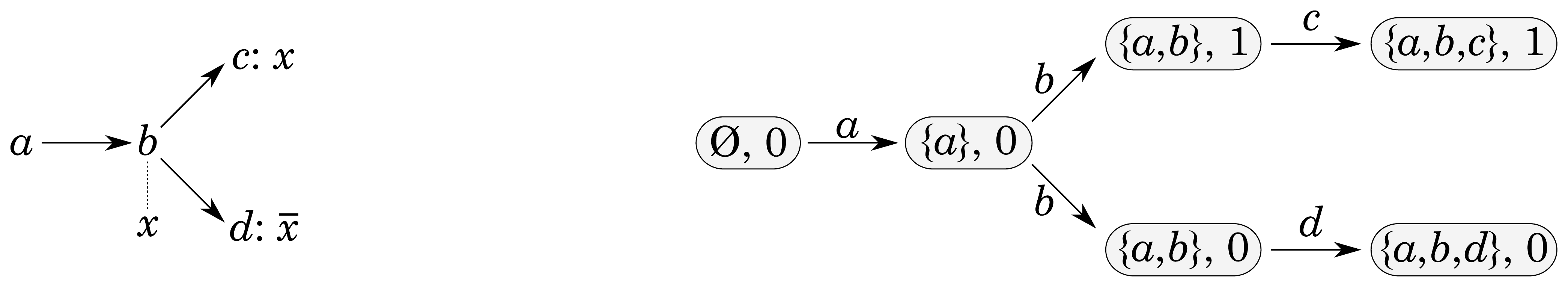}
\par\end{centering}

}
\par\end{centering}
\begin{centering}
\subfloat[Late choice between sequential and concurrent behaviours]{\begin{centering}
\includegraphics[scale=0.3]{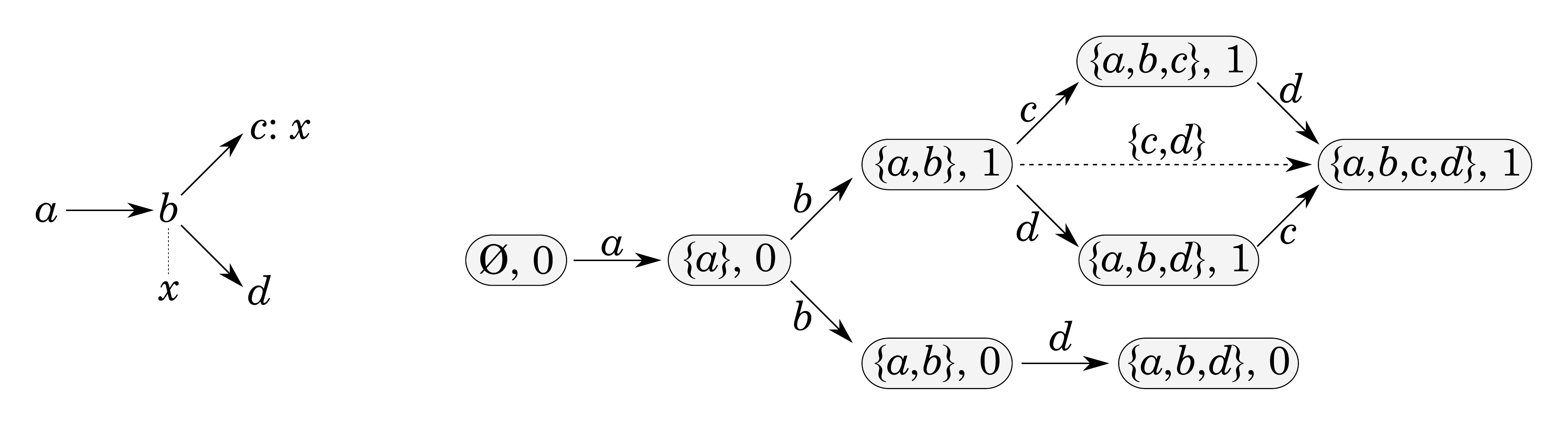}
\par\end{centering}

}
\par\end{centering}
\caption{Examples of CPOG LTS semantics\label{fig:lts-semantics}}
\end{figure}

Figure~\ref{fig:lts-semantics} shows three CPOG examples and the corresponding
LTSs. The states are denoted by grey ovals with the history/assignment pairs,
for example $(\{a\}, 0)$ means $H=\{a\}$ and $\psi(x)=0$.
The first CPOG corresponds to a system with two possible sequential behaviours:
$a\rightarrow b \rightarrow c$ (when $x=1$) and $a\rightarrow b \rightarrow d$
(when $x=0$), where the choice between $c$ and $d$ is made during the execution
of action $a$ by appropriately setting variable $x$, as denoted by the dotted
line between $a$ and $x$ (hence, $v_x=a$ using the above notation). The
corresponding LTS (on the right-hand side) contains a non-deterministic choice in the initial
state $(\emptyset,\mathbf{0})$. The second CPOG shown in
Figure~\ref{fig:lts-semantics}(b) contains the same set of behaviours, but
the choice is delayed until action $b$.
Notice that the corresponding LTS is not bisimilar to the previous one.
Finally, the example in Figure~\ref{fig:lts-semantics}(c) shows a CPOG containing a sequential
behaviour $a\rightarrow b \rightarrow d$ (when $x=0$) and a concurrent behaviour
$a\rightarrow b \rightarrow (c + d)$ (when $x=1$). The corresponding LTS shows
the effect of combinatorial state explosion due to concurrency;
the single step transition $\{c,d\}$ is shown with a dotted arc.

\subsubsection*{Reconfiguration example}

\begin{figure}
\begin{centering}
\includegraphics[scale=0.3]{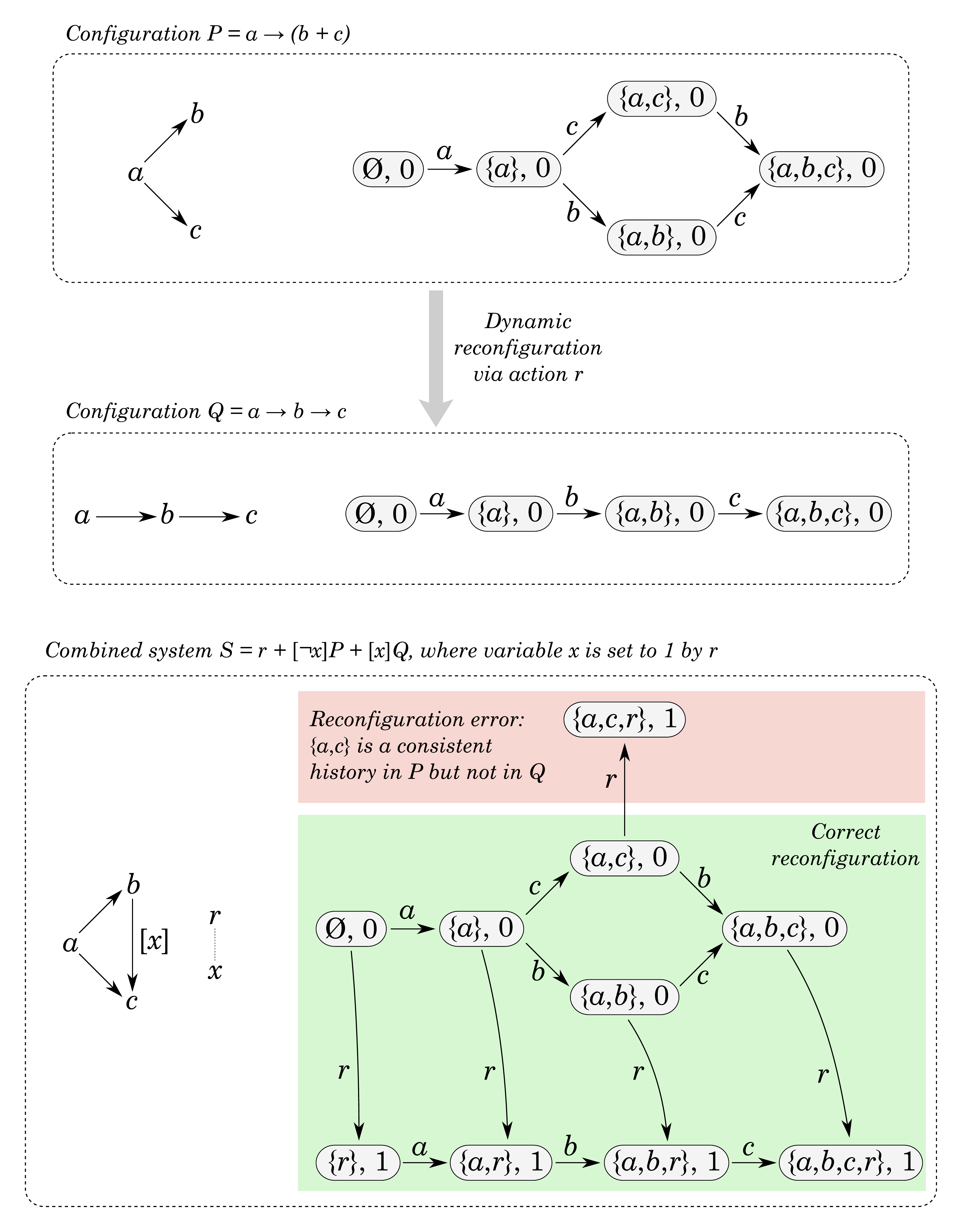}
\caption{CPOG LTS semantics and dynamic reconfiguration\label{fig:lts-semantics-reconfiguration}}
\end{centering}
\end{figure}

We now give an example illustrating CPOG LTS semantics in the presence of dynamic reconfiguration.
Figure~\ref{fig:lts-semantics-reconfiguration}~(top) shows two system configurations:
$P = a \rightarrow (b+c)$, where actions $b$ and $c$ can execute in parallel,
and $Q = a \rightarrow b \rightarrow c$, where $b$ and $c$ are executed in sequence.
This example is based on our case study,
with $b$ and $c$ corresponding to $\mathsf{Shipping}$ and $\mathsf{Billing}$ respectively.

We algebraically combine $P$ and $Q$ to form $S = r + [\neg x]P + [x]Q$,
where variable $x$ indicates the completion of the \emph{reconfiguration action}
$r$ that reconfigures the system from $P$ to $Q$. The LTS semantics of the complete
system $S$ is shown in Figure~\ref{fig:lts-semantics-reconfiguration}~(bottom).
Notice that actions $\{a, b, c\}$ and the reconfiguration action $r$ are concurrent.
The LTS shows the \emph{interference} between the reconfiguration and computation actions,
specifically, if $r$ is executed after $c$ in state $(\{a, c\}, 0)$ of $P$
then the history $\{a, c\}$ of the resulting state $(\{a, c, r\}, 1)$ with $r$ elided is an inconsistent history of $Q$,
since $c$ cannot execute before $b$ in $Q$.
In the next section, we discuss interference in more detail
and show how we can derive \emph{reconfiguration guidelines}
that restrict concurrency and guarantee the system does not reach an inconsistent state due to reconfiguration.
In this example, such a guideline is: \emph{reconfiguration action $r$ should be executed before action $c$},
which makes the inconsistent state $(\{a, c, r\}, 1)$ unreachable.

The above example illustrates that CPOG theory and LTS semantics apply to both computation and reconfiguration actions.
Therefore, a special reconfiguration rule is not required.

\subsection{Dynamic reconfiguration\label{cpog-dynamic}}

In this section, we show how CPOGs can be used to reason about dynamic reconfiguration,
in particular, how to prove reconfiguration requirement R2 in Section~\ref{sec:wfreqts12}.
We start by elaborating the important notion of \emph{history} from the previous section,
which allows reasoning about states of a system whose behaviour is described by a CPOG specification.

A \emph{history} $H \!\subseteq\! \mathcal{A}$ is a set of actions that have
occured in a system up to a certain moment in time.
The set must be \emph{causally closed}, that is, if an action has occured then
all the actions it depends on must also have occurred.
However, since CPOGs are capable of describing not just single workflows but
families of workflows, the notion of causality must be clarified.
In fact, we can only talk about \emph{conditional causality},
where an action~$b$ depends on another action~$a$ under a condition~$x$, or
algebraically: $[x](a \rightarrow b)$.
This leads to the following notion of consistency.

Given a history $H$ and a CPOG specification $S$,
we define the \emph{consistency condition} $C(H,S)$ under which all actions in $H$ could have occurred
without violating the specification $S$ as follows:
\begin{equation}\label{eq:consistent_history}
C(H,S) \triangleq \bigwedge_{a\in H}{f_a} ~\wedge \!\!\!\bigwedge_{a\notin H,\  b\in H}{\!\!\!\!\!\overline{f_{ab}}}
\end{equation}
where $f_a$ and $f_{ab}$ are from the canonical form of the specification $S$
(see Proposition~\ref{prop-canonical}). In other words, a history $H$ is
consistent with a specification $S$ if and only if:
\begin{itemize}
  \item Every $a \!\in\! H$ must be allowed by the specification, that is, its condition $f_a$ must be satisfied.
  \item For each pair of actions $a$ and $b$ such that $a$ is not in the history but $b$ is,
        the dependency $a \rightarrow b$ must not be in the specification, that is, $f_{ab}$ must not be satisfied.
\end{itemize}

To clarify the above, consider the following example.
Let $H=\{\mathsf{Start}, \mathsf{OrderReceipt}, \mathsf{InventoryCheck}, \mathsf{Reject}\}$.
The consistency condition for $H$ with respect to Configuration~1 is $C(H, c_1) = \overline{x}$,
which means that history~$H$ can only be consistent if the
inventory check failed, as otherwise action \textsf{Reject} should causally
depend on \textsf{CreditCheck}, but the latter is not present in $H$.
The consistency condition with respect to Configuration~2 is $C(H, c_2) = \textit{false}$,
because to be rejected an order must either fail the supplier check or fail the credit check, but neither of them is in $H$.
Therefore, we call history $H$ \emph{inconsistent} with the specification $c_2$.
The reader may recognise that the notion of a history is related to (and in fact is quite similar to)
the notion of a \emph{conflict free subset of an event structure}~\cite{1981_winskel_tcs}.

Having defined the notion of consistency, we can now formulate the circumstances
under which a system can be reconfigured from one configuration to another.
If a system's state is described by a history $H$ then it can
be dynamically reconfigured from $S_1$ to $S_2$ under the condition that $H$
is consistent with both $S_1$ and $S_2$, that is:
\[
C(H, S_1) \wedge C(H, S_2)
\]
In other words, both $S_1$ and $S_2$ must be \emph{compatible} with respect to history $H$.
Referring to our previous example, we can say that specifications
$c_1$ and $c_2$ are not compatible with respect to history \\
$\{\mathsf{Start}, \mathsf{OrderReceipt}, \mathsf{InventoryCheck}, \mathsf{Reject}\}$,
because $C(H, c_1) \wedge C(H, c_2) = \overline{x} \wedge \textit{false} = \textit{false}$.

To model unplanned reconfiguration we introduce new \emph{reconfiguration actions}
that can add and remove workflow actions and/or requirements on their order
by modifying the graph family
(i.e. adding new graphs and removing existing graphs that are no longer relevant).
Notice that reconfiguration actions can
occur concurrently with workflow actions and can also have requirements imposed on their order.

Consider a concurrent reconfiguration action $r$ that changes the system's
configuration from $c_1$ to $c_2$. The combined family of graphs that contains
$c_1$, $c_2$, and $r$ can be specified as follows:
\begin{equation}\label{eq:reconfiguration_model}
S \triangleq r + [\text{\textlnot}\mathrm{r\_done}]c_{1}+[\mathrm{r\_done}]c_{2}
\end{equation}
where predicate `$\mathrm{r\_done}$' is \textit{true} after the execution of
reconfiguration action $r$, and \textit{false} before that. Notice the initial
specification $c_1$ is defined independently of the reconfiguration action $r$
and the new specification $c_2$. Therefore, the reconfiguration is unplanned.

We can now compute a set $\mathcal{R}(r,S)$ of \emph{safe reconfiguration histories}
by finding consistent histories $H$ that remain consistent after the execution of action $r$:
\begin{equation}
\mathcal{R}(r,S) \triangleq \{ H\ |\ C(H, S)\wedge C(H \cup \{r\}, S) \neq \textit{false} \}\label{eq:reconfiguration}
\end{equation}
For example, we know that:
\[
\{\mathsf{Start}, \mathsf{OrderReceipt}, \mathsf{InventoryCheck}, \mathsf{Reject}\} \!\notin\! \mathcal{R}(r,S)
\]
However, if we drop action \textsf{Reject} the result is a safe reconfiguration history:
\[
\{\mathsf{Start}, \mathsf{OrderReceipt}, \mathsf{InventoryCheck}\} \!\in\! \mathcal{R}(r,S)
\]

An important practical question arises at this point:
is it possible to derive \emph{reconfiguration guidelines} from $\mathcal{R}(r,S)$
such that implementing the guidelines will ensure the safe reconfiguration of the system?
We give a positive and constructive answer below, but with no claim for optimality.

To derive reconfiguration guidelines for the specification $S$ defined above,
notice that all histories which are consistent with $S$ and do not contain the
actions $\mathsf{Reject}$ or $\mathsf{Confirmation}$ also belong to $\mathcal{R}(r,S)$:
\begin{equation}
\forall H (H \cap RC = \emptyset \wedge C(H, S) \neq \textit{false} \Rightarrow
H \!\in\! \mathcal{R}(r,S))\label{eq:forbidden-actions}
\end{equation}
where $RC=\{\mathsf{Reject},\mathsf{Confirmation}\}$ is a set of \emph{forbidden actions}.
It is easy to check that any history that does not contain the forbidden actions is consistent with Configurations 1 and 2
by inspecting Figures~\ref{OfficeWorkflowConfig1} and~\ref{OfficeWorkflowConfig2} respectively
(Appendix~\ref{sec:appendixcpogs} illustrates this using graphical representations of the consistent histories of $c_1$ and $c_2$).
Therefore, we can formulate the following guideline:
the reconfiguration should only be permitted when no forbidden action has occured.
This guideline can be enforced by transforming the specification $S$ into $S_\text{safe}$ as follows:
\begin{equation}\label{eq:reconfiguration_safe}
S_\text{safe} \triangleq S + r \rightarrow (\mathsf{Reject} + \mathsf{Confirmation})
\end{equation}
That is, we require action $r$ to occur before the forbidden actions.  

Similarly, we can ensure that the reverse reconfiguration (from $c_2$ to $c_1$)
is safe by the following specification:
\begin{equation}\label{eq:reconfiguration_safe_rev}
S^{\text{rev}}_\text{safe} \triangleq r \rightarrow (\mathsf{SupplierCheck}
+\mathsf{Reject} + \mathsf{Billing})\ + [\text{\textlnot}\mathrm{r\_done}]c_{2}+[\mathrm{r\_done}]c_{1}
\end{equation}
The reason for forbidding action $\mathsf{Billing}$, which seems innocuous,
is that $\mathsf{Billing}$ must occur after $\mathsf{Shipping}$ in Configuration~1.
If we do not forbid $\mathsf{Billing}$, then a reconfiguration from $c_2$ to $c_1$ can bring the
system to a state in which $\mathsf{Billing}$ has occurred but $\mathsf{Shipping}$ has not, which is inconsistent with $c_1$.

%
%

\subsection{Implementation}
\label{cpog-implementation}

We have automated CPOG transformation and verification procedures described in
this section in a prototype implementation as an embedded domain-specific
language~(DSL) in Haskell~\cite{1996_hudak_dsl}, although the implementation is not yet publicly available.
We have also successfully cross-checked the results using Maude, a well-known
environment for rewriting logic~\cite{2014_maude_www}. However, implementing
CPOG axioms and transformations as a collection of rewrite rules in Maude
was a fragile and challenging process: any new rewrite rule could
trigger non-termination of the tool's rewrite engine, requiring its manual shutdown and restart.
Therefore, we decided to focus our implementation effort on the DSL,
where we had full control over the internal CPOG-specific rewrite engine.
An additional benefit of embedding our DSL in Haskell was that
the DSL acquired abstraction, sharing, and parameterisation capabilities essentially for free
due to the purity and rich type system of Haskell.

As shown in~\cite{2009_mokhov_phd}, most interesting properties that can be
defined on CPOGs are reducible to the Boolean Satisfiability~(SAT)~problem~\cite{2004_miniSAT_lncs}.
For example, checking the equivalence of two CPOGs requires a pairwise comparison of
conditions $f_a$ and $f_{ab}$ of their canonical forms for equality, and each such comparison is trivially a SAT problem.
When working with Maude we relied on its built-in SAT solver, but with our Haskell-based implementation
we decided to use Binary Decision Diagrams~(BDDs)~\cite{1959_lee_bdd}
to store CPOG conditions in a canonical form and to share of their common subexpressions.
We noticed that in practice, actions and dependencies between actions typically have the same or similar conditions;
canonical forms of the expressions $c_1$ and $c_2$ are good examples of this phenomenon.

\section{Basic ${\bf CCS^{dp}}$}\label{sec:bccsdp}


Basic $\mathrm{CCS^{dp}}$ is a two-sorted process algebra based on basic $\mathrm{CCS}$ \cite{kn:Mil89},
which is extended with a single construct -- the fraction process $\frac{P'}{P}$ -- in order to describe unplanned process reconfiguration.
Therefore, basic $\mathrm{CCS^{dp}}$ has a behavioural approach to describing a system.
One sort is used to represent general purpose computation actions, including communication,
the other sort is used to represent process reconfiguration actions,
and interference between the two kinds of action is represented by interleaving actions.
The process expressions are amenable to equational reasoning using the theory developed in \cite{kn:Bha13} and to model checking.
We proceed by defining the syntax and semantics of basic $\mathrm{CCS^{dp}}$ briefly,
modelling the case study, then attempting to verify reconfiguration requirement R3;
an overall evaluation of the formalism is given in Section \ref{sec:comp}.

It is important to notice that basic $\mathrm{CCS^{dp}}$ is an experimental process algebra
developed as a test bed in order to explore the properties of the fraction process, define its semantics, and develop its theory.
Having done this in \cite{kn:Bha13}, it should be possible to import the fraction process and its transition rules into any process algebra
that enables reaction between processes using the parallel composition operator,
and thereby support the modelling of unplanned process reconfiguration.
Thus, the fraction process is intended as a 'plug-in' for process algebras, and basic $\mathrm{CCS^{dp}}$ is the first demonstrator.

\subsection{Syntax}\label{sec:bccsdpsyntax}

Let $\mathcal{N}$ be the countable set of names (e.g. $a$, $b$, $c$)
that represent both input ports and input actions of the processes in basic $\mathrm{CCS^{dp}}$;
and let $\mathcal{\overline{N}}$ be the countable set of complementary names (e.g. $\overline{a}$, $\overline{b}$, $\overline{c}$)
that represent both output ports and output actions of the processes in basic $\mathrm{CCS^{dp}}$,
where $\mathcal{\overline{N}} \triangleq \{ \overline{l} ~|~ l \!\in\! \mathcal{N} \}$.
Let $\mathcal{PN}$ be the countable set of names (e.g. $A$, $B$, $C$) of the processes in basic $\mathrm{CCS^{dp}}$.
The sets $\mathcal{N}$, $\overline{\mathcal{N}}$, and $\mathcal{PN}$ are assumed to be pairwise disjoint.

\noindent Thus, given $a \!\in\! \mathcal{N}$, $a$ represents the input action on the input port $a$ of a process;
and $\overline{a}$ represents the complementary output action on the output port $\overline{a}$ of a process.
Internal action of a process, such as the interaction between complementary actions $a$ and $\overline{a}$,
is represented by the special action $\tau$.

\noindent Let $\mathcal{L}$ be the set of names that represent both ports and actions of the processes in basic $\mathrm{CCS^{dp}}$, where \\
$\mathcal{L} \triangleq \mathcal{N} \cup \mathcal{\overline{N}}$. As usual in CCS, $\forall l \!\in\! \mathcal{N}(\overline{\overline{l}} = l)$.

\noindent Let $\mathcal{I}$ be the set of input and output ports/actions of the processes in basic $\mathrm{CCS^{dp}}$
and their internal action, where $\mathcal{I} \triangleq \mathcal{L} \cup \{\tau\}$.

\noindent Let $\mathcal{P}$ be the set of processes in basic $\mathrm{CCS^{dp}}$.

\noindent The syntax of a process $P$ in $\mathcal{P}$ is defined as follows:
\begin{equation*}\label{eq:ccsdpnonusyndef}
P \;::=\; PN\!<\!\wt{\beta}\!> \;\mid\; 0 \;\mid\; \sum_{i \in I}\alpha_i.P \;\mid\; P|P \;\mid\; \mbox{$\frac{P}{P}$}
\end{equation*}
where $PN \!\in\! \mathcal{PN}$,
$\wt{\beta}$ is a tuple of elements of $\mathcal{L}$, $\alpha \!\in\! \mathcal{I}$, and $I$ is a finite indexing set.

\noindent Thus, the syntax of basic $\mathrm{CCS^{dp}}$ is the syntax of basic $\mathrm{CCS}$ without the restriction operator ($\nu$)
and extended with the $\frac{P'}{P}$ construct.

\noindent As in $\mathrm{CCS}$, $0$ is the $NIL$ process, which has no behaviour.
Prefix (e.g. $\alpha.P$) models sequential action.
Summation (e.g. $\alpha_1.P_1 + \alpha_2.P_2$) models non-deterministic choice of actions by a process.
Notice that a non-$0$ term in a summation is guarded by a prefix action in order to prevent
the creation of an infinite number of processes, which complicates reasoning.
$A\!\!<\!\!\wt{\beta}\!\!>$ models the invocation of a constant process named $A$,
instantiated with a tuple of port/action names $\wt{\beta}$.
$A(\wt{\beta})$ has a unique definition, which can be recursive.
As usual in CCS, $A(\wt{\beta})$ is used to define $A$, and $A\!\!<\!\!\wt{\beta}\!\!>$ is used to represent an instance of $A$
(e.g. $A(a, b) \triangleq a.b.A\!\!<\!\!a, b\!\!>$).
Parallel composition (e.g. $P|P'$) models the execution of concurrent processes and their direct functional interaction,
as well as process composition and decomposition. Interaction between processes is synchronous and point-to-point.

\noindent A \emph{fraction} (e.g. $\frac{P'}{P}$) is a process that models process replacement and deletion.
On creation, the fraction $\frac{P'}{P}$ identifies any instance of a process
matching its denominator process $P$ with which it is composed in parallel,
and replaces that process atomically with the numerator process $P'$.
If no such process instance exists,
the fraction continues to exist until such a process is created
(or the fraction is itself deleted or replaced).
If there is more than one such process instance,
a non-deterministic choice is made as to which process is replaced.
Similarly, if more than one fraction can replace a process instance,
a non-deterministic choice is made as to which fraction replaces the process.
Deletion of a process P is achieved by parallel composition with \mbox{$\frac{0}{P}$}.
If $P$ progresses to $Q$, then \mbox{$\frac{P'}{P}$} will not replace $Q$ by $P'$
(unless $Q$ matches $P$). Notice that a fraction has no communication behaviour;
its only behaviour is to replace a process with which it is composed in parallel that matches its denominator.
The matching is done by behaviour using a bisimulation, as explained in the following section.

The precedence of the operators (in decreasing order) is:
fraction formation, relabelling, prefix, summation, parallel composition.

\subsection{LTS Semantics}\label{sec:bccsdpsemantics}

Let $\mathcal{R}$ be the countable set of reconfiguration actions of the processes in $\mathcal{P}$
(e.g. $\rho_X$, $\rho_Y$, $\rho_Z$) that create a process in $\mathcal{P}$;
and let $\mathcal{\overline{R}}$ be the countable set of complementary reconfiguration actions of the processes in $\mathcal{P}$
(e.g. $\overline{\rho}_X$, $\overline{\rho}_Y$, $\overline{\rho}_Z$) that delete a process in $\mathcal{P}$, where
$\mathcal{\overline{R}} \triangleq \{ \overline{\rho}_X ~|~ \rho_X \!\in\! \mathcal{R} \}$
(see the $Creat$ and $Delet$ rules below).
Each action in $\mathcal{R}$ is represented by $\rho_X$, with $X \!\in\! \mathcal{P}$.
The sets $\mathcal{N}$, $\overline{\mathcal{N}}$, $\{\tau\}$,
$\mathcal{R}$, $\overline{\mathcal{R}}$, and $\mathcal{PN}$ are assumed to be pairwise disjoint\protect\footnote{The reconfiguration actions
$\rho_X$, $\overline{\rho}_X$ are written as $\tau_{r_X}$, $\overline{\tau}_{r_X}$ respectively in \cite{kn:Bha13}.
The change to $\rho_X$, $\overline{\rho}_X$ simplifies the notation, and is due to advice from Kohei Honda.}.

\noindent The interaction between complementary reconfiguration actions (such as $\rho_X$ and $\overline{\rho}_X$)
results in the replacement of one or more processes (see the $Creat$, $Delet$, and $React$ rules below).

\noindent Let $\mathcal{C}$ be the set of reconfiguration actions of the processes in $\mathcal{P}$,
where $\mathcal{C} \triangleq \mathcal{R} \cup \mathcal{\overline{R}}$.
As before, $\forall \lambda \!\in\! \mathcal{L} \cup \mathcal{C} \,(\overline{\overline{\lambda}} = \lambda)$.

\noindent Let $\mathcal{A}$ be the set of actions of the processes in $\mathcal{P}$,
where $\mathcal{A} \triangleq \mathcal{I} \cup \mathcal{C}$.

\noindent The labelled transition system (LTS) rules for basic $\mathrm{CCS^{dp}}$ are a superset of the LTS rules for basic $\mathrm{CCS}$
without the restriction operator ($\nu$), consisting of an unchanged rule of basic $\mathrm{CCS}$ (i.e. $Sum$)
plus basic $\mathrm{CCS}$ rules applicable to reconfiguration transitions (i.e. $React$, \textit{L-Par}, \textit{R-Par}, and $Ident$)
plus additional rules to describe new reconfiguration behaviour (i.e. $Creat$, $Delet$, $CompDelet$, \textit{L-React}, and \textit{R-React}).
See Table \ref{tab:ccsdpnonultss}.

\renewcommand{\arraystretch}{3.0}
\begin{table}
\[
\begin{array}{|llll|}
\hline
\;\; \mbox{Sum} &
\frac{k \in I}
{\sum_{i \in I} \alpha_{i}.P_{i} \trans{\alpha_{k}} P_{k}}
\;\; \textrm{where $I$ is a finite indexing set} && \\
\;\; \mbox{React} &
\frac{\lambda \in \mathcal{L} \cup \mathcal{C} \;\; \wedge \;\;
P \trans{\lambda} P' \;\; \wedge \;\;
Q \trans{\overline{\lambda}} Q'}
{P | Q \trans{\tau} P' | Q'} && \\
\;\; \mbox{L-Par} &
\frac{\mu \in \mathcal{A} \;\; \wedge \;\; P \trans{\mu} P'}
{P | Q \trans{\mu} P' | Q} &
\;\;\;\; \mbox{R-Par} &
\frac{\mu \in \mathcal{A} \;\; \wedge \;\; Q \trans{\mu} Q'}
{P | Q \trans{\mu} P | Q'} \;\; \\
\;\; \mbox{Ident} &
\frac{|\wt{b}| = |\wt{a}| \;\; \wedge \;\;
\mu \in \mathcal{A} \;\; \wedge \;\;
P[\frac{\wt{b}}{\wt{a}}] \trans{\mu} P'}
{A<\wt{b}> \trans{\mu} P'}
\;\; \textrm{where $A(\wt{a}) \triangleq P$} && \\
\;\; \mbox{Creat} &
\frac{P \sim_{of} Q \;\; \wedge \;\; P \in \mathcal{P}^{+}}
{\frac{P'}{P} \trans{\rho_Q} P'} &
\;\;\;\; \mbox{Delet} &
\frac{P \sim_{of} Q \;\; \wedge \;\; P \in \mathcal{P}^{+}}
{P \trans{\overline{\rho}_Q} 0} \;\; \\
\;\; \mbox{CompDelet} &
\frac{R \sim_{of} R_1 | R_2 \;\; \wedge \;\;
P \stackrel{\overline{\rho}_{R_1}}{\longrightarrow} P' \;\; \wedge \;\;
P' \stackrel{\overline{\rho}_{R_2}}{\longrightarrow} P''}
{P \stackrel {\overline{\rho}_R}{\longrightarrow} P''} && \\
\;\; \mbox{L-React} &
\frac{R \sim_{of} R_1 | R_2 \;\; \wedge \;\;
P \stackrel{\overline{\rho}_{R_1}}{\longrightarrow} P' \;\; \wedge \;\;
P' \stackrel{\rho_R}{\longrightarrow} P'' \;\; \wedge \;\;
Q \stackrel{\overline{\rho}_{R_2}}{\longrightarrow} Q'}
{P | Q \stackrel {\tau}{\longrightarrow} P'' | Q'} &
\;\;\;\; \mbox{R-React} &
\frac{R \sim_{of} R_1 | R_2 \;\; \wedge \;\;
P \stackrel{\overline{\rho}_{R_1}}{\longrightarrow} P' \;\; \wedge \;\;
Q \stackrel{\overline{\rho}_{R_2}}{\longrightarrow} Q' \;\; \wedge \;\;
Q' \stackrel{\rho_R}{\longrightarrow} Q''}
{P | Q \stackrel {\tau}{\longrightarrow} P' | Q''} \;\; \\[2ex]
\hline
\end{array}
\]
\caption{Labelled transition system semantics of basic $\mathrm{CCS^{dp}}$}\label{tab:ccsdpnonultss}
\end{table}

\noindent The $Sum$ rule states that summation preserves the transitions of constituent processes
as a non-deterministic choice of alternative transitions.
For example, the process \small $\overline{InventoryCheckNotOK}_o.0 + \overline{InventoryCheckOK}_o.0$ \normalsize
can either output \small $\overline{InventoryCheckNotOK}_o$ \normalsize or output \small $\overline{InventoryCheckOK}_o$ \normalsize
non-deterministically then terminate.

\noindent The $React$ rule states that if two processes can perform complementary transitions,
then their parallel composition can result in a $\tau$ transition in which both processes
undergo their respective complementary transitions atomically.
Notice that this rule results from overloading the semantics of the parallel composition operator ($|$).
The rule has been extended from $\mathcal{L}$ to $\mathcal{L} \cup \mathcal{C}$
in order to be applicable to processes that can perform complementary reconfiguration transitions.
For example, consider the processes \small $CONFIG^1$ \normalsize and \small $CONFIG^2$ \normalsize defined in Section \ref{sec:bccsdpmodelling}.
The fraction process \small $\frac{CONFIG^2}{CONFIG^1}$ \normalsize can perform \small $\rho_{CONFIG^1}$ \normalsize
to become \small $CONFIG^2$ \normalsize by the $Creat$ rule because \small $CONFIG^1$ \normalsize is a positive process (defined below),
and \small $CONFIG^1$ \normalsize can perform \small $\overline{\rho}_{CONFIG^1}$ \normalsize to become \small $0$ \normalsize by the $Delet$ rule.
Therefore, \small $CONFIG^1 ~|~ \frac{CONFIG^2}{CONFIG^1}$ \normalsize can become \small $0 ~|~ CONFIG^2$ \normalsize by the $React$ rule
because \small $\overline{\rho}_{CONFIG^1}$ \normalsize and \small $\rho_{CONFIG^1}$ \normalsize are complementary actions in $\mathcal{C}$.
Thus, the fraction process \small $\frac{CONFIG^2}{CONFIG^1}$ \normalsize reconfigures \small $CONFIG^1$ \normalsize to \small$CONFIG^2$\normalsize,
since $0$ is the identity process.
Notice also that the reconfiguration actions are \textbf{not} port names, since $\mathcal{L}$ and $\mathcal{C}$ are disjoint.
Therefore, the reconfiguration reaction does not require a communication channel.
Hence, fraction processes specify the effects of process reconfiguration rather than describe mechanisms for implementing process reconfiguration.

\noindent The \textit{L-Par} and \textit{R-Par} rules state that
parallel composition preserves the transitions of constituent processes, including reconfiguration transitions.
For example, consider the processes \small $ARC^1$\normalsize, \small $ARC^2$\normalsize, and \small $ARCH^1$ \normalsize
defined in Section \ref{sec:bccsdpmodelling}. The process \small $\frac{ARC^2}{ARC^1} ~|~ \frac{0}{ARCH^1}$ \normalsize
can perform \small $\rho_{ARC^1}$ \normalsize to become \small $ARC^2 ~|~ \frac{0}{ARCH^1}$ \normalsize by the $Creat$ and \textit{L-Par} rules,
or it can perform \small $\rho_{ARCH^1}$ \normalsize to become \small $\frac{ARC^2}{ARC^1} ~|~ 0$ \normalsize by the $Creat$ and \textit{R-Par} rules.
The relationship between the \textit{L-Par} and \textit{R-Par} rules and the other LTS rules
indicates that the granularity of process reconfiguration using fraction processes is a single concurrent process.


\noindent $Creat$ and $Delet$ are the key rules of dynamic process reconfiguration.
The $Creat$ rule states that if $P$ is a positive process ($P \!\in\! \mathcal{P}^+$, see Definition \ref{def:posprocs} below)
that matches $Q$ using strong of-bisimulation ($P \sim_{of} Q$, see Definition \ref{def:strofbisim} below),
then the fraction process $\frac{P'}{P}$ can perform the reconfiguration transition $\rho_Q$ that results in the creation of $P'$.
The $Delet$ rule is complementary to the $Creat$ rule,
and states that if $P$ is a positive process that matches $Q$ using strong of-bisimulation,
then $P$ can be deleted by performing the reconfiguration transition $\overline{\rho}_Q$
that is complementary to the reconfiguration transition $\rho_Q$ performed by some fraction that creates a process.
Thus, $P'$ replaces $P$ as a result of the reaction between $\frac{P'}{P}$ performing $\rho_Q$ and
$P$ performing $\overline{\rho}_Q$ (defined by the $React$ rule).

\noindent The hypotheses of $Creat$ and $Delet$
use the notions of positive process ($P \!\in\! \mathcal{P}^{+}$) and strong of-bisimulation ($\sim_{of}$).
The reconfiguration transitions are restricted to positive processes
in order to retain the identity property of $0$ in parallel compositions up to $\sim_{of}$,
so that $0$ and processes similar to $0$ (i.e. processes with no behaviour) can be ignored, which simplifies process matching.
In contrast, a positive process is a process with behaviour, that is,
a process that can communicate, or can perform an internal action, or can perform a reconfiguration action on another process.
The notion is defined as follows: \vspace{-0.1cm}

\begin{defn}\label{def:posprocs}
\noindent $\mathcal{P}^{+}$ denotes the set of \emph{positive processes} of $\mathcal{P}$,
which is the smallest subset of $\mathcal{P}$ that satisties the following conditions: \\
\renewcommand{\arraystretch}{1.5}
\(
\begin{array}{ll}
1. & \forall \alpha \!\in\! \mathcal{I} ~\forall p \!\in\! \mathcal{P} ~(\alpha.p \!\in\! \mathcal{P}^{+}) \\
2. & \forall p, q \!\in\! \mathcal{P}
~(p + q \!\in\! \mathcal{P} \wedge (p \!\in\! \mathcal{P}^{+} \vee q \!\in\! \mathcal{P}^{+}) \Longrightarrow p + q \!\in\! \mathcal{P}^{+}) \\
3. & \forall p, q \!\in\! \mathcal{P}
~(p \!\in\! \mathcal{P}^{+} \vee q \!\in\! \mathcal{P}^{+} \Longrightarrow p|q \!\in\! \mathcal{P}^{+}) \\
4. & \forall p \!\in\! \mathcal{P} ~\forall q \!\in\! \mathcal{P}^{+} \left(\frac{p}{q} \!\in\! \mathcal{P}^{+}\right) \\
5. & \forall \beta \!\in\! \mathcal{I} ~\forall X \!\in\! \mathcal{PN} ~(\beta.X \!\in\! \mathcal{P}^{+})
\end{array}
\)
\end{defn} \vspace{-0.1cm}

\noindent Rule 4 above enables recursive fraction processes to be defined that can reconfigure other fraction processes using,
for example, the $Creat$, $Delet$, and $React$ rules.
Therefore, the reconfiguration of a reconfiguration system can be modelled,
so that the dynamic evolution of a system throughout its lifetime can be modelled in a uniform manner. \vspace{-0.1cm}

\begin{defn}\label{def:strofbisim}
\noindent \emph{Strong of-bisimulation} ($\sim_{of}$) is the largest symmetric binary relation on $\mathcal{P}$ such that
the following condition holds $\forall (p, q) \in\, \sim_{of}$ \vspace{0.1cm} \\
\(
\forall \alpha \!\in\! \mcIsb{I}{p} \cup \mathcal{R}_p ~\forall p' \!\in\! \mathcal{P}
    ~(p \stackrel{\alpha}{\longrightarrow} p' \Longrightarrow
     \alpha \!\in\! \mcIsb{I}{q} \cup \mathcal{R}_q \wedge
     \exists q' \!\in\! \mathcal{P} ~(q \stackrel{\alpha}{\longrightarrow} q' \wedge (p', q') \in\, \sim_{of}))
\) \\[0.2cm]
where $\mcIsb{I}{p} \cup \mathcal{R}_p$ and $\mcIsb{I}{q} \cup \mathcal{R}_q$ are the sets of actions in $\mathcal{I} \cup \mathcal{R}$
that $p$ and $q$ can perform respectively.
\end{defn} \vspace{-0.1cm}

\noindent Strong of-bisimulation is used for process matching for two reasons that have several advantages.
First, strong of-bisimulation is a relation between processes.
One advantage of using a relation is that modelling reconfiguration mechanisms is avoided, which simplifies models through abstraction.
A second advantage is that special reconfiguration operators that require syntactic proximity of their operands are avoided,
so that the reconfiguring processes can be located in the context of the system model,
which enables dynamic evolution of a system due to changes in its environment to be modelled abstractly.
A third advantage of syntactic separation of the reconfiguring processes and the configuration model is modularity:
syntactically separate models of a configuration and its reconfiguring processes can be produced
that interact when composed in parallel and result in the required reconfiguration.
A fourth advantage is that a relation defines a pre-condition that allows a process to be reconfigured only when it is in a specified state,
which is an important requirement for reconfigurable systems.
The second reason is that strong of-bisimulation (rather than structural congruence or syntactic equality)
helps to maximize the terseness of expressions modelling reconfiguration, which simplifies models through abstraction.
Notice that restriction is not currently used in basic $\mathrm{CCS^{dp}}$ because restriction renders strong of-bisimuation undecidable \cite{kn:ChrHirMol94},
which complicates tool support for process matching based on strong of-bisimuation.

\noindent The mutual dependency between LTS transitions and strong of-bisimulation suggests the dependency is circular, which is problematic.
However, the dependency is an inductive relationship if the depth of fractional recursion of a process is bounded suitably.
Therefore, we restrict $\mathcal{P}$ to the domain of the \textit{sfdrdepth} function (defined below),
which creates an inductive relationship between LTS transitions and strong of-bisimulation, and thereby avoids a circular dependency.
\textit{sfdrdepth} is defined as follows:

\noindent \( succ : \mathcal{P} \; \text{x} \; \mathbb{N} \longrightarrow \mathbb{P} ~\mathcal{P} ~\textrm{ such that }~
succ(p,i) \triangleq
\left\{ \begin{array}{ll}
\!\!\!\{p\}                                                                                                \!\!\! & \textrm{if} \,\, i = 0 \\
\!\!\!\{q' \!\in\! \mathcal{P} \,|\,
   \exists q \!\in\! succ(p, i-1)
      (\exists \alpha \!\in\! \mathcal{I}_q \cup \mathcal{R}_q (q \stackrel{\alpha}{\longrightarrow} q'))\}\!\!\! & \textrm{else}
\end{array} \right. \)

\noindent $succ(p, i)$ is the set of $i^{th}$ \emph{successor processes} (or equivalently, \emph{$i^{th}$ successors}) of $p$.
That is, the set of processes reached after $i$ consecutive transitions in $\mathcal{I} \!\cup \mathcal{R}$
starting from $p$, with $succ(p, 0) = \{p\}$.

\noindent \( successors : \mathcal{P} \longrightarrow \mathbb{P} ~\mathcal{P} ~\textrm{ such that }~
successors(p) \triangleq \bigcup_{i \in \mathbb{N}} succ(p, i) \)

\noindent $successors(p)$ is the set of all the successors of $p$, including $p$.
That is, the set of all the processes reached after zero, one or more consecutive transitions in $\mathcal{I} \!\cup \mathcal{R}$
starting from $p$.

\noindent \( \textit{sfdrdepth} : \mathcal{P} \longrightarrow \mathbb{N} ~\textrm{ such that }~
\textit{sfdrdepth}(p) \triangleq max \{ \textit{fdrdepth}(s) ~|~ s \!\in\! successors(p) \} \) with

\noindent \( \textit{fdrdepth} : \mathcal{P} \longrightarrow \mathbb{N} ~\textrm{ such that }~
\textit{fdrdepth}(s) \triangleq \left\{ \begin{array}{ll}
\!\!\!0                                                                    & \textrm{if} \; \mathcal{R}_s = \emptyset \\
\!\!\!1 + max\{ \textit{sfdrdepth}(X) ~|~ \rho_X \!\in\! \mathcal{R}_s\}   & \textrm{else}
\end{array} \right. \)

\noindent Thus, for any process $p$ in basic $\mathrm{CCS^{dp}}$,
$\textit{sfdrdepth}(p)$ is the maximum depth of fractional recursion of $p$ and its successors. \\
For example, for any process $p$ in basic CCS, any successor $p'$ of $p$ is also a process in basic CCS \\
   (by definition of the LTS rules of basic CCS) \\
$\Longrightarrow \textit{fdrdepth}(p') = 0$
   (by definition of \textit{fdrdepth}, $\because \mathcal{R}_{p'} = \emptyset$) \\
$\Longrightarrow \textit{sfdrdepth}(p) = 0$
   (by definition of \textit{sfdrdepth}, because $p'$ is arbitrary). \\
For the fraction process $\frac{b.0}{a.0}$, $successors(\frac{b.0}{a.0}) = \{\frac{b.0}{a.0}\} \cup successors(b.0)$
   (by definition of $successors$) \\
$\Longrightarrow \textit{sfdrdepth}(\frac{b.0}{a.0}) = max\{\textit{fdrdepth}(s) ~|~ s \!\in\! \{\frac{b.0}{a.0}\} \cup successors(b.0)\}$
   (by definition of \textit{sfdrdepth}) \\
$\Longrightarrow \textit{sfdrdepth}(\frac{b.0}{a.0}) = \textit{fdrdepth}(\frac{b.0}{a.0})$
   (by the previous example, because $b.0$ is a process in basic CCS) \\
$\Longrightarrow \textit{sfdrdepth}(\frac{b.0}{a.0}) = 1 + max\{\textit{sfdrdepth}(a.0)\}$
   (by definition of \textit{fdrdepth}, $\because \rho_{a.0} \!\in\! \mathcal{R}_{\frac{b.0}{a.0}}$) \\
$\Longrightarrow \textit{sfdrdepth}(\frac{b.0}{a.0}) = 1$ ($\because \textit{sfdrdepth}(a.0) = 0$, by the previous example).

\noindent The remaining three LTS rules facilitate the reconfiguration of multiple concurrent processes by a single fraction process through a single transition.
$CompDelet$ states that consecutive delete transitions of a process can be composed into a single delete transition of the process.
The rule is applicable only if it is used in combination with \textit{L-Par} or \textit{R-Par}.
For example, consider the processes \small $ICH^1$\normalsize, \small $CC^1$\normalsize, and \small $CCH^1$ \normalsize defined in the following section.
The transitions
\small \(
ICH^1 ~|~ CC^1 ~|~ CCH^1 \stackrel{\overline{\rho}_{ICH^1}}{\longrightarrow} 0 ~|~ CC^1 ~|~ CCH^1
                         \stackrel{\overline{\rho}_{CCH^1}}{\longrightarrow} 0 ~|~ CC^1 ~|~ 0
\) \normalsize
can be composed into the single transition \small $ICH^1 ~|~ CC^1 ~|~ CCH^1 \stackrel{\overline{\rho}_{ICH^1 | CCH^1}}{\longrightarrow} 0 ~|~ CC^1 ~|~ 0$\normalsize.
The \textit{L-React} and \textit{R-React} rules enable a fraction process to reconfigure processes on both sides of the fraction through a single transition.
Collectively, $CompDelet$, \textit{L-React}, and \textit{R-React} ensure that parallel composition involving fraction processes
is commutative and associative with respect to strong of-bisimulation, which are necessary in order to model unplanned dynamic reconfiguration abstractly.

\noindent Regarding the soundness of the semantics of basic $\mathrm{CCS^{dp}}$,
the LTS transitions are defined to be the smallest relation on $\mathcal{P}$ that satisfies the LTS rules.
Therefore, a process $p \!\in\! \mathcal{P}$
performs a transition $p \trans{\mu} p'$ with $\mu \!\in\! \mathcal{A}$ and $p' \!\in\! \mathcal{P}$
if and only if the hypothesis of some LTS rule that determines the $p \trans{\mu} p'$ transition is satisfied.
Furthermore, the LTS rules do not contain any negative premise, nor any negative transition.
Therefore, the LTS rules contain no contradiction.
Therefore, the LTS semantics of basic $\mathrm{CCS^{dp}}$ is sound.

\subsection{Modelling}\label{sec:bccsdpmodelling}

The granularity of reconfiguration using the fraction process is a single concurrent process,
and multiple concurrent processes can also be reconfigured by a single fraction process through a single transition.
Thus, it is possible to model all three cases of dynamic reconfiguration identified in Section \ref{sec:intro}
(see Appendix \ref{prf:reconfelabs} for examples of Case 1 and Case 2 reconfigurations using fraction processes).
The scope of reconfiguration of Configuration 1 and of functional interference between application and reconfiguration tasks
is maximum when the model consists of the maximum number of concurrent processes (i.e. model of Design 4 in Section \ref{sec:designs}).
However, it is easiest to demonstrate both the strengths and the weaknesses of basic $\mathrm{CCS^{dp}}$ using Design 1.
Furthermore, the models of all four designs in Section \ref{sec:designs} are similar, and differ only in the location of the recursion.
Therefore, we proceed by modelling Design 1 of Configuration 1 and Configuration 2 and Case 3 of the reconfiguration,
and attempt to verify reconfiguration requirement R3 in the following section;
the models of Designs 2, 3, and 4 of Configuration 1 are given in Appendix \ref{mdl:designs234}.

\subsection*{Modelling Configuration 1}\label{sec:bccsdpmodel1}

Let $O$ be the set of possible order identifiers, which can be infinite.

Configuration 1 consists of a set of tasks.
Each task consists of a set of subtasks, which are modelled as actions of processes in $\mathcal{P}$.
For example, the subtasks of the \texttt{Order Receipt} task are modelled as the actions of the \small $REC$ \normalsize process.
The subtasks of the \texttt{Evaluation} task are modelled as actions of the processes
\small $IC$\normalsize, \small $ICH$\normalsize, \small $CC$\normalsize, and \small $CCH$\normalsize.
The \texttt{Rejection} task is modelled as the actions \small $\overline{RejectIC}_o$ \normalsize and \small $\overline{RejectCC}_o$ \normalsize
in the processes \small $ICH$ \normalsize and \small $CCH$ \normalsize respectively.
The \texttt{Confirmation} task is modelled as the action \small $\overline{Confirm}_o$ \normalsize in the \small $ARCH$ \normalsize process.
Notice that Configuration 1 is a cyclic executive, which is modelled by the recursive definition of \small $CONFIG^1$ \normalsize
following \small $\overline{RejectIC}_o$\normalsize, \small $\overline{RejectCC}_o$\normalsize, and \small $\overline{Confirm}_o$\normalsize.

Configuration 1 of the workflow is denoted by the process \small $CONFIG^1$\normalsize, and \\
\small $CONFIG^1 \triangleq REC^1 ~|~ IC^1 ~|~ ICH^1 ~|~ CC^1 ~|~ CCH^1 ~|~ SHIP^1 ~|~ BILL^1 ~|~ ARC^1 ~|~ ARCH^1$ \normalsize

\small $REC^1 \triangleq \sum_{o \in O} Receipt_o.\overline{InventoryCheck}_o$ \normalsize
~and denotes the \texttt{Order Receipt} task. By convention, we omit the $0$ process at the end of a trace of actions by a process.

\small $IC^1 \triangleq \sum_{o \in O} InventoryCheck_o.(\overline{InventoryCheckNotOK}_o + \overline{InventoryCheckOK}_o)$ \normalsize
~and denotes the \texttt{Inventory Check} subtask of \texttt{Evaluation}.

\small $ICH^1 \triangleq \sum_{o \in O} InventoryCheckNotOK_o.\overline{RejectIC}_o.CONFIG^1 + InventoryCheckOK_o.\overline{CreditCheck}_o$ \normalsize
~and denotes subtasks of \texttt{Evaluation} and \texttt{Rejection}.

\small $CC^1 \triangleq \sum_{o \in O} CreditCheck_o.(\overline{CreditCheckNotOK}_o + \overline{CreditCheckOK}_o)$ \normalsize
~and denotes the \texttt{Credit Check} subtask of \texttt{Evaluation}.

\small $CCH^1 \triangleq \sum_{o \in O} CreditCheckNotOK_o.\overline{RejectCC}_o.CONFIG^1 + CreditCheckOK_o.\overline{Ship}_o$ \normalsize
~and denotes subtasks of \texttt{Evaluation} and \texttt{Rejection}.

\small $SHIP^1 \triangleq \sum_{o \in O} Ship_o.\overline{Bill}_o$ \normalsize
~and denotes the \texttt{Shipping} task,
~\small $BILL^1 \triangleq \sum_{o \in O} Bill_o.\overline{Archive}_o$ \normalsize
~and denotes the \texttt{Billing} task.

\small $ARC^1 \triangleq \sum_{o \in O} Archive_o.\overline{ArchiveOK}_o$ \normalsize
~and denotes the \texttt{Archiving} task.

\small $ARCH^1 \triangleq \sum_{o \in O} ArchiveOK_o.\overline{Confirm}_o.CONFIG^1$ \normalsize
~and denotes the \texttt{Confirmation} task.


The execution of Configuration 1 of the workflow is modelled as transitions of the \small $CONFIG^1$ \normalsize process.
If the action \small $\overline{RejectIC}_o$ \normalsize is performed by \small $ICH^1$ \normalsize or
\small $\overline{RejectCC}_o$ \normalsize is performed by \small $CCH^1$\normalsize,
the processes to be executed subsequently in \small $CONFIG^1$ \normalsize are deleted implicitly (i.e. garbage collected).
The process deletion can be represented explicitly
(e.g. as \small $\overline{RejectIC}_o.\frac{CONFIG^1}{CC^1 ~|~ CCH^1 ~|~ SHIP^1 ~|~ BILL^1 ~|~ ARC^1 ~|~ ARCH^1}$ in $ICH^1$\normalsize),
but this has the disadvantage of encoding information about the workflow's structure within a workflow process,
which (in general) complicates reconfiguration of workflows.

\subsection*{Modelling Configuration 2}\label{sec:bccsdpmodel2}

Configuration 2 is different in structure from Configuration 1,
although some of the tasks are unchanged (such as \texttt{Inventory Check} and \texttt{Credit Check}),
and this difference is reflected in the processes used to model Configuration 2.
For example, a new process \small $SC$ \normalsize is needed in order to model the new subtask (\texttt{Supplier Check}) of the \texttt{Evaluation} task.
The \small $CCH$ \normalsize process must be different in order to ensure that \texttt{Shipping} and \texttt{Billing} are performed concurrently.
The removal of the \texttt{Confirmation} task implies the \small $ARCH$ \normalsize process is no longer needed.

Configuration 2 of the workflow is denoted by the process \small $CONFIG^2$\normalsize, and \\
\small $CONFIG^2 \triangleq REC^1 ~|~ IC^1 ~|~ ICH^2 ~|~ CC^1 ~|~ CCH^2 ~|~ SHIP^2 ~|~ BILL^2 ~|~ ARC^2$ \normalsize

\small $REC^1 \triangleq \sum_{o \in O} Receipt_o.\overline{InventoryCheck}_o$ \normalsize
~and~
\small $IC^1 \triangleq \sum_{o \in O} InventoryCheck_o.(\overline{InventoryCheckNotOK}_o + \overline{InventoryCheckOK}_o)$ \normalsize

\small 
$ICH^2 \triangleq \sum_{o \in O} InventoryCheckNotOK_o.(\overline{SupplierCheck}_o ~|~ SC) ~+~ InventoryCheckOK_o.\overline{CreditCheck}_o$ \normalsize
~and denotes subtasks in \texttt{Evaluation} that initiate a \texttt{Supplier Check} or a \texttt{Credit Check}.

\small $SC \triangleq \sum_{o \in O} SupplierCheckNotOK_o.\overline{RejectIC}_o.CONFIG^2 + SupplierCheckOK_o.\overline{CreditCheck}_o$ \normalsize
~and denotes the new \texttt{Supplier Check} handling subtask in \texttt{Evaluation} followed by either \texttt{Reject}
or initiation of a \texttt{Credit Check}.
Notice that \small $SC$ \normalsize represents a stock check external to the organisation,
and (therefore) communicates with a process external to \small $CONFIG^2$ \normalsize
(unlike \small $IC^1$ \normalsize and \small $CC^1$ \normalsize in both \small $CONFIG^1$ \normalsize and \small $CONFIG^2$\normalsize).
Thus, there is no \small $SCH$ \normalsize process in \small $CONFIG^2$ \normalsize that corresponds to \small $ICH^2$ \normalsize or \small $CCH^2$\normalsize,
such a process would be located in the context of \small $CONFIG^2$\normalsize,
and (therefore) the external behaviours of \small $CONFIG^1$ \normalsize and \small $CONFIG^2$ \normalsize are different.

\small $CC^1 \triangleq \sum_{o \in O} CreditCheck_o.(\overline{CreditCheckNotOK}_o + \overline{CreditCheckOK}_o)$ \normalsize

\small $CCH^2 \triangleq \sum_{o \in O} CreditCheckNotOK_o.\overline{RejectCC}_o.CONFIG^2 + CreditCheckOK_o.(\overline{Ship}_o ~|~ \overline{Bill}_o)$ \normalsize
~and denotes subtasks in \texttt{Evaluation} and \texttt{Rejection},
including the concurrent initiation of \texttt{Shipping} and \texttt{Billing}.

\small $SHIP^2 \triangleq \sum_{o \in O} Ship_o.\overline{ShipOK}_o$ \normalsize
~and denotes the changed \texttt{Shipping} task that allows \texttt{Shipping} and \texttt{Billing} to be performed concurrently.

\small $BILL^2 \triangleq \sum_{o \in O} Bill_o.\overline{BillOK}_o$ \normalsize
~and denotes the changed \texttt{Billing} task that allows \texttt{Shipping} and \texttt{Billing} to be performed concurrently.

\small $ARC^2 \triangleq \sum_{o \in O} ShipOK_o.BillOK_o.\overline{Archive}_o.CONFIG^2 + BillOK_o.ShipOK_o.\overline{Archive}_o.CONFIG^2$ \normalsize
~and denotes the changed \texttt{Archiving} task.


The execution of Configuration 2 of the workflow is modelled as transitions of the \small $CONFIG^2$ \normalsize process.
Processes are deleted implicitly following the execution of the
\small $\overline{RejectIC}_o$ \normalsize action or the \small $\overline{RejectCC}_o$ \normalsize action (as in \small $CONFIG^1$\normalsize).

\subsection*{Modelling the Reconfiguration}\label{sec:bccsdprecmodel}

The workflow is reconfigured by a reconfiguration manager (modelled by the process \small $RM$ \normalsize defined below)
that is activated after receiving a triggering message from an observer in the environment of the workflow
(e.g. a human operator or a reconfiguration tool) and reconfigures the workflow from Configuration 1 to Configuration 2.
There are two ways of reconfiguring the workflow (depending on its state of execution),
and they are triggered by different messages.
The \small $trigger1$ \normalsize guard models receipt of the message that is used to trigger reconfiguration of the workflow
if it has \textbf{not} yet started to execute.
After the release of \small $trigger1$\normalsize,
\small $RM$ \normalsize replaces the process \small $CONFIG^1$ \normalsize with the process \small $CONFIG^2$\normalsize, see Appendix \ref{prf:reconfelabs}.
The \small $trigger2$ \normalsize guard models receipt of the message that is used to trigger reconfiguration of the workflow
if it has completed \texttt{Order Receipt} and \texttt{Inventory Check}
but not yet determined the subtask to be performed after the \texttt{Inventory Check}.
After the release of \small $trigger2$\normalsize, \small $RM$ \normalsize deletes the process \small $ARCH^1$\normalsize,
replaces the processes \small $ICH^1$\normalsize, \small $CCH^1$\normalsize, \small $SHIP^1$\normalsize, \small $BILL^1$\normalsize,
and \small $ARC^1$ \normalsize with the processes
\small $ICH^2$\normalsize, \small $CCH^2$\normalsize, \small $SHIP^2$\normalsize, \small $BILL^2$\normalsize, and \small $ARC^2$ \normalsize respectively,
as shown in Appendix \ref{prf:reconfelabs}.

\small \( RM \triangleq trigger1.\frac{CONFIG^2}{CONFIG^1} \, + \,
                        trigger2.\big(\frac{ICH^2}{ICH^1} ~|~ \frac{CCH^2}{CCH^1} ~|~
                                      \frac{SHIP^2}{SHIP^1} ~|~ \frac{BILL^2}{BILL^1} ~|~ \frac{ARC^2}{ARC^1} ~|~ \frac{0}{ARCH^1}\big) \) \normalsize

\noindent Thus, \small $RM$ \normalsize performs two operations of unplanned process reconfiguration, namely,
the deletion and replacement of processes that are not designed to be reconfigured.
The definitions of \small $CONFIG^1$ \normalsize and \small $RM$ \normalsize are modular, that is, syntactically separate,
and \small $CONFIG^1$ \normalsize is not syntactically altered in order to be reconfigured by \small $RM$\normalsize.
The reconfiguration of \small $CONFIG^1$ \normalsize occurs through its reactions with
\small $RM$ \normalsize in the expression \small $CONFIG^1 ~|~ RM$\normalsize.
The step through which \small $RM$ \normalsize is added to the context of \small $CONFIG^1$\normalsize, that is,
the step through which \small $CONFIG^1$ \normalsize becomes \small $CONFIG^1 ~|~ RM$\normalsize, is performed outside basic $\mathrm{CCS^{dp}}$,
and thereby captures the fact that the reconfiguration is unplanned.
Notice that because \small $RM$ \normalsize is located in the context of \small $CONFIG^1$ \normalsize and contains \small $CONFIG^2$\normalsize,
Configuration 2 is located in the environment of the workflow. Therefore, Configuration 2 is \textbf{not} pre-defined within the workflow.
Hence, an arbitrary number of configurations and reconfigurations can be represented for the workflow in an incremental way,
which can be used to represent the dynamic evolution of the workflow throughout its lifetime.

The concurrent execution of the workflow and reconfiguration tasks is represented as the set of total orders (i.e. sequences)
of the transitions of the processes modelling the tasks,
and the functional interference between the tasks is represented using interleaved transitions of the processes.

\subsection{Analysis}\label{sec:bccsdpanalysis}

Reconfiguration requirement R3 states that
any order received after the start of the reconfiguration must satisfy all the requirements on Configuration 2.
Clearly, \small $CONFIG^2$ \normalsize satisfies all the requirements on Configuration 2.
Therefore, after the execution of either of the two actions that start the reconfiguration process \small $RM$ \normalsize in the expression
\small $CONFIG^1 ~|~ RM ~|~ \overline{trigger1} + \overline{trigger2}$ \normalsize
~the resulting expressions must be weakly observationally bisimilar to \small $CONFIG^2$\normalsize, that is, the following property must hold: \\[1mm]
\small
\(CONFIG^1 ~|~ \frac{CONFIG^2}{CONFIG^1} \approx_o CONFIG^2 ~~\wedge~~
  CONFIG^1 ~|~ \frac{ICH^2}{ICH^1} ~|~
               \frac{CCH^2}{CCH^1} ~|~ \frac{SHIP^2}{SHIP^1} ~|~ \frac{BILL^2}{BILL^1} ~|~ \frac{ARC^2}{ARC^1} ~|~ \frac{0}{ARCH^1} \approx_o CONFIG^2\)
\normalsize

Weak observational bisimulation is defined as follows, based on \cite{kn:Mil99}.

Let $\mathcal{T}_{\!\!\!trans}^*$ be the transitive reflexive closure of the set of $\tau$ transitions of the processes in $\mathcal{P}$, \\
where $\mathcal{T}_{\!\!\!trans}^* \triangleq \{(r, s) \!\in\! \mathcal{P} ~\text{x}~ \mathcal{P} ~|~ r \stackrel{\tau}{\rightarrow} s\}^*$,
~let $q \Rightarrow q'$ ~denote~ $(q, q') \!\in\! \mathcal{T}_{\!\!\!trans}^*$, \\
let $\wt{\beta}$ be a tuple of elements of $\mathcal{L}$ with $|\wt{\beta}| \!\in\! \mathbb{N}^+$, \\
and let $q \stackrel{\wt{\beta}}{\Rightarrow} q''$ ~denote~
\( \forall i \!\in\! [1..|\wt{\beta}|] ~\exists q_{i-1}, q_{i-1, 1}, q_i \!\in\! \mathcal{P}
   ~(q = q_0 ~\wedge~ q_{i-1} \Rightarrow q_{i-1, 1} ~\wedge~
     \beta_i \!\in\! \mathcal{L}_{q_{i-1, 1}} ~\wedge~ q_{i-1, 1} \stackrel{\beta_i}{\rightarrow} q_i ~\wedge~ q_{|\wt{\beta}|} \Rightarrow q'') \).

\emph{Weak observational bisimulation} ($\approx_o$) is the largest symmetric binary relation on $\mathcal{P}$ such that
the following condition holds $\forall (p, q) \in\, \approx_o$ \\
\(
\forall p' \!\in\! \mathcal{P}
   ~(\textrm{if } p \stackrel{\wt{\beta}}{\Rightarrow} p' \textrm{ then }
    \exists q' \!\in\! \mathcal{P} ~(q \stackrel{\wt{\beta}}{\Rightarrow} q' \wedge (p', q') \!\in\, \approx_o))
\)

Both reconfigurations of \small $CONFIG^1$ \normalsize by the fraction processes in \small $RM$ \normalsize can result in processes
that are weakly observationally bisimilar to \small $CONFIG^2$\normalsize, as shown in Appendix \ref{prf:reconfelabs}.
However, neither of the left process expressions of the two conjuncts is weakly observationally bisimilar to \small $CONFIG^2$\normalsize, that is, \\[1mm]
\small
\(CONFIG^1 ~|~ \frac{CONFIG^2}{CONFIG^1} \napprox_o CONFIG^2 ~~\wedge~~
  CONFIG^1 ~|~ \frac{ICH^2}{ICH^1} ~|~
               \frac{CCH^2}{CCH^1} ~|~ \frac{SHIP^2}{SHIP^1} ~|~ \frac{BILL^2}{BILL^1} ~|~ \frac{ARC^2}{ARC^1} ~|~ \frac{0}{ARCH^1} \napprox_o CONFIG^2\)
\normalsize

The proof is given in Appendices \ref{prf:nonweakobsbisimgen} and \ref{prf:nonweakobsbisimspec}
and consists of finding a sequence of transitions of \small $CONFIG^1$ \normalsize in a parallel composition that cannot be matched by \small $CONFIG^2$ \normalsize
in the context containing only the process $\overline{Receipt}_o \,|\, RejectIC_o$.
The problem is caused by lack of control of non-deterministic transitions of process expressions in basic $\mathrm{CCS^{dp}}$,
which is inherited from $\mathrm{CCS}$.
The purpose of restricting the context of the process expressions in Appendix \ref{prf:nonweakobsbisimgen}
is to prevent behaviour of the expressions that would unnecessarily complicate the proof.


\subsection{Extensions}\label{sec:bccsdpextensions}

Basic CCS was used as the host process algebra in which to explore the properties of the fraction process 'plug-in',
define its semantics, and develop its theory, because basic CCS is very simple.
However, in order to \textbf{use} the fraction process effectively, a hosting process algebra must meet certain requirements.
As the above analysis shows, one of the requirements is a facility to control non-deterministic transitions, which is absent in CCS.

One way of controlling problematic non-deterministic transitions is to use a priority scheme for transitions (see \cite{kn:Bra02})
that is designed to satisfy requirements on workflows and on their reconfiguration.
Since the requirements can be application specific, different priority schemes may be necessary.
Therefore, the semantics of basic $\mathrm{CCS^{dp}}$ should be extended with a generic notion of transition priority
such that different system models can be produced using different priority schemes.
Notice that the use of a priority scheme raises the issue of compositionality of process expressions,
but that this issue exists whenever there is functional interference between processes.
One solution is to use rely and guarantee conditions \cite{kn:Jon81} on process transitions,
which (if satisfied) implicitly define a set of partial orders on the transitions,
which can be explicitly defined by a priority scheme.
Thus, the compositionality of the processes is ensured by the satisfaction of their rely and guarantee conditions
and is implemented using the priority scheme.
Use of process identifiers in process matching and the restriction operator (see below) also help to achieve compositionality of processes.

Appendix \ref{mdl:designs234} describes models of Configuration 1 with multiple executing workflows.
In order to describe the unplanned reconfiguration of such models,
a dynamic binding between \small $CONFIG^1$ \normalsize instances and \small $RM$ \normalsize instances is necessary
that will support the selective reconfiguration of specific process instances.
Such a binding can be achieved by extending the semantics of process matching to use process identifiers.
If a process identifier is passed as a parameter to a fraction process, the fraction can reconfigure
different process instances in a flexible and controlled manner.
Furthermore, the identification of a specific process for reconfiguration precludes the matching of other processes,
and thereby significantly reduces the computational complexity of process matching.
For example, in the expression $p_1 ~|~ p_2 ~|~ p_3 ~|~ x(i).\frac{p'}{p}(i)$
where $p_1$, $p_2$, and $p_3$ are different instances of the same process $p$,
$\frac{p'}{p}(1)$ will be able to reconfigure only $p_1$, and the processes
$p_2$, $p_3$, $p_1 ~|~ p_2$, $p_1 ~|~ p_3$, $p_2 ~|~ p_3$, and $p_1 ~|~ p_2 ~|~ p_3$ will not be matched.
Notice that basic $\mathrm{CCS^{dp}}$ is a class-based process algebra;
that is, like numbers in arithmetic, the processes in basic $\mathrm{CCS^{dp}}$ are classes, and different instances of a process
can be used interchangeably in any context with identical results.
However, the use of process identifiers in process matching makes the modification of basic $\mathrm{CCS^{dp}}$ an instance-based process algebra,
so that different instances of a process with different identifiers in identical contexts can produce different results.

Process matching based on strong of-bisimulation produces very terse models, but is computationally very complex.
The computational complexity of process matching can be reduced significantly by using syntax-based process matching with process identifiers,
discussed in \cite{kn:Bha13}. Moreover, the restriction operator ($\nu$) can be added to basic $\mathrm{CCS^{dp}}$ to enable scoping of names,
since restriction does not affect the decidability of syntactic process matching.

We briefly consider the reverse reconfiguration of a workflow (from Configuration 2 to Configuration 1) for the sake of completeness.
The reconfiguration is performed by the reconfiguration manager denoted by the process \small $MR$\normalsize:

\small
\( MR \triangleq trigger3.\big(\frac{CONFIG^1}{CONFIG^2}\big) \, + \,
                 trigger4.\big(\frac{ICH^1}{ICH^2} ~|~ \frac{CCH^1}{CCH^2} ~|~
                               \frac{SHIP^1}{SHIP^2} ~|~ \frac{BILL^1}{BILL^2} ~|~ \frac{ARC^1 ~|~ ARCH^1}{ARC^2}\big) \)
\normalsize

Transposing Configuration 1 and Configuration 2 in the reconfiguration requirements defined in Section 3.2, 
the reconfiguration from Configuration 2 to Configuration 1 is restricted by the existence of \texttt{Supplier Check}.
Consider an executing workflow that started before the reconfiguration (see Figures \ref{OfficeWorkflowConfig2} and \ref{OfficeWorkflowConfig1}).
If the outcome of \texttt{Inventory Check} is negative then \texttt{Supplier Check} is performed, which cannot be done in Configuration 1.
Therefore, even if the \texttt{Supplier Check} is positive, the reconfigured workflow cannot meet Requirement C1.2 of Configuration 1.
Hence, the reconfiguration should not be performed.
If the outcome of \texttt{Inventory Check} is positive then the workflow can be reconfigured just after \texttt{Credit Check}.
However, there is no mechanism in basic $\mathrm{CCS^{dp}}$ for testing the history of transitions of a workflow in order to determine a reconfiguration transition.
Such testing can be performed by extending the syntax of a process to include the history of transitions that produced the process,
and extending the semantics of process matching to include matching of traces of transitions,
which will increase the potential for reconfiguration of workflows.

\section{Comparison of VDM, CPOGs, and basic ${\bf CCS^{dp}}$\label{sec:comp}}

\let\oldthesubsection\thesubsection
\renewcommand{\thesubsection}{F\arabic{subsection}}

In this section, we evaluate and compare the strengths and weaknesses of VDM, CPOGs, and basic $\mathrm{CCS^{dp}}$
with respect to the requirements on an ideal formalism defined in Section \ref{sec:reqts},
using material presented in Sections \ref{sec:vdm}, \ref{sec:cpog}, and \ref{sec:bccsdp} to justify our findings,
which are summarised in Table 3.
Notice that what is currently done using the formalisms is indicated in the table in normal font,
and what is feasible (albeit with extensions) is indicated in italics.
Recollect that for the case study, the formalisms were used according to their respective `idioms'.

\subsection{It should be possible to model, and to identify instances of, software components and tasks, and their communication links}

VDM-SL allows workflows to be represented as data types, including their actions, configurations, and traces.
In this paper, VDM-SL was used to model tasks as simple data types, and to model workflows as linked tasks.
An invariant was defined that excludes invalid workflows.
VDM-SL does not have specific mechanisms for modelling communication links, but these can be modelled where appropriate.
Below we describe an extended development process using other dialects of VDM that have built-in mechanisms
(including object-orientation) for modelling software components, tasks, and their communication links more completely.

%
%
%

A CPOG represents a software component or a task as a set of graphs of actions,
where each vertex and arc of a graph can be guarded by a predicate.
Hence, Configuration 1 of the case study workflow is represented as the CPOG $c1$.
Different instances of a software component or of a task can be identified using CPOGs with different identifiers
and subscripting the actions with their CPOG identifier.
There is no facility for value passing between actions,
but a value to be passed can be encoded into the name of a receiving action (as in basic $\mathrm{CCS^{dp}}$)
that is guarded by a Boolean variable corresponding to the value.
There is no communication facility for actions, but the behaviour of a communication link can be represented by a CPOG,
which enables synchronous and asynchronous point-to-multipoint communication to be modelled.
An instance of a communication link can be represented as a CPOG instance.
There is a close relationship between CPOGs and a basic process algebra, because a CPOG is the unfolding of a process.

In basic $\mathrm{CCS^{dp}}$, software components and tasks are represented as processes.
Hence, the case study workflow in Configuration 1 is represented as the process \small $CONFIG^1$\normalsize.
Different instances of a software component or of a task can be identified using processes with different identifiers,
for example, \small $A \triangleq REC$ \normalsize and \small $B \triangleq REC$\normalsize, or
\small $REC_A$ \normalsize and \small $REC_B$ \normalsize (as in CCS).
Notice that basic $\mathrm{CCS^{dp}}$ (as in basic CCS) has no facility for value passing,
but that a value to be passed (e.g. $o$) can be encoded into a process identifier (e.g. \small $REC$\normalsize),
so that a unique value passed to a process can be used to identify the process instance uniquely (e.g. \small $REC_o$\normalsize).
A communication link is represented as a pair of complementary port names on two processes, and is used to model synchronous point-to-point communication.
Instances of communication links can be identified indirectly using process identifiers and port names that are unique to the linked processes.
Alternatively, a communication link can be represented as a process (not used in the case study),
which enables both synchronous and asynchronous point-to-multipoint communication to be modelled,
and different instances of a communication link can be identified using processes with different identifiers.

\subsection{It should be possible to model the creation, deletion, and replacement of software components and tasks,
and the creation and deletion of their communication links}

The VDM model described in this paper used the VDM-SL dialect.
Following a standard VDM-SL paradigm to keep the model small, a single workflow was modelled,
with concurrency of the parallel tasks modelled as non-deterministic interleaving.
With sufficient work, the model could be extended to include these features, but it is common to use another dialect of VDM as a development continues.
As described in Section~\ref{sec:vdm_extensions},
there are two other dialects of VDM that extend the language with features for modelling object-orientation (\vpp) and real-time (VDM-RT); these dialects form a family.
Guidance exists~\cite{Larsen&09b} for a development process that begins with VDM-SL and moves through \vpp and finally VDM-RT,
adding complexity at each stage and moving closer to implementation.
The features of the \vpp dialect~\cite{Fitzgerald&05} could be used to extend the existing model to include dynamic creation, deletion,
and replacement of software components and tasks. Each would be represented by a class, which would include definition of a thread of control.
These classes can then be instantiated as objects dynamically.
A new thread can be spawned using the \texttt{\textbf{start}} keyword, which can be called on an object whose class defines a thread.
Objects can be deleted once their threads have completed, or when all references to them are removed (in which case they are cleaned up by the garbage collector).

The creation and deletion of software components and tasks is represented respectively as the creation and deletion of CPOGs,
and replacement is a combination of CPOG creation and deletion.
The target CPOG (modelling a software component or a task) is guarded by one or more predicates that are set by the reconfiguration actions
and determine which actions and action dependencies of the CPOG are active (can be performed) or inactive (cannot be performed).
For example, $c1$ is reconfigured to $c2$ by embedding $c1$ in the expression $r + [\neg r\_done]c1 + [r\_done]c2$.
The guards $[\neg r\_done]$ and $[r\_done]$ penetrate into $c1$ and $c2$ respectively,
by Proposition~\ref{prop-canonical} and the four properties of conditional workflows in Section \ref{cpog-axioms},
and thereby guard individual actions and their dependencies.
Therefore, before the execution of reconfiguration action $r$ only actions in $c1$ can be performed,
and after the execution of $r$ has set the guard value only the remaining actions in $c2$ can be performed.
A new CPOG ($c_{new}$) can be added to an existing CPOG ($c_{old}$) by parallel composition ($c_{old} + c_{new}$)
or by sequential composition ($c_{old} \rightarrow c_{new}$),
and an existing CPOG ($[x]c_{old}$) can be deleted by setting its guard ($[x]$) to false.
The granularity of reconfiguration is a single action and a single dependency between actions,
and an entire CPOG can also be reconfigured by setting a Boolean variable.
Since communication links can be represented as CPOGs,
the creation and deletion of communication links can be expressed respectively as the creation and deletion of CPOGs.

In basic $\mathrm{CCS^{dp}}$, the replacement, deletion, and creation of software components and tasks can be modelled using fraction processes.
A target process (modelling a software component or a task) is reconfigured by a fraction process
with the denominator of the fraction binding dynamically to the target using the strong of-bisimulation relation
and the numerator of the fraction replacing the target through a reaction transition.
Thus, \small $ARC^1 ~|~ \frac{ARC^2}{ARC^1} \stackrel{\tau}{\longrightarrow} ARC^2$\normalsize.
The fraction and target processes perform complementary reconfiguration actions that combine to produce the reaction transition ($\tau$)
that results in the replacement of the target.
The granularity of reconfiguration is a single concurrent process (e.g. \small $ARC^1$\normalsize),
and multiple concurrent processes can also be reconfigured through a single reaction transition
(e.g. \small $CONFIG^1 ~|~ \frac{CONFIG^2}{CONFIG^1} \stackrel{\tau}{\longrightarrow} CONFIG^2$\normalsize).
Deletion of a process is expressed as replacement with the identity process ($0$).
Thus, \small $ARCH^1 ~|~ \frac{0}{ARCH^1} \stackrel{\tau}{\longrightarrow} 0$\normalsize.
A process is created either by including it in the numerator of a fraction
(e.g. \small $ARC^2 ~|~ \frac{ARC^1 ~|~ ARCH^1}{ARC^2} \stackrel{\tau}{\longrightarrow} ARC^1 ~|~ ARCH^1$\normalsize),
or by using a guarded parallel composition of processes that creates a new process after the guard is released (as in $\mathrm{CCS}$).
Since communication links between software components and between tasks can be represented as processes,
the creation and deletion (and replacement) of communication links can be expressed as process replacement using fractions.
Alternatively, basic $\mathrm{CCS^{dp}}$ can be extended to enable link passing, as in $\pi$-calculi \cite{kn:Mil99},
or a $\pi$-calculus can be extended with the fraction process and its semantics.

\subsection{It should be possible to model the relocation of software components and tasks on physical nodes}

In VDM-SL, physical nodes are not modelled.
However, the VDM-RT dialect~\cite{Verhoef&06b} has features that support such modelling.
The VDM-RT dialect extends \vpp with abstract models of compute nodes and communication buses.
A VDM-RT model must describe one or more compute nodes (e.g.\ CPUs) to which objects are deployed.
Objects on the same node compete for computation time. In order to call an operation of an object deployed to a different CPU,
a communication link between them must be defined. Currently, the graph of compute nodes and communication links
must be declared statically before simulation, and object deployment cannot be changed dynamically.
However, multiple simulations with different configuration can be run and compared.

In CPOGs, physical nodes are not modelled.
However, it is possible to represent the location of an action using a set of Boolean variables specific to the action
that represent all possible locations of the action.
Thus, the location of a CPOG can be represented, which models the location of a software component or a task.
Relocation is represented as a change in the value of location-indicating Boolean variables by their controlling actions.

In basic $\mathrm{CCS^{dp}}$, physical nodes are not modelled.
Hence, the relocation of software components or tasks on physical nodes cannot be modelled.
However, if the process syntax is extended with a location attribute and the semantics of communication is extended with process passing,
then relocation can be modelled simply as communication with process passing
in which the location of the process being passed changes from the location of the sending process to the location of the receiving process.

\subsection{It should be possible to model state transfer between software components and between tasks}

VDM-SL is state-based, with models using persistent state and operations over that state.
State transfer can be modelled through parameter passing in operation calls.
The object-orientation features of \vpp and VDM-RT dialects extend this by allowing objects to be passed as parameters that contain both state and functionality.
Additionally, the concurrency features of these two dialects allow for synchronisation between threads, which can model state transfer where appropriate.

In a CPOG, there is no facility for data communication between actions.
However, it is possible to use a set of action specific Boolean variables to represent communication data.
The Boolean variables can collectively represent all possible data values to be communicated by an action, including the state of a CPOG,
that can be used in predicates of actions in another CPOG,
which enables data communication and state transfer between software components and between tasks to be modelled.
A more expressive variant of CPOGs
enables Boolean variables to be controlled by more than one action and their
value to be changed more than once~\cite{2012_mokhov_async}, which simplifies the modelling
of data communication.

In basic $\mathrm{CCS^{dp}}$, state transfer between software components and between tasks can be modelled by
encoding information about the state in the names of the complementary communicating actions
and (thereby) selecting the process with the transferred state in a summation,
or by replacing the receiving process with the process that has received the transferred state using a fraction process.
Alternatively, basic $\mathrm{CCS^{dp}}$ can be extended to express value passing communication.

\subsection{It should be possible to model both planned and unplanned reconfiguration}

In VDM-SL, both planned and unplanned reconfiguration can be modelled. In the model of the case study,
the \texttt{Reconfiguration} operation allows the workflow to be changed in the \texttt{Interpreter} module during the execution of the interpreter.
The interpreter is not in control of when the reconfiguration operation is called,
and Configuration1 is defined independently of the \texttt{Reconfiguration} operation and Configuration2.
Therefore, the reconfiguration is unplanned. A pre-condition ensures that any requested reconfiguration is valid.
Notice that VDM can pass functions as parameters to operations.
Therefore, it is possible to pass a predicate (expressing a pre-condition or an invariant) as well as a workflow
as parameters to a reconfiguration operation (although this was not modelled).
This is necessary because the invariants and pre-conditions that must be satisfied in order to ensure the correctness of a reconfiguration
cannot always be pre-defined for a dynamically evolving system.
Planned reconfiguration was not modelled for this case study,
but a simple extension can be envisioned where the \texttt{Reconfiguration} operation is called internally by the interpreter at a planned time,
or in response to planned stimuli.

CPOGs can be used to model both planned and unplanned reconfiguration.
In the planned reconfiguration of a CPOG, the CPOG consists of different configurations (each represented as a CPOG)
with pre-defined reconfiguration predicates and reconfiguration actions that determine the value of the predicates.
The execution of the reconfiguration actions deactivates actions and action dependencies in the target CPOG
and activates actions and action dependencies in the destination CPOG, and thereby reconfigures the target to the destination.
In the unplanned reconfiguration of a CPOG, the target CPOG (e.g. $c1$)
is typically embedded within a reconfiguration CPOG (e.g. $r + [\neg r\_done]c1 + [r\_done]c2$),
which performs reconfiguration actions (e.g. $r$) that control the actions and action dependencies of the target
through reconfiguration predicates (e.g. $[\neg r\_done]$)
and thereby replace the target with the destination CPOG (e.g. $c2$) located in the target's environment.
The reconfiguration actions constitute a CPOG and (therefore) can themselves be reconfigured.
Hence, CPOG reconfiguration is recursive.

In basic $\mathrm{CCS^{dp}}$, planned and unplanned reconfiguration can both be modelled using fraction processes.
For modelling planned reconfiguration, the reconfiguring fractions are located within the system model,
whereas for unplanned reconfiguration, the fractions are located in the context of the system model.
The target process to be reconfigured does not require any syntactic modification or syntactic proximity to a reconfiguring fraction.
Therefore, the modelling of reconfiguration can be modular.
Thus, a system with $n$ configurations can be represented by $n$ syntactically separate process expressions (each modelling a different configuration),
and the $n(n-1)$ reconfigurations of the system can be represented by $n(n-1)$ syntactically separate process expressions that contain fraction processes.
Furthermore, the reconfiguration of the reconfiguration software can be represented by the replacement of one or more fraction processes by other fractions,
since the notion of fraction process is recursive. The selective reconfiguration of specific process instances
requires the extension of the semantics of process matching to use process identifiers.

\subsection{It should be possible to model the functional interference between application tasks and reconfiguration tasks}

The tasks modelled in VDM-SL are atomic, and (therefore) cannot be interrupted during their execution.
However, the \texttt{Reconfiguration} operation can be called during workflow execution,
and (therefore) can interfere with the workflow as a whole.
An extended concurrent model in \vpp or VDM-RT could represent true functional interference,
if the reconfiguration tasks are run in a separate thread to application tasks and true race conditions can occur.

In CPOGs, functional interference between application tasks and reconfiguration tasks
is represented explicitly as a CPOG resulting either from an interleaving of computation and reconfiguration actions
or from the simultaneous execution of the actions.
The interference can be controlled using the predicates on the interfering actions and on their dependencies.

In basic $\mathrm{CCS^{dp}}$, functional interference between application tasks and reconfiguration tasks is represented explicitly
as a process expression resulting from an interleaving of communication, internal, and reconfiguration transitions of concurrently executing processes.
The control of interference can be achieved by extending the semantics of transitions
with rely and guarantee conditions on transitions or with a priority scheme for transitions derived from such conditions.

\subsection{It should be possible to express and to verify the functional correctness requirements of application tasks and reconfiguration tasks}

The verification of the case study in VDM-SL used simulation and testing. These techniques are weaker than model checking and proof.
The two VDM tools do not currently support model checking due to the generality of the formalism.
However, experimental coupling to SPIN has been reported~\cite{Lin16}.
While a proof theory exists for core VDM functionality, there is currently a lack of tool support for discharging proof obligations,
although proof obligations can be generated automatically.
In addition, the object-oriented and real-time extensions to VDM are currently not covered by the proof theory.
However, this is an active area of research.

Tasks and functional requirements on tasks can both be expressed as CPOGs.
The verification of functional correctness of a task can be done in different ways.
The axioms support equational reasoning and can transform the CPOG of a task into the CPOG of its requirement,
assuming the two CPOGs are algebraically equivalent.
Alternatively, we can show that both CPOGs have the same set of consistent histories.
Finally, the LTS semantics of CPOGs supports model
checking~\cite{2008_mokhov_acsd,2009_mokhov_phd}.

In basic $\mathrm{CCS^{dp}}$, the equational theory uses congruence based on strong of-bisimulation ($\sim_{of}$), developed in \cite{kn:Bha13}.
However, strong of-bisimulation is too strong for the case study.
Therefore, weak observational bisimulation ($\approx_o$) was used to express requirement R3 and to attempt its verification.
However, equational reasoning requires an invariant to be verified,
which can be lacking in a reconfiguration where the source and destination configurations are significantly different.
In these situations, temporal logic and model checking can be used.

\subsection{It should be possible to model the concurrent execution of tasks}

VDM-SL does not have built-in abstractions for modelling concurrency.
In this paper, concurrent execution of tasks was achieved using non-deterministic interleaving of tasks defined using the \texttt{Par} type.
Modelling of fine-grained concurrency or true parallelism is of course possible~\cite{ColemanJones07}, but takes more effort.
The \vpp and VDM-RT dialects permit modelling of true concurrency and, as described above,
can be used to continue the modelling as part of a more complete development process.

In CPOGs, concurrently executing tasks are represented as concurrently executing CPOGs.
The LTS rules of CPOGs show that a CPOG can perform either one transition at a time or a set of multiple transitions simultaneously.
Therefore, CPOGs have both an interleaving semantics of concurrency and a true concurrency semantics,
and thereby can model both pseudo-concurrency and true concurrency.
Hence, the concurrent execution of tasks is modelled as the set of partial orders of the transitions of the concurrent CPOGs representing the tasks.
The granularity of concurrency in CPOGs is a single action.

In basic $\mathrm{CCS^{dp}}$, concurrently executing tasks are represented as concurrently executing processes.
The LTS rules of the algebra show that a process expression can perform only one transition at a time.
Therefore, basic $\mathrm{CCS^{dp}}$ has an interleaving semantics of concurrency,
and thereby models pseudo-concurrency rather than true concurrency (as in CCS).
Hence, the concurrent execution of tasks is modelled as the set of total orders of the transitions of the concurrent processes representing the tasks.
The preemption of actions is not modelled. Notice that the granularity of reconfiguration is the same as the granularity of concurrency.

\subsection{It should be possible to model state transitions of software components and tasks}

State transitions in VDM are modelled using operations acting on global state (in VDM-SL) or on object state (in \vpp and VDM-RT).

In CPOGs, state transitions of software components and of tasks are modelled as transitions of CPOGs,
see the LTS rules in Section \ref{cpog-lts-semantics}.
The state of a CPOG $G$ is given by $(H, \psi)$, where $H$ is the set of completed actions performed by $G$,
and $\psi$ is the assignment of values to Boolean variables performed by actions in $H$ after initialisation of the variables.

In basic $\mathrm{CCS^{dp}}$, state transitions of software components and of tasks are modelled as transitions of processes,
see Table \ref{tab:ccsdpnonultss}. The state of a process $P$ after performing one or more transitions is given by process $P'$.

\subsection{The formalism should be as terse as possible}

The VDM dialects are general purpose languages, with many similarities to imperative programming languages.
Hence, they lack built-in abstractions for modelling processes (for example)
and (therefore) are in general less terse than process algebras in this regard.
However, the VDM dialects allow for abstract data type definition, permitting terse description of data types relative to implementations.
Similarly, implicit definition of operations using only post-conditions permits terse definition of functionality.

CPOGs are a compact representation of graphs, and graphs are a useful formalism for model checking.
Thus, CPOGs are useful for verification by model checking using SAT solvers and by equational reasoning using algebraic manipulation.
CPOGs are a less expressive formalism than basic $\mathrm{CCS^{dp}}$,
and their key limitation is the inability to represent cyclic processes.
However, CPOGs can be obtained from cyclic process descriptions using a standard unfolding procedure~\cite{1993_mcmillan_unfolding}.
Furthermore, more expressive variants of CPOGs enable predicates to be defined over non-Boolean variables,
and Boolean variables to be controlled by more than one action and their value to be changed more than once,
which increases the terseness of CPOG models.

Basic $\mathrm{CCS^{dp}}$ is a terse formalism for several reasons. First, CCS is terse.
Second, fraction processes do not contain implementation detail.
Third, overloading the parallel composition operator avoids the use of a new operator for performing reconfiguration,
such as the interrupt operator in CSP.
Fourth, process matching uses behaviour to match processes rather than structural congruence or process syntax,
which enables the denominator of a fraction to match a larger set of processes than is possible with
structural congruence or syntactic equality.
The terseness of basic $\mathrm{CCS^{dp}}$ models can be increased by extending the algebra to enable value passing,
which avoids the summation of action names encoded with values (used in the case study).

\subsection{The formalism should be supported by tools}

VDM is supported by two industrial-strength tools
(Overture\footnote{\url{http://overturetool.org/}} and VDMTools\footnote{\url{http://www.vdmtools.jp/en/}}) that are both under active development.
These tools provide syntax highlighting and type checking facilities, and include interpreters that allow simulation of the workflow case study.
The tools also provide more advanced features such as unit and combinatorial testing, proof obligation generation, and code generation.

The algebraic manipulation of CPOGs can be automated either by reusing standard term rewriting engines like Maude
or by developing a custom proof assistant embedded in a high-level language like Haskell;
the authors eventually followed the latter approach. However, the tools are not yet ready for public release.

Basic $\mathrm{CCS^{dp}}$ is a new process algebra, and the notions of fraction process and process matching are extremely novel.
Therefore, the algebra does not have tool support at present, although there are plans to develop tools for modelling and verification.
The computational complexity of process matching based on behaviour suggests that a simpler form of matching is required,
for example, based on syntactic equality or a decidable structural congruence,
that will also allow the restriction operator to be added to the algebra to enable scoping of names.

\begin{figure}
\begin{center}
\scalebox{0.8}{\includegraphics{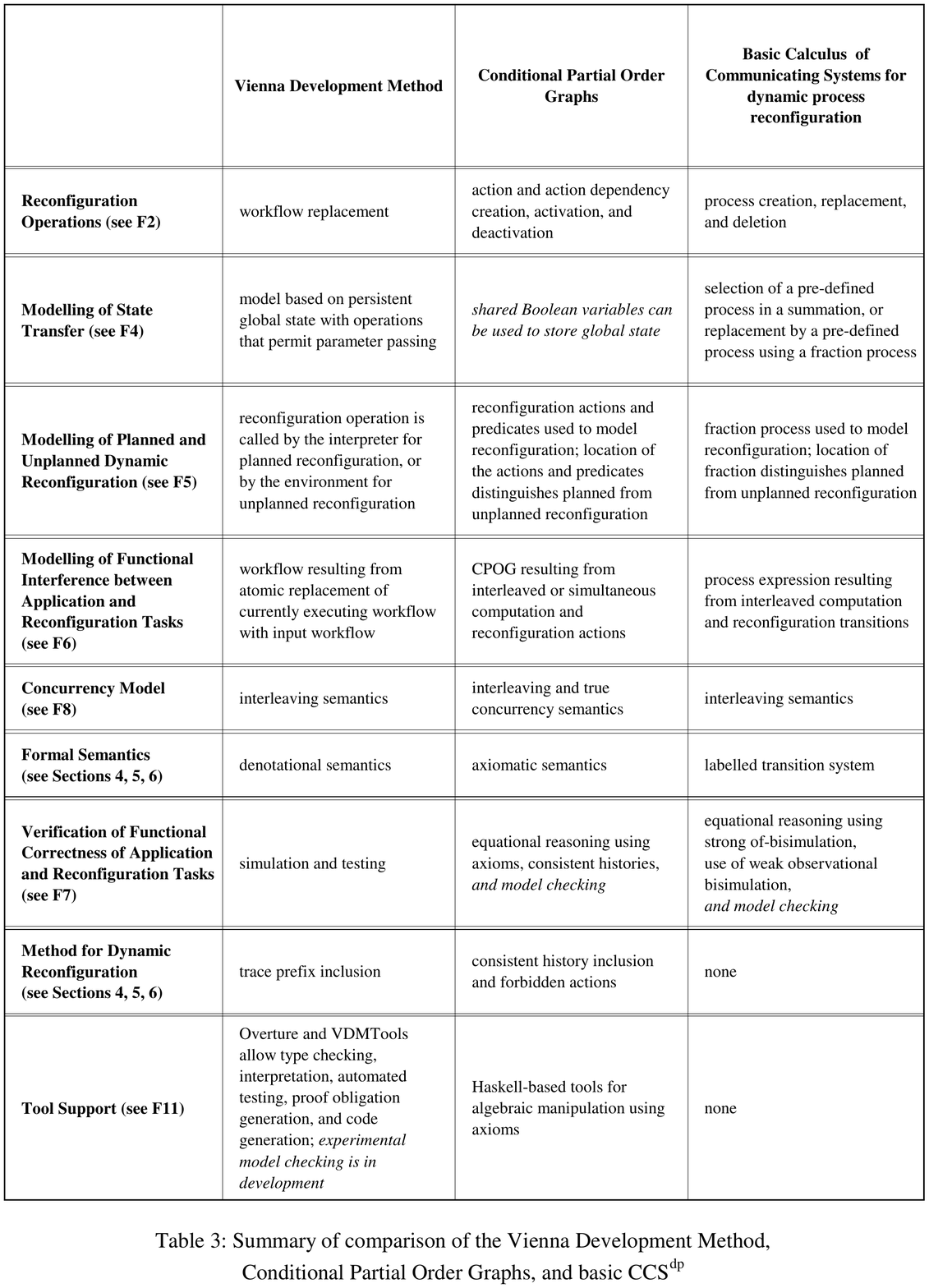}}
\end{center}
\end{figure}

\subsection{The formalism should be easy to learn and to use}

VDM is a well-established formal method with a history of industrial use,
with a variety of materials available including books~\cite{Jones90a,Fitzgerald&05,Fitzgerald&09}
and examples that are included with the Overture tool\footnote{See Overture examples repository: \url{http://overturetool.org/download/examples/}.}.
Typically, engineers can begin modelling with only a few days training. Agerholm et al.~\cite{Agerholm:1998:FSV:298595.298861} support this claim,
suggesting that this is because `VDM supports a range of abstraction levels and its concepts are easy to learn' and that
`validation based on testing (animation/prototyping) is already well-known to engineers,
in contrast to other techniques typically associated with formal methods such as refinement and formal proof.'~\cite[83]{Agerholm:1998:FSV:298595.298861}.
So the adaptability of VDM as a general language is seen as a strength in this instance.

CPOGs are as easy to learn as graphs, which are a very simple formalism.
However, CPOGs were designed for hardware
and (therefore) are relatively low-level and less easy to use than formalisms designed for software, such as VDM.
In using CPOGs, the designer is expected to operate with low-level events and conditions,
and at present it is not known whether any higher level concept has a meaningful interpretation in CPOG theory.
Another limitation is the lack of mature tool support.
We were able to employ generic tools like Maude and to implement a prototype domain-specific language in Haskell.
However, interoperability with other existing tool-kits is very limited.
It is unrealistic to expect CPOGs to be used to specify a complete system.
Therefore, it is essential to develop tools that enable conversion between well-established system design methods,
such as VDM, and CPOGs in order to use the verification capabilites offered by CPOGs.

Basic $\mathrm{CCS^{dp}}$ is as easy to learn and to use as basic CCS, which is a simple formalism.
Furthermore, the reaction transition through which a fraction process reconfigures a target process
(e.g. \small $p\!\!\!/ ~|~ \frac{P'}{P\!\!\!/} \stackrel{\tau}{\longrightarrow} P'$ \normalsize)
resembles the cancellation of numbers in arithmetic (e.g. $3\!\!\!/ ~.~ \frac{2}{3\!\!\!/} ~=~ 2$).
Therefore, the behaviour a fraction process is likely to be surprisingly familiar to users.
Moreover, the complexity of the theory behind the fraction process will be encapsulated within analysis tools.
It is also quite possible that the tools of the algebra will be embedded in a tool chain with a graphical front end,
so that the end users will not need to use $\mathrm{CCS^{dp}}$ directly.

\let\thesubsection\oldthesubsection

\section{Related Work}\label{sec:relwork}


There is a considerable amount of research into formalisms for the dynamic reconfiguration of software \cite{kn:WeyIftIglAhm12}.
The research can be categorised into approaches based on process algebras, graphs, logics, and control theory \cite{kn:BraEtAl04}.
We review a selection of formalisms and then summarize the key findings.

\subsection{Process Algebras}\label{sec:rw-procs}

$\pi$-calculi are extensions of the Calculus of Communicating Systems (CCS) \cite{kn:Mil89},
which form a large and diverse family of process algebras and are widely used in the study of dynamic reconfiguration,
see \cite{kn:MilParWal92}, \cite{kn:Tho90}, \cite{kn:HonTok91}, \cite{kn:Bou92}, \cite{kn:ParVic98}, and \cite{kn:RouSon03}.
As in basic $\mathrm{CCS^{dp}}$, software components and tasks are represented as processes,
and their communication links are represented as pairs of complementary port/action names (e.g. $\overline{a}$ and $a$).
An individual process can be identified by a unique process identifier (e.g. $A(a, b) \triangleq \overline{a}.b.A\!<\!a, b\!>$),
but a communication link has no identifier. As in most process algebras, a process can be easily created (e.g. $a.(P_1 | P_2)$)
or deleted (e.g. $a.0$) if designed to do so, but modelling the deletion of non-terminating processes is problematic.
The special feature of $\pi$-calculi is the passing of port/action names as parameters and
the uniform treatment of parameter values and variables, which in combination enable link creation and deletion to be modelled very simply.
For example, if $P \triangleq \overline{x}\!<\!y\!>\!.P'$ and
$Q \triangleq x(u).u(v).Q'$ and $R \triangleq \overline{y}\!<\!w\!>\!.R'$,
then $P$ in the expression $P|Q|R$ passes the port name $y$ to $Q$ (thereby substituting $u$ by $y$)
so that $Q$ can communicate with $R$ (and thereby receive $w$), which is expressed by the following transitions:

\( \overline{x}\!<\!y\!>\!.P' ~|~ x(u).u(v).Q' ~|~ \overline{y}\!<\!w\!>\!.R' \stackrel{\tau}{\longrightarrow}
   P' ~|~ y(v).\{\frac{y}{u}\}Q' ~|~ \overline{y}\!<\!w\!>\!.R' \stackrel{\tau}{\longrightarrow}
   P' ~|~ \{\frac{w}{v}\}\{\frac{y}{u}\}Q' ~|~ R'
\)

The communication link between $Q$ and $R$ can be deleted by a subsequent substitution of $u$.
Relocation of processes on physical nodes is easily modelled in higher-order $\pi$-calculi \cite{kn:Tho90} using process passing,
which can be encoded in first-order $\pi$-calculi \cite{kn:San93}.
This is important, because the theory of first-order $\pi$-calculi is simpler than the theory of higher-order $\pi$-calculi.
State transfer is modelled as communication with value passing. Thus, planned reconfiguration is easily modelled.
Notice that the modelling of reconfiguration is based on communication, which cannot be unplanned.
Therefore, the modelling of unplanned reconfiguration is problematic.
Functional interference between processes is modelled using interleaved transitions,
since $\pi$-calculi have an interleaving semantics of concurrency. 
Verification of functional correctness is based on equational reasoning using structural congruence or
process congruence based on strong or weak bisimulation; the LTS of $\pi$-calculi enables model checking of processes.
However, there is no control over non-deterministic transitions, no method for managing reconfiguration, and bisimulation is undecidable.
The expressivity of $\pi$-calculi complicates verification of requirements, which requires processes to be restricted,
for example, to finite control processes for model checking \cite{kn:MeyKhoStr09}.
$\pi$-calculi are reasonably terse because they describe only communication and its reconfiguration.
Tool support for Milner's, Parrow's, and Walker's synchronous $\pi$-calculus \cite{kn:MilParWal92}
includes the Mobility Workbench \cite{kn:VicMol94}, which checks for open bisimilarity between processes and for deadlocks,
and TyPiCal \cite{kn:Kob06}, which is a type-based static analyzer for checking deadlock freedom and termination;
and tool support for the asynchronous $\pi$-calculus $A\pi$ \cite{kn:HonTok91}\cite{kn:Bou92} includes Pict \cite{kn:PieTur00},
which is a strongly-typed programming language.
The tools facilitate the use of $\pi$-calculi by researchers, but are not designed for use by system designers.

Another large family of process algebras is designed specifically for workflows and service-oriented computing,
see \cite{kn:BocLanZav03}, \cite{kn:ButHoaFer05}, \cite{kn:BBCetal06} and \cite{kn:GuiLucGorBusZav06},
and includes \webpii \cite{kn:LucMaz07} and the Calculus for Orchestration of Web Services (COWS) \cite{kn:PugTie12}.
\webpii~is a conservative extension of $A\pi$ designed to model web service orchestration \cite{kn:Maz06}.
The process syntax of $A\pi$ is extended with the construct $\trs{P}{Q}{x}$ (termed a \emph{workunit}) in order to model error handling \cite{kn:LucMaz07}.
The workunit executes $P$ until either $P$ terminates (whereupon the workunit terminates) or
an interrupt is received on channel $x$ during the execution of $P$.
The interrupt can be sent either by $P$ or by a process in the context of the workunit,
and causes the premature termination of $P$ (without rollback) and the execution of $Q$.
Thus, workunits can be used to model event-triggered planned process reconfiguration.
However, unplanned process deletion and process replacement cannot be modelled,
and specific instances of processes cannot be identified for reconfiguration.
Currently, \webpii has no tool support.
COWS represents a system as a composition of \emph{services} provided by components (termed \emph{partners}) \cite{kn:PugTie12}.
Services consist of other services or basic operations,
and are loosely coupled and reusable computational units that communicate asynchronously using messages.
There is no notion of a transaction or of a session between communicating services, but messages contain sufficient header information
for a session to be inferred by a service using pattern matching between message headers (termed \emph{message correlation}).
Link reconfiguration is restricted to output capability, as in the localised $\pi$-calculus $L\pi$ \cite{kn:SanWal01},
that is, a received name can be used only for invoking a service.
Process reconfiguration is achieved by creating a service instance, and by terminating a service instance using the \texttt{kill} operator.
The combination of asynchronous communication, message correlation, output capability,
the \texttt{kill} operator, and the publish-discover-bind \& invoke paradigm of service-oriented computing
enables a high degree of reconfiguration of a workflow to be expressed.
Tool support for COWS consists of the CMC 'on-the-fly' model checker that can verify formulae
expressed in the \texttt{SocL} branching-time temporal logic using a COWS expression converted into a doubly labelled transition system,
that is, an LTS in which each transition is labelled with a set of actions \cite{kn:FanGneLapMazPugTie12}.
However, as with \webpii, unplanned process deletion and process replacement cannot be modelled,
since the \texttt{kill} operator is embedded within the system model and is statically bound to the service instance to be terminated;
true concurrency cannot be represented, since COWS has only an interleaving semantics;
and there is no development tool for refining a COWS expression to an implementation.

Paradigm is a coordination modelling language that can represent dynamic adaptation in distributed component-based systems \cite{kn:AndGroVin14}.
Each component of a system is represented as a state transition diagram (STD)
with sub-STDs (termed \emph{phases}) used to represent the internal transitions of the component.
An individual component can be identified by a unique identifier (e.g. $\mathit{McPal}$ or $\mathit{McPhil_i}$).
The phases of a component are connected by sets of states they have in common (termed \emph{traps})
that enable transitions between two phases of the component (termed \emph{phase transfers})
to be synchronized with other phase transfers in concurrently executing components of the system.
Each transition of the system is defined by a \emph{consistency rule}
that determines the synchronised phase transfers of the components involved in the transition.
Thus, communication between components is not explicitly represented; hence, link reconfiguration is not represented.
Dynamic reconfiguration is expressed as coordination involving a special reconfiguration component ($\mathit{McPal}$) to create and delete STDs.
Thus, unplanned component creation and deletion is represented.
$\mathit{McPal}$ uses shared variables (e.g. $C_{rs}$ and $C_{rs_i}$)
that store STDs, phases, traps, and consistency rules to delegate (and thereby distribute)
the reconfiguration of components to subsidiary reconfiguration components (e.g. $\mathit{McPhil_i}$),
and controls the reconfiguration using a combination of orchestration and choreography.
The use of traps and consistency rules to define system transitions (including reconfiguration transitions)
enables both interleaving concurrency and true concurrency to be handled,
enables functional interference to be controlled,
and in combination with the use of shared variables facilitates state transfer between components.
Tool support is provided by the mCRL2 model checker (based on ACP \cite{kn:BerKlo84}) that can verify formulae
expressed in a variant of the modal $\mu$-calculus \cite{kn:BraSti07} using a Paradigm model converted into an mCRL2 process model.
However, the Paradigm models are not terse, since each system transition must be defined explicitly by a consistency rule,
and there is no development tool for refining a Paradigm model to an implementation.

\subsection{Graphs}\label{sec:rw-graphs}

Graph-based formalisms for dynamic reconfiguration include graph grammars,
such as Graph Abstractions for Concurrent Programming (GARP) \cite{kn:KapKai88} and $\Delta\textrm{-grammars}$ \cite{kn:KapGoeCam89},
rewriting systems, such as the Chemical Abstract Machine (CHAM) \cite{kn:BerBou92} and Maude \cite{2014_maude_www},
and formalisms based on category theory, such as Reo \cite{kn:KraMarLazArb11} and \COMMUNITY~\cite{kn:WerLopFia01}.

GARP models a system as a directed graph,
in which named vertices (termed \emph{agents}) represent tasks that communicate asynchronously by message passing through ports.
Agents perform computation,
and graph rewrites that reconfigure the model by replacing a vertex with a subgraph defined in a production rule.
Thus, GARP models planned task reconfiguration, but the reconfiguration of communication links is not modelled.
True concurrency is represented,
but interference between computation and reconfiguration actions is not represented, since graph rewrites are atomic.
Hence, GARP models the effect of dynamic reconfiguration rather than the process of reconfiguration.
State transfer is modelled through parameter passing to an agent.
The similarity of graph grammars to string grammars enables a GARP model to be converted to a program.
However, there is no method for formally verifying a requirement using a GARP model.

CHAM is based on the GAMMA formalism defined in \cite{kn:BanMet90}.
GAMMA models a data value as a molecule,
the system's state as a solution (i.e. a finite multiset) of molecules,
and a computation as a sequence of reactions between molecules
defined by transformation rules between solutions and guarded by reaction conditions.
Different reactions can run with true concurrency if their source multisets are disjoint;
otherwise, a non-deterministic choice is made as to which reaction will occur.
GAMMA uses multisets in order to avoid unnecessary ordering restrictions
in the specification of an algorithm caused by the use of list-based data structures.
CHAM extends GAMMA by allowing the user to define the syntax of a molecule;
a membrane construct is used to encapsulate a solution, so that it behaves like a single molecule,
thereby enabling a large system to be structured as a hierarchy of solutions;
and an airlock construct is used to control reactions between a given solution and its environment.
System reconfiguration is expressed as rewrites of multisets of molecules \cite{kn:Met96}.
CHAM has been used to specify software architectures \cite{kn:InvWol95},
and to specify the dynamic reconfiguration of software architectures \cite{kn:Wer99}.
However, as with GARP, CHAM does not model the process of reconfiguration.
Furthermore, the concepts underlying the CHAM constructs are very different from those normally used by architects to design systems \cite{kn:Ore98},
so that ensuring a CHAM description is an abstraction of an architect's description becomes an issue.
In contrast, the `conceptual gap' between the architect's description and a process algebraic description is much less.
There is no development tool for refining a CHAM model to an implementation.

Reo is a channel-based coordination language that uses \emph{connectors} to model the composition and reconfiguration of component-based systems \cite{kn:KraMarLazArb11}.
A \emph{channel} is a means of communication with an associated protocol (e.g. FIFO) and can be synchronous or asynchronous.
A channel has two endpoints (i.e. \emph{nodes}) that can be connected to communication ports of components or can be used to compose channels.
A connector is a set of components and channels, and is represented by a typed hypergraph whose vertices and edges denote nodes and channels/components respectively.
Notice that a component is treated as a type of channel in Reo.
Reconfiguration consists of the atomic replacement of one or more connectors, and is performed using double pushout (DPO) graph rewriting \cite{kn:EhrPfeSch73}.
A reconfiguration is defined by a rule consisting of a pattern (formulated using a graph grammar) that must be matched by one or more connectors,
and a template that describes the rewrite to be performed on the matched connectors.
Additional conditions can be defined to restrict further the application of the rules, such as conditions on the state of a component or of a channel.
Structural and behavioural invariants can be defined on a connector to ensure consistency of the state of the reconfigured connector.
However, there is no common formal semantics for connector execution and graph rewriting.
Therefore, functional interference between connectors and the reconfiguration engine cannot be formally expressed or analyzed.
Tool support is provided on the Eclipse platform and includes centralized and distributed execution engines that run Reo models,
a reconfiguration engine to perform pattern matching and DPO graph rewriting, graphical editors,
and a code generator that can be invoked from an editor to produce Java code from a Reo model \cite{kn:SEN08}.
The authors are not aware of any proof of correctness of the code generator.
Third party tools are also used, namely,
the GRaphs for Object-Oriented VErification (GROOVE) tool \cite{kn:Ren03}
for symbolic model checking using computation tree logic (CTL) formulae,
and the Attributed Graph Grammar (AGG) system \cite{kn:Tae03}
for checking whether or not two rules are in conflict in the reconfiguration of a connector.

\subsection{Logics}\label{sec:rw-logics}

Logic-based formalisms for dynamic reconfiguration include the Generic Reconfiguration Language (Gerel) \cite{kn:EndWei92},
Aguirre-Maibaum's specification language \cite{kn:AguMai02},
the half-order dynamic temporal logic (HDTL) \cite{kn:ChoEtAl01},
and linear temporal logic (LTL) \cite{kn:Pnu77}, \cite{kn:Maz14}.

The Aguirre-Maibaum language is a declarative specification language for component-based systems
that uses a combination of first-order logic (to reason about data types) and temporal logic (to reason about actions).
A software component is represented as an instance of a \emph{class},
a communication link between components is represented as an instance of an \emph{association},
and classes are combined using associations to produce a \emph{subsystem}.
A class definition consists of:
\emph{attributes} that represent variables of basic data types, \emph{actions} that represent the methods of the class,
\emph{exports} that represent the public methods offered by the class to its environment,
\emph{read variables} that are used to obtain information from the environment,
and \emph{axioms} that define the effect of the actions on the attributes.
Communication between components is represented using synchronised actions and shared attributes,
and the behaviour of associations is defined by axioms in the subsystem definition.
State transition, state transfer, and component relocation can all be described as effects of actions on attributes.
Reconfiguration is represented as the planned creation and deletion of instances of classes and of associations in a subsystem,
but unplanned reconfiguration is not represented. The declarative nature of the language enables the effect of reconfiguration to be described easily,
but describing the process of reconfiguration with functional interference between application and reconfiguration tasks is problematic.
The language supports verification of functional correctness by model checking, but there is no tool support,
and there is no development method for refining a model to an implementation.

\subsection{Control Theory}\label{sec:rw-control}

The application of control theory to dynamic software reconfiguration is relatively new in comparison to the other formal approaches,
and is focused on modelling control loops and synthesizing controllers in self-adaptive software systems \cite{kn:PatColHanWan12}.
One example is given in \cite{kn:FilHofMag15}, which presents an automated method of synthesizing cascade controller systems with multiple actuators
and multiple controlled variables (termed \emph{dimensions}) based on a prioritization of the dimensions.
Each actuator has an associated control variable (termed a \emph{knob}) that can affect one or more dimensions simultaneously,
and the knobs are partitioned according to the highest priority of the dimensions they control.
Thus, a knob associated with a given priority does not control any dimension with a higher priority.
The method accepts as input a set of quantified goals (termed \emph{setpoints}) to be achieved for the dimensions, a set of knobs,
and possibly a prioritization scheme for the dimensions,
and outputs a synthesized control system.
No assumption is made about the relationship between a knob and a setpoint,
which is determined empirically at runtime using existing parametric modelling techniques,
and the setpoints are achieved by calculating values of the knobs in decreasing order of priority,
which reduces the computational complexity of the calculations.
The method assumes that each knob has default value,
the number of knobs is greater than or equal to the number of setpoints,
and that there is at least one dimension that is \emph{free}
(i.e. the dimension can be maximized or minimized without guaranteeing a specific value).
Dynamic reconfiguration consists of the replacement of one set of controllers by another set of dynamically synthesized controllers.
However, the research does not address the process through which the replacement occurs,
and (therefore) the interference between controllers and reconfiguration tasks is not addressed.
Nevertheless, the absence of any pre-defined relationship between knobs and setpoints supports unplanned reconfiguration.
The method is amenable to automation that would render it highly usable by system designers, but it currently lacks tool support.

\subsection{Summary}\label{sec:rw-summary}

The review of related work shows that no single formalism or category of formalisms is ideal,
since none of them meets all the requirements on an ideal formalism for dynamic software reconfiguration defined in Section \ref{sec:reqts}.

$\pi$-calculi represent the planned creation, deletion, and replacement of components/tasks using processes,
and the planned creation and deletion of communication links using port/action names,
but they cannot represent the unplanned creation, deletion, and replacement of components/tasks (unlike basic $\mathrm{CCS^{dp}}$).
Functional interference between concurrently executing tasks is represented using interleaved transitions, since true concurrency cannot be represented (unlike CPOGs).
The same limitations exist in \webpii~and in COWS.
Paradigm can express both planned and unplanned creation and deletion of components/tasks (but not reconfiguration of communication links),
and also functional interference due to interleaving concurrency and true concurrency,
but its models are not as terse as those of basic $\mathrm{CCS^{dp}}$ or of CPOGs, and it lacks development tools (unlike VDM).

Graph-based formalisms such as GARP and Reo represent reconfiguration in terms of graph rewrites,
and can represent both interleaving concurrency and true concurrency.
However, they have no formal semantics for representing the execution of both application tasks and reconfiguration tasks (unlike $\mathrm{CCS^{dp}}$ and CPOGs),
and (therefore) they cannot formally express or analyze functional interference between the two kinds of task (unlike $\mathrm{CCS^{dp}}$ and CPOGs).
Like GARP, CHAM represents the effect of reconfiguration rather than the process of reconfiguration,
and (therefore) expressing interference between application and reconfiguration tasks is problematic.
Furthermore, the CHAM notion of reconfiguration as the application of rewrite rules to rewrite rules can be problematic for a system designer to understand
(unlike the fraction process in basic $\mathrm{CCS^{dp}}$, a reconfiguration action in a CPOG, and a reconfiguration operation in a VDM model).

The Aguirre-Maibaum language can represent the planned creation and deletion of tasks (using class instances) and of communication links (using association instances),
but not in an unplanned manner.
Furthermore, representing interference between application and reconfiguration tasks is problematic, since the language is declarative.
However, the language supports model checking, which is also supported by VDM, CPOGs, and basic $\mathrm{CCS^{dp}}$.
In fact, temporal logics are complementary to VDM, CPOGs, and basic $\mathrm{CCS^{dp}}$,
because requirements on application tasks, reconfiguration tasks, and on their interference can be formulated using temporal logics,
and the formulae can be verified by model checking the VDM, CPOG, and basic $\mathrm{CCS^{dp}}$ expressions that represent the execution of the tasks.

The reconfiguration presented in \cite{kn:FilHofMag15} does not discuss the process of reconfiguration,
and (therefore) the issue of interference between application and reconfiguration tasks is not addressed.
However, the research is complementary to our use of VDM, CPOG, and basic $\mathrm{CCS^{dp}}$,
because VDM can be used to verify formally that the synthesized controllers are a correct implementation of their specification,
and basic $\mathrm{CCS^{dp}}$ and CPOGs can be used to construct and to verify correct reconfiguration paths between the old set of controllers and the new set of controllers
that take into account interference between the concurrent execution of old controllers, new controllers, and reconfiguration tasks.

\section{Concluding Remarks}\label{sec:concl}


This paper has used the dynamic reconfiguration of a simple office workflow for order processing as a case study
in order to compare empirically the modelling and analysis capabilities of three formalisms of different kinds,
namely, VDM, CPOGs, and basic $\mathrm{CCS^{dp}}$.

The evaluations of the three formalisms show that none of them is ideal,
since none of them meets all the requirements on an ideal formalism for dynamic software reconfiguration defined in Section \ref{sec:reqts}.
For example, key requirements include:
the ability to express tersely change in the composition and structure of software components and tasks
for both planned and unplanned dynamic reconfiguration;
the ability to express tersely the concurrent execution of tasks and their functional interference;
and the ability to verify the functional correctness requirements of tasks,
which includes verifying the functional correctness of refinements.
Neither VDM, nor CPOGs, nor basic $\mathrm{CCS^{dp}}$ meets all three requirements.
However, the formalisms meet the requirements collectively.
Furthermore, all three formalisms can easily express traces of actions.
Therefore, the formalisms are complementary, and it should be possible to combine them
using basic $\mathrm{CCS^{dp}}$ for modelling, CPOGs for verification, and VDM for type checking and refinement.

The main strengths of basic $\mathrm{CCS^{dp}}$ are its ability:
to model abstractly and tersely the composition and concurrent execution of application and reconfiguration tasks using concurrent processes,
to model their functional interference using interleaved transitions,
and to model their planned and unplanned reconfiguration using fraction processes.
Furthermore, cyclic processes can be modelled using recursion,
and fraction processes can themselves be reconfigured using other fractions.
The main weaknesses of basic $\mathrm{CCS^{dp}}$ are its inability
to control non-deterministic transitions
and inability to reconfigure selectively specific process instances,
the computational complexity of process matching based on strong of-bisimulation,
the computational complexity and restrictiveness of process congruence
that severely limits the use of equational reasoning to verify requirements \cite{kn:Bha13},
and lack of tools.
In contrast, the main strength of a CPOG is its ability to verify requirements efficiently.
A model and its requirement can each be transformed into a canonical form and then compared using a Boolean SAT solver,
and the predicates on actions and on action dependencies and the acyclic topology of a CPOG support efficient model checking.
The correctness of a reconfiguration from one configuration to another can be proved
using consistent histories of actions of the two configurations and by restricting interference through forbidden actions.
Furthermore, functional interference between tasks can be modelled using either interleaved actions or simultaneous actions.
The main weaknesses of a CPOG are:
its inability to model composition and structure of software components and tasks,
its low level of abstraction for modelling,
its inability to model cyclic processes,
and lack of available tools.
In contrast to both basic $\mathrm{CCS^{dp}}$ and CPOGs, VDM was designed for formal development of software.
The main strengths of VDM-SL are:
its ability to model workflows, software components, and tasks as data types, which facilitates their refinement to an implementation,
its mature and available tools for development, simulation, and testing,
and its ease of use by system designers.
The main weaknesses of VDM-SL are lack of constructs for modelling concurrency and interference,
and lack of formal verification tools.

All three formalisms can represent traces of actions or transitions. Furthermore, a CPOG is the unfolding of a process.
Therefore, it is possible to map a process to a CPOG.
A variant of CPOGs with variables whose value can be changed more than once by actions can be used to simplify the mapping.
A recursively defined process can be unfolded to a CPOG using a standard unfolding procedure.
Rely and guarantee conditions on the transitions of a process can be used to define a priority scheme for the transitions
in order to control their non-determinism and to define a partial order between their corresponding actions in the CPOG of the process.
As with CPOGs, it should be possible to convert a process in basic $\mathrm{CCS^{dp}}$ into a graph of actions in VDM-RT
for type checking and refinement into an executable form.
Thus, it should be possible to construct an integrated approach to the formal modelling, verification, and development
of dynamically reconfigurable dependable systems based on VDM, CPOGs, and basic $\mathrm{CCS^{dp}}$
or a combination of similar formalisms. We intend to demonstrate this hypothesis in our future work.

\section{Acknowledgements}

The research leading to this paper was funded from several sources:
Bhattacharyya's research was funded by the UK Engineering and Physical Sciences Research Council (EPSRC)
under the terms of a graduate studentship,
the Platform Grant on Trustworthy Ambient Systems (TrAmS), and the UNCOVER project,
and by the European Community's Seventh Framework (FP7) Deploy project;
Mokhov's research was funded by the UNCOVER project and by a Newcastle University School Fellowship;
and Pierce's research was funded from the FP7 and Horizon 2020 programmes (287829 COMPASS, 644047 INTO-CPS, and 644400 CPSE Labs).
The authors acknowledge the help given by numerous colleagues, in particular:
Jeremy Bryans, John Fitzgerald, Regina Frei, Kohei Honda, Alexei Iliasov, Cliff Jones, Victor Khomenko, Maciej Koutny,
Manuel Mazzara, Richard Payne, Traian Florin Serbanuta, Giovanna Di Marzo Serugendo, and Chris Woodford.

This is a revised version of the paper.
The authors thank the anonymous reviewers for their constructive feedback and helpful suggestions,
the implementation of which undoubtedly improved the quality of the paper.

\bibliographystyle{alpha}
\bibliography{mainpaper,cpogs}

\appendices

\counterwithin{lstlisting}{section}
\newpage
\section{VDM}\label{sec:appendix:vdm}
The following three listings give the full specification of the VDM-SL model described in Section~\ref{sec:vdm}.

\begin{lstlisting}[style=styleVDM,basicstyle=\ttfamily\scriptsize,caption=\texttt{Configurations} module,label=lst:app_configurations]
module Configurations

exports all

definitions

types

-- actions in a workflow
Action = <OrderReceipt>
       | <InventoryCheck>
       | <Reject>
       | <CreditCheck>
       | <SupplierCheck>
       | <Shipping>
       | <Billing>
       | <Archiving>
       | <Confirmation>;

-- workflow with invariant
Workflow = Element
inv w == forall tr in set tracesof(w) & card elems tr = len tr;

-- elements that make up a workflow
Element =  [Simple | Branch | Par];

-- a simple Workflow
Simple :: a : Action
          w : Workflow;

-- a conditional Workflow
Branch :: a : Action
          t : Workflow
          f : Workflow;

-- parallel Workflows
Par :: b1 : Action
       b2 : Action
        w : Workflow;

-- record of an action or termination
Event = Action | <TERMINATE>;

-- trace of events
Trace = seq of Event
inv t == (<TERMINATE> in set elems t and len t > 1) =>
  <TERMINATE> not in set elems t(1, ..., len t - 1);

functions

-- compute all traces of an workflow
tracesof: Element -> set of Trace
tracesof(el) ==
    cases el:
        mk_Simple(a,e) -> {[a] ^ tr | tr in set tracesof(e)},
      mk_Branch(a,t,f) -> {[a] ^ tr | tr in set tracesof(t)} union
                          {[a] ^ tr | tr in set tracesof(f)},
       mk_Par(b1,b2,e) -> {[b1,b2] ^ tr | tr in set tracesof(e)} union
                          {[b2,b1] ^ tr | tr in set tracesof(e)},
                   nil -> {[<TERMINATE>]},
                others -> {}
   end;

-- true if a is a prefix of b, false otherwise
prefixof: Trace * Trace -> bool
prefixof(a, b) ==
	if a = [] then true
	elseif b = [] then false
	else if hd a <> hd b then false else prefixof(tl a, tl b)

values

-- first configuration
Configuration1: Workflow =
    mk_Simple(<OrderReceipt>,
        mk_Branch(<InventoryCheck>,
            mk_Branch(<CreditCheck>,
                mk_Simple(<Shipping>,
                    mk_Simple(<Billing>,
                        mk_Simple(<Archiving>,
                            mk_Simple(<Confirmation>,nil)
                        )
                    )
                ),
                mk_Simple(<Reject>, nil)
            ),
            mk_Simple(<Reject>, nil)
        )
    );

-- second configuration
Configuration2: Workflow =
    mk_Simple(<OrderReceipt>,
        mk_Branch(<InventoryCheck>,
            mk_Branch(<CreditCheck>,
                mk_Par(<Billing>, <Shipping>,
                    mk_Simple(<Archiving>, nil)),
                mk_Simple(<Reject>, nil)
            ),
            mk_Branch(<SupplierCheck>,
                mk_Branch(<CreditCheck>,
                    mk_Par(<Billing>, <Shipping>,
                        mk_Simple(<Archiving>, nil)),
                    mk_Simple(<Reject>, nil)
                ),
                mk_Simple(<Reject>, nil)
            )
        )
    );

end Configurations
\end{lstlisting}

\begin{lstlisting}[style=styleVDM,basicstyle=\ttfamily\scriptsize,caption=\texttt{Interpreter} module,label=lst:app_interpreter]
module Interpreter

exports all

imports from Configurations
            types Action renamed Action;
                  Workflow renamed Workflow;
                  Simple renamed Simple;
                  Branch renamed Branch;
                  Par renamed Par;
                  Event renamed Event;
                  Trace renamed Trace
            functions tracesof renamed tracesof;
            					prefixof renamed prefixof
            values Configuration1 renamed Configuration1
                   Configuration2 renamed Configuration2,
        from MATH
        	operations rand,
        from IO
        	functions println

definitions

types

-- collapse probabilities
Choices = map Action to bool
inv c == dom c = {<InventoryCheck>, <CreditCheck>, <SupplierCheck>}

-- interpreter state
state S of
       trace : Trace
    workflow : Workflow
init s == s = mk_S([], nil)
inv mk_S(trace, workflow) ==
workflow <> nil =>
  (forall t in set
    {trace ^ tr | tr in set tracesof(workflow)} &
	    (exists x in set tracesof(Configuration1) union tracesof(Configuration2) &
        prefixof(t, x)))
end;

operations

-- reset interpreter
Init: Workflow ==> ()
Init(w) == (
    trace := [];
    workflow := w
);

-- return: current trace
GetTrace: () ==> Trace
GetTrace() == return trace;

-- execute workflow in one go
Execute: Choices ==> ()
Execute(c) == (
    while workflow <> nil do
    	let - = Step(c) in skip;
    trace := trace ^ [<TERMINATE>]
);

-- perform a single step of execution
-- return: last action that occurred
Step: Choices ==> Event
Step(choices) ==
(
	-- simple element
	if is_Simple(workflow) then
	(
		let temp = workflow in (
			workflow := nil;
			trace := trace ^ [temp.a];
			workflow := temp.w;		
		)
	)
	-- branching element
	elseif is_Branch(workflow) then
	(
	  let temp = workflow in (
			workflow := nil;
	    trace := trace ^ [temp.a];	
			-- brance based on choices
			let test = choices(temp.a) in
				if test then workflow := temp.t
				else workflow := temp.f
		)
	)
	-- parallel element
	elseif is_Par(workflow) then
	(
	  let temp = workflow in (
			workflow := nil;
		  if MATH`rand(1) = 1 then
			  trace := trace ^ [temp.b1,temp.b2]
		  else
				trace := trace ^ [temp.b2,temp.b1];		
		  workflow := temp.w
		)
	)
	else error;

	-- return latest action
	return trace(len trace)
)
pre workflow <> nil;

-- reconfigure, replacing current workflow
Reconfigure: Workflow ==> ()
Reconfigure(w) ==
	workflow := w
pre w <> nil and
    branch_check(trace, workflow, w) and
    forall t in set {trace ^ ftr | ftr in set tracesof(w)} &
	  	(exists x in set tracesof(Configuration2) & t = x);

functions

branch_check: Trace * Workflow * Workflow -> bool
branch_check(tr, w, w') ==
  (last(tr) = <InventoryCheck> and first(w) = {<Reject>}
    => {<SupplierCheck>} subset first(w')) and
  (last(tr) = <InventoryCheck> and first(w) = {<CreditCheck>}
    => {<CreditCheck>} subset first(w')) and
  (last(tr) = <CreditCheck> and first(w) = {<Reject>} 	
    => {<Reject>} subset first(w')) and
  (last(tr) = <CreditCheck> and first(w) = {<Shipping>}
    => first(w') subset {<Billing>, <Shipping>});

-- last action of a trace
last: Trace -> Action
last(tr) == tr(len tr);

-- first action(s) of a workflow
first: Workflow -> set of Action
first(w) ==
	cases w:
	  mk_Simple(a,e)		-> {a},
    mk_Branch(a,t,f)	-> {a},
    mk_Par(b1,b2,e)		-> {b1,b2}
	end;

end Interpreter
\end{lstlisting}

\begin{lstlisting}[style=styleVDM,basicstyle=\ttfamily\scriptsize,caption=\texttt{Test} module,label=lst:app_test]
module Test

imports from Configurations
						types Action renamed Action;
									Trace renamed Trace;
									Event renamed Event;
									Workflow renamed Workflow;
									Simple renamed Simple;
									Branch renamed Branch;
									Par renamed Par
						functions tracesof renamed tracesof
											prefixof renamed prefixof
						-- remove need to fully qualify these names
						values Configuration1 renamed Configuration1;
									 Configuration2 renamed Configuration2,
			  from Interpreter
			  		types Choices renamed Choices
			  		operations Execute; Init; Step; Reconfigure; GetTrace,
        from IO
        	  operations printf; println

definitions

operations

TestReconfig: Choices * Action * Workflow ==> ()
TestReconfig(c, rp, w) == (
	-- initialise interpreter
	dcl a: [Action] := nil;
	dcl valid: bool := true;
	Interpreter`Init(Configuration1);
	-- execute until rp reached
	while a <> rp do a := Interpreter`Step(c);
	IO`println(Interpreter`GetTrace());
	IO`println("Reconfiguring Configuration1 to Configuration2...");
	-- check if pre-condition will fail
 	for all t in set {Interpreter`GetTrace() ^ ftr | ftr in set tracesof(w)} do (
 		dcl atleastonevalid: bool := false;
 		for all x in set tracesof(Configuration2) do
 			if prefixof(t, x) then atleastonevalid := true;
		if not atleastonevalid then (
			IO`printf("This potential trace is not valid under Configuration2: %s\n", [t]);
			valid := false
		);
 	);
 	if not valid then IO`println(
		"Reconfiguration could generate invalid traces; pre-condition will fail."
	);		
	-- perform reconfiguration
	Interpreter`Reconfigure(w);
	Interpreter`Execute(c);
	IO`println(Interpreter`GetTrace());
);

TestReconfigSuccess: () ==> ()
TestReconfigSuccess() ==
	let w =
		mk_Branch(<SupplierCheck>,
		  mk_Branch(<CreditCheck>,
		      mk_Par(<Billing>, <Shipping>,
		          mk_Simple(<Archiving>, nil)),
		      mk_Simple(<Reject>, nil)
		  ),
		  mk_Simple(<Reject>, nil))
  in TestReconfig(ExternalStock, <InventoryCheck>, w);

TestReconfigFail: () ==> ()
TestReconfigFail() == let w =
		mk_Par(<Billing>, <Shipping>,
    	mk_Simple(<Archiving>, nil))
  in TestReconfig(NoProblems, <Shipping>, w);

operations

-- test all basic configurations
TestAll: () ==> ()
TestAll() == (
	-- Configuration1
	IO`printf("Config1NoStock():\t\t%s\n", 							[Config1NoLocalStock()]);
	IO`printf("Config1ExternalStock():\t\t%s\n",        [Config1ExternalStock()]);
	IO`printf("Config1ExternalStockNoCredit():\t%s\n",	[Config1ExternalStockNoCredit()]);
	IO`printf("Config1NoCredit():\t\t%s\n", 		 				[Config1NoCredit()]);
	IO`printf("Config1NoProblems():\t\t%s\n\n", 				[Config1NoProblems()]);

	-- Configuration2
	IO`printf("Config2NoStock():\t\t%s\n", 							[Config2NoLocalStock()]);
	IO`printf("Config2ExternalStock():\t\t%s\n",				[Config2ExternalStock()]);
	IO`printf("Config2ExternalStockNoCredit():\t%s\n", 	[Config2ExternalStockNoCredit()]);
	IO`printf("Config2NoCredit():\t\t%s\n", 						[Config2NoCredit()]);
	IO`printf("Config2NoProblems():\t\t%s\n", 					[Config2NoProblems()]);
);

-- Test Configuration1 / NoProblems
Config1NoProblems: () ==> Trace
Config1NoProblems() == (
	Interpreter`Init(Configuration1);
	Interpreter`Execute(NoProblems);
	return Interpreter`GetTrace()
);

-- Test Configuration1 / NoLocalStock
Config1NoLocalStock: () ==> Trace
Config1NoLocalStock() == (
	Interpreter`Init(Configuration1);
	Interpreter`Execute(NoStock);
	return Interpreter`GetTrace()
);

-- Test Configuration1 / NoCredit	
Config1NoCredit: () ==> Trace
Config1NoCredit() == (
	Interpreter`Init(Configuration1);
	Interpreter`Execute(NoCredit);
	return Interpreter`GetTrace()
);
	
-- Test Configuration1 / NoLocalButExternalStock	
Config1ExternalStock: () ==> Trace
Config1ExternalStock() == (
	Interpreter`Init(Configuration1);
	Interpreter`Execute(ExternalStock);		
	return Interpreter`GetTrace()
);

-- Test Configuration1 / ExternalStockNoCredit	
Config1ExternalStockNoCredit: () ==> Trace
Config1ExternalStockNoCredit() == (
	Interpreter`Init(Configuration1);
	Interpreter`Execute(ExternalStockNoCredit);	
	return Interpreter`GetTrace()
);

-- Test Configuration2 / NoProblems
Config2NoProblems: () ==> Trace
Config2NoProblems() == (
	Interpreter`Init(Configuration2);
	Interpreter`Execute(NoProblems);
	return Interpreter`GetTrace()
);

-- Test Configuration2 / NoLocalStock
Config2NoLocalStock: () ==> Trace
Config2NoLocalStock() == (
	Interpreter`Init(Configuration2);
	Interpreter`Execute(NoStock);
	return Interpreter`GetTrace()
);

-- Test Configuration2 / NoCredit	
Config2NoCredit: () ==> Trace
Config2NoCredit() == (
	Interpreter`Init(Configuration2);
	Interpreter`Execute(NoCredit);
	return Interpreter`GetTrace()
);

-- Test Configuration2 / ExternalStock	
Config2ExternalStock: () ==> Trace
Config2ExternalStock() == (
	Interpreter`Init(Configuration2);
	Interpreter`Execute(ExternalStock);	
	return Interpreter`GetTrace()
);

-- Test Configuration2 / ExternalStockNoCredit	
Config2ExternalStockNoCredit: () ==> Trace
Config2ExternalStockNoCredit() == (
	Interpreter`Init(Configuration2);
	Interpreter`Execute(ExternalStockNoCredit);	
	return Interpreter`GetTrace()
);

values

-- all branches true
NoProblems = {
	<InventoryCheck> |-> true,
   <SupplierCheck> |-> true,
     <CreditCheck> |-> true
};

-- fail inventory check
NoStock = {
	<InventoryCheck> |-> false,
   <SupplierCheck> |-> false,
     <CreditCheck> |-> false
};

-- have local inventory, fail credit
NoCredit = {
	<InventoryCheck> |-> true,
   <SupplierCheck> |-> false,
     <CreditCheck> |-> false
};

-- have external inventory
ExternalStock = {
	<InventoryCheck> |-> false,
   <SupplierCheck> |-> true,
     <CreditCheck> |-> true
};

-- have external inventory
ExternalStockNoCredit = {
	<InventoryCheck> |-> false,
   <SupplierCheck> |-> true,
     <CreditCheck> |-> false
};

end Test
\end{lstlisting}

\newpage
\section{Model Checking VDM Traces of Actions to Verify Requirement R2}\label{sec:appendix:vdmmc}
Our approach to verifying reconfiguration requirement R2 by model checking is:
to identify the VDM action traces of \texttt{Configuration1} and \texttt{Configuration2},
map each action trace to an infinite trace of states in a Kripke structure
such that each state records the most recent action of the workflow executed by the interpreter (in the \textit{action} attribute)
and the LTS rule application that justifies the action (in the \textit{rule} attribute),
then verify that the requirements on Configuration 1 and on Configuration 2 expressed in linear temporal logic (LTL)
are satisfied by the state traces of \texttt{Configuration1} and \texttt{Configuration2} respectively.
LTL was chosen because of reports of a VDM interpreter front end to the SPIN model checker \cite{Lin16}, which verifies LTL formulae.
The Kripke structures of \texttt{Configuration1} and \texttt{Configuration2} are used
to derive Kripke structures of the reconfigured workflows that are relevant to the R2 requirement,
from which it is clear that the requirement is satisfied.
Invocation of the \textit{Reconfigure} operation (which is not a workflow action) is recorded in the \textit{reconfigure} Boolean attribute,
and is indicated by~\CheckedBox~in the Kripke structure diagrams.
Notice that recording LTS rule applications helps to represent, and to verify the correctness of, conditional branching within workflows.

There are exactly three action traces of Configuration 1, by definition of \texttt{Configuration1} and the application of LTS rules:

$AT_{1, 1} \triangleq$
$[\langle{}OrderReceipt\rangle, \langle{}InventoryCheck\rangle, \langle{}CreditCheck\rangle, \langle{}Shipping\rangle, \langle{}Billing\rangle,$
$\langle{}Archiving\rangle, \langle{}Confirmation\rangle,$ \\ $\langle{}TERMINATE\rangle]$

$AT_{1, 2} \triangleq$
$[\langle{}OrderReceipt\rangle, \langle{}InventoryCheck\rangle, \langle{}Reject\rangle, \langle{}TERMINATE\rangle]$

$AT_{1, 3} \triangleq$
$[\langle{}OrderReceipt\rangle, \langle{}InventoryCheck\rangle, \langle{}CreditCheck\rangle, \langle{}Reject\rangle, \langle{}TERMINATE\rangle]$

\begin{figure*}
\begin{center}
\includegraphics[width=1\textwidth]{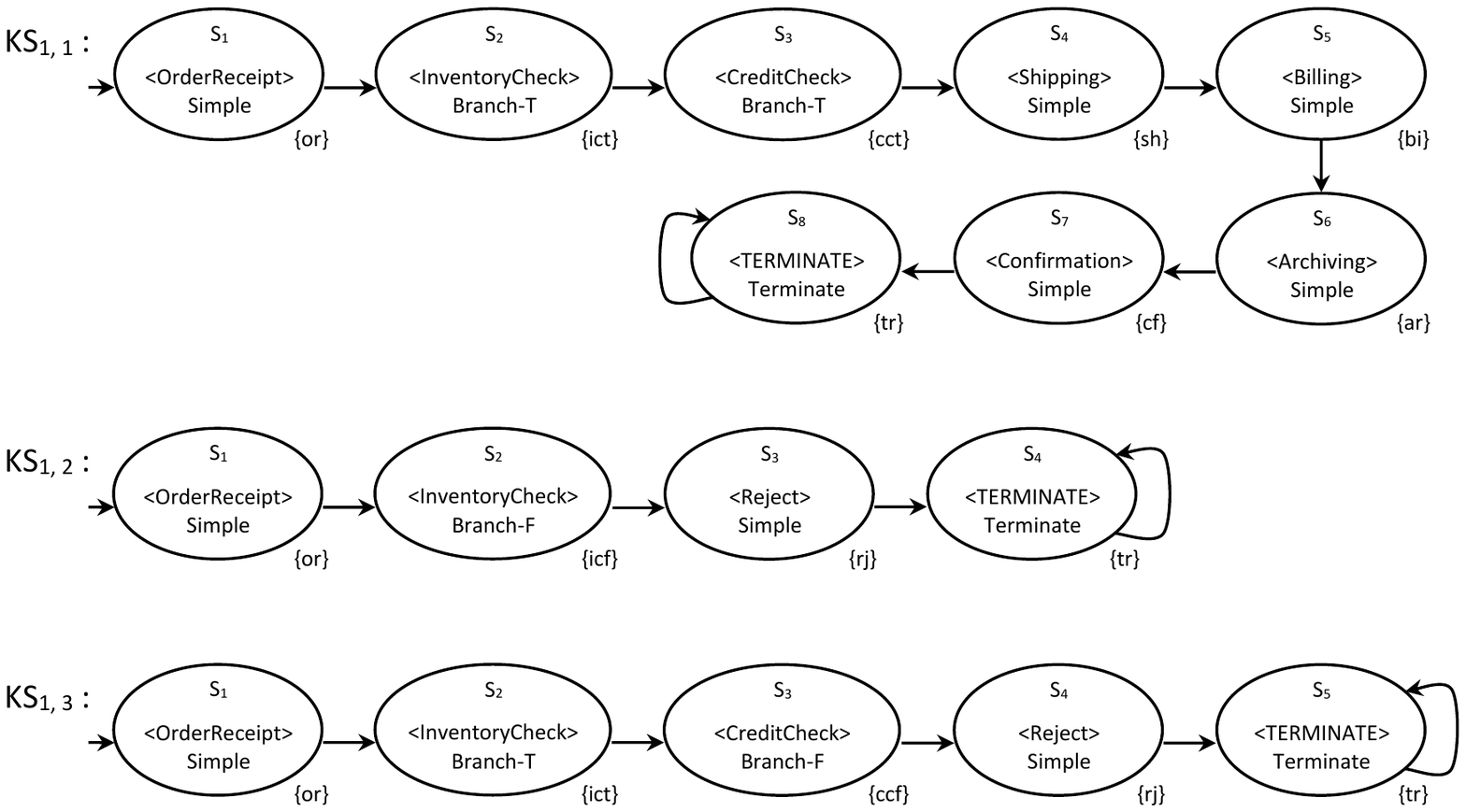} \vspace{-15.6cm}
\caption{Kripke structures of the action traces of \texttt{Configuration1}\label{fig:KSconfig1}}
\end{center}
\end{figure*}

The above three action traces can be mapped to the three Kripke structures shown in Figure \ref{fig:KSconfig1},
where $AT_{1, 1}$, $AT_{1, 2}$, and $AT_{1, 3}$ correspond to $KS_{1, 1}$, $KS_{1, 2}$, and $KS_{1, 3}$ respectively.

The atomic propositions of the Kripke structures are defined below.

\(
\begin{array}{ll}
\mathit{or   \,\triangleq\, action = \langle{}OrderReceipt\rangle} & \\
\mathit{ict \,\triangleq\, action = \langle{}InventoryCheck\rangle \,\wedge\, rule = Branch\!-\!T}\;\;\; &
\mathit{icf \,\triangleq\, action = \langle{}InventoryCheck\rangle \,\wedge\, rule = Branch\!-\!F} \\
\mathit{cct \,\triangleq\, action = \langle{}CreditCheck\rangle \,\wedge\, rule = Branch\!-\!T}\;\;\; &
\mathit{ccf \,\triangleq\, action = \langle{}CreditCheck\rangle \,\wedge\, rule = Branch\!-\!F} \\
\mathit{rj   \,\triangleq\, action = \langle{}Reject\rangle} &
\mathit{tr   \,\triangleq\, action = \langle{}TERMINATE\rangle} \\
\mathit{sh   \,\triangleq\, action = \langle{}Shipping\rangle} &
\mathit{bi   \,\triangleq\, action = \langle{}Billing\rangle} \\
\mathit{ar   \,\triangleq\, action = \langle{}Archiving\rangle} &
\mathit{cf   \,\triangleq\, action = \langle{}Confirmation\rangle} \\
\mathit{sct \,\triangleq\, action = \langle{}SupplierCheck\rangle \,\wedge\, rule = Branch\!-\!T}\;\;\; &
\mathit{scf \,\triangleq\, action = \langle{}SupplierCheck\rangle \,\wedge\, rule = Branch\!-\!F} \\
\mathit{rc   \,\triangleq\, reconfigure = \mathit{true}} & \\
\end{array}
\)

The LTL formulae for verifying requirements on Configuration 1 given in Section \ref{sec:wfreqts1} are defined below.
Notice that the requirements C1.5.5 (the customer must not receive more than one shipment of an order -- the safety requirement)
and C1.6 (each task must be performed at most once) are met through the invariant condition on workflows defined in VDM (see Listing \ref{lst:element}),
and (therefore) are not model checked.

\(\mathit{CF_{1, 1} \triangleq
or ~\textbf{U}~ (ict ~\textbf{U}~ (cct ~\textbf{U}~ (sh ~\textbf{U}~ (bi ~\textbf{U}~ (ar ~\textbf{U}~ (cf ~\textbf{U}~ \square tr)))))) ~\wedge~ \square \neg rj}
\) \\
This formula verifies requirements if the order is successfully completed, see Figure \ref{OfficeWorkflowConfig1}.

\(\mathit{CF_{1, 2} \triangleq
or ~\textbf{U}~ (icf ~\textbf{U}~ (rj ~\textbf{U}~ \square tr)) ~\wedge~ \square \neg cf}
\) \\
This formula verifies requirements if \texttt{Inventory Check} fails.

\(\mathit{CF_{1, 3} \triangleq
or ~\textbf{U}~ (ict ~\textbf{U}~ (ccf ~\textbf{U}~ (rj ~\textbf{U}~ \square tr))) ~\wedge~ \square \neg cf}
\) \\
This formula verifies requirements if \texttt{Credit Check} fails.

$\mathit{CF_1 \triangleq CF_{1, 1} \,\vee\, CF_{1, 2} \,\vee\, CF_{1, 3}}$ \\
This formula verifies requirements on Configuration 1,
and is satisfied by the paths (starting at state $S_1$) of the Kripke structures of the action traces of \texttt{Configuration1}
(by inspection of Figure \ref{fig:KSconfig1}).

There are exactly seven action traces of Configuration 2, by definition of \texttt{Configuration2} and the application of LTS rules:

$AT_{2, 1} \triangleq$
$[\langle{}OrderReceipt\rangle, \langle{}InventoryCheck\rangle, \langle{}CreditCheck\rangle, \langle{}Billing\rangle, \langle{}Shipping\rangle,$
$\langle{}Archiving\rangle, \langle{}TERMINATE\rangle]$

$AT_{2, 2} \triangleq$
$[\langle{}OrderReceipt\rangle, \langle{}InventoryCheck\rangle, \langle{}CreditCheck\rangle, \langle{}Shipping\rangle, \langle{}Billing\rangle,$
$\langle{}Archiving\rangle, \langle{}TERMINATE\rangle]$

$AT_{2, 3} \triangleq$
$[\langle{}OrderReceipt\rangle, \langle{}InventoryCheck\rangle, \langle{}SupplierCheck\rangle, \langle{}Reject\rangle, \langle{}TERMINATE\rangle]$

$AT_{2, 4} \triangleq$
$[\langle{}OrderReceipt\rangle, \langle{}InventoryCheck\rangle, \langle{}SupplierCheck\rangle, \langle{}CreditCheck\rangle, \langle{}Reject\rangle, \langle{}TERMINATE\rangle]$

$AT_{2, 5} \triangleq$
$[\langle{}OrderReceipt\rangle, \langle{}InventoryCheck\rangle, \langle{}SupplierCheck\rangle, \langle{}CreditCheck\rangle, \langle{}Billing\rangle, \langle{}Shipping\rangle,$
$\langle{}Archiving\rangle,$ \\ $\langle{}TERMINATE\rangle]$

$AT_{2, 6} \triangleq$
$[\langle{}OrderReceipt\rangle, \langle{}InventoryCheck\rangle, \langle{}SupplierCheck\rangle, \langle{}CreditCheck\rangle, \langle{}Shipping\rangle, \langle{}Billing\rangle,$
$\langle{}Archiving\rangle,$ \\ $\langle{}TERMINATE\rangle]$

$AT_{2, 7} \triangleq$
$[\langle{}OrderReceipt\rangle, \langle{}InventoryCheck\rangle, \langle{}CreditCheck\rangle, \langle{}Reject\rangle, \langle{}TERMINATE\rangle]$

\begin{figure*}
\begin{center}
\includegraphics[width=1\textwidth]{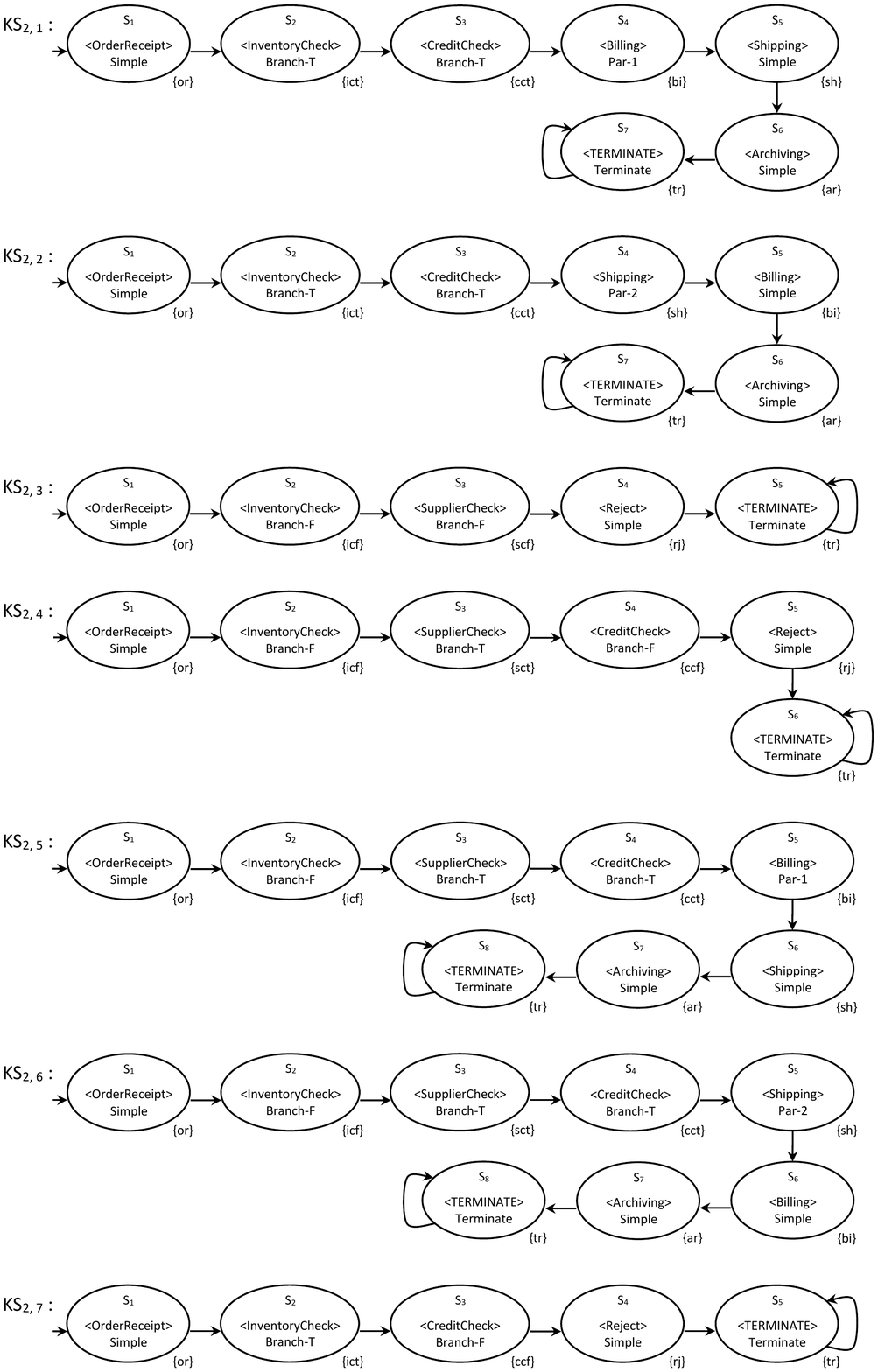} \vspace{-1cm}
\caption{Kripke structures of the action traces of \texttt{Configuration2}\label{fig:KSconfig2}}
\end{center}
\end{figure*}

\begin{figure*}
\begin{center}
\includegraphics[width=1\textwidth]{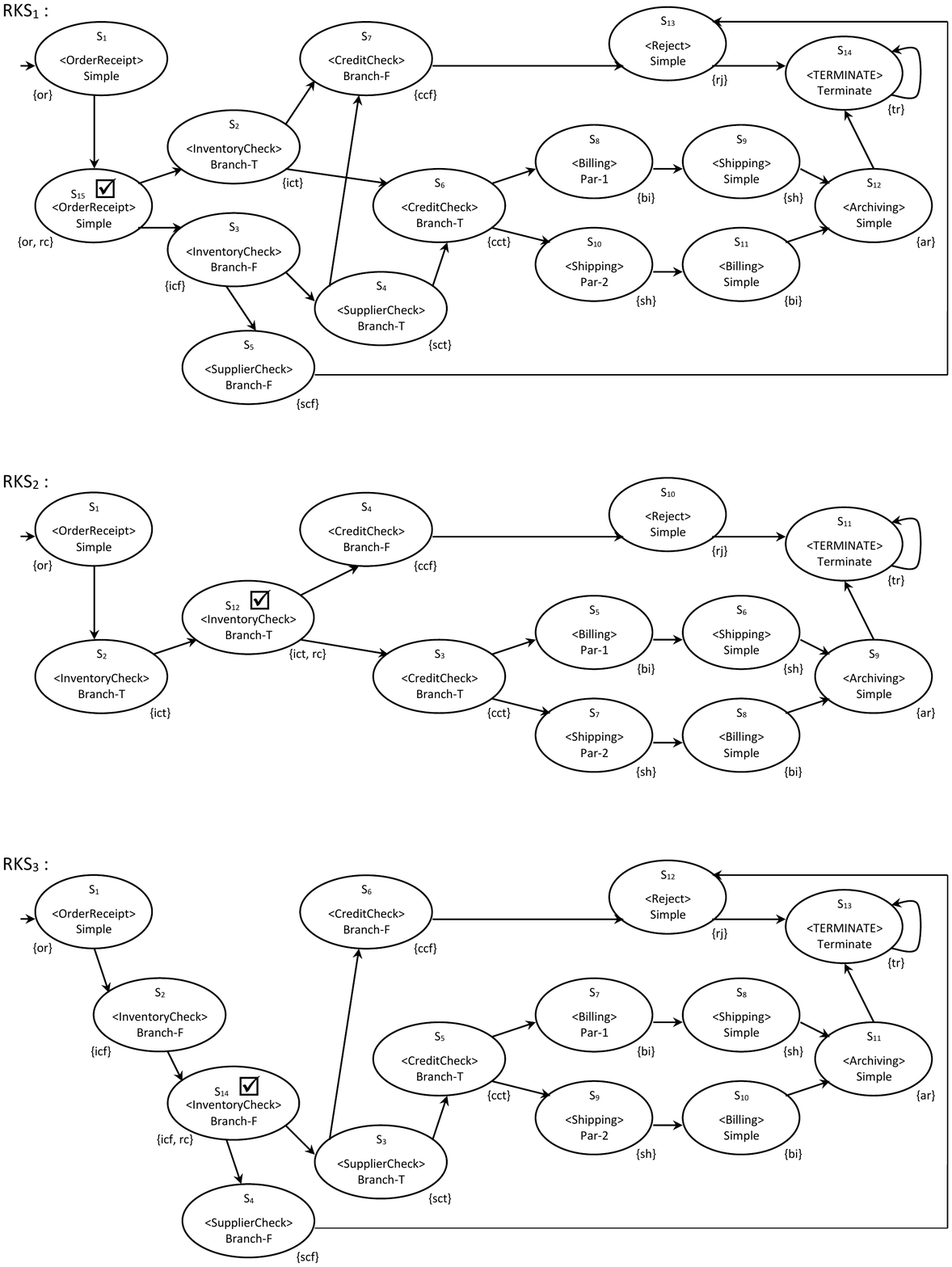} \vspace{-1.5cm}
\caption{Kripke structures of action traces of workflows reconfigured immediately after execution of $\mathit{OrderReceipt}$ or $\mathit{InventoryCheck}$
\label{fig:KSconfig12a}}
\end{center}
\end{figure*}

\begin{figure*}
\begin{center}
\includegraphics[width=1\textwidth]{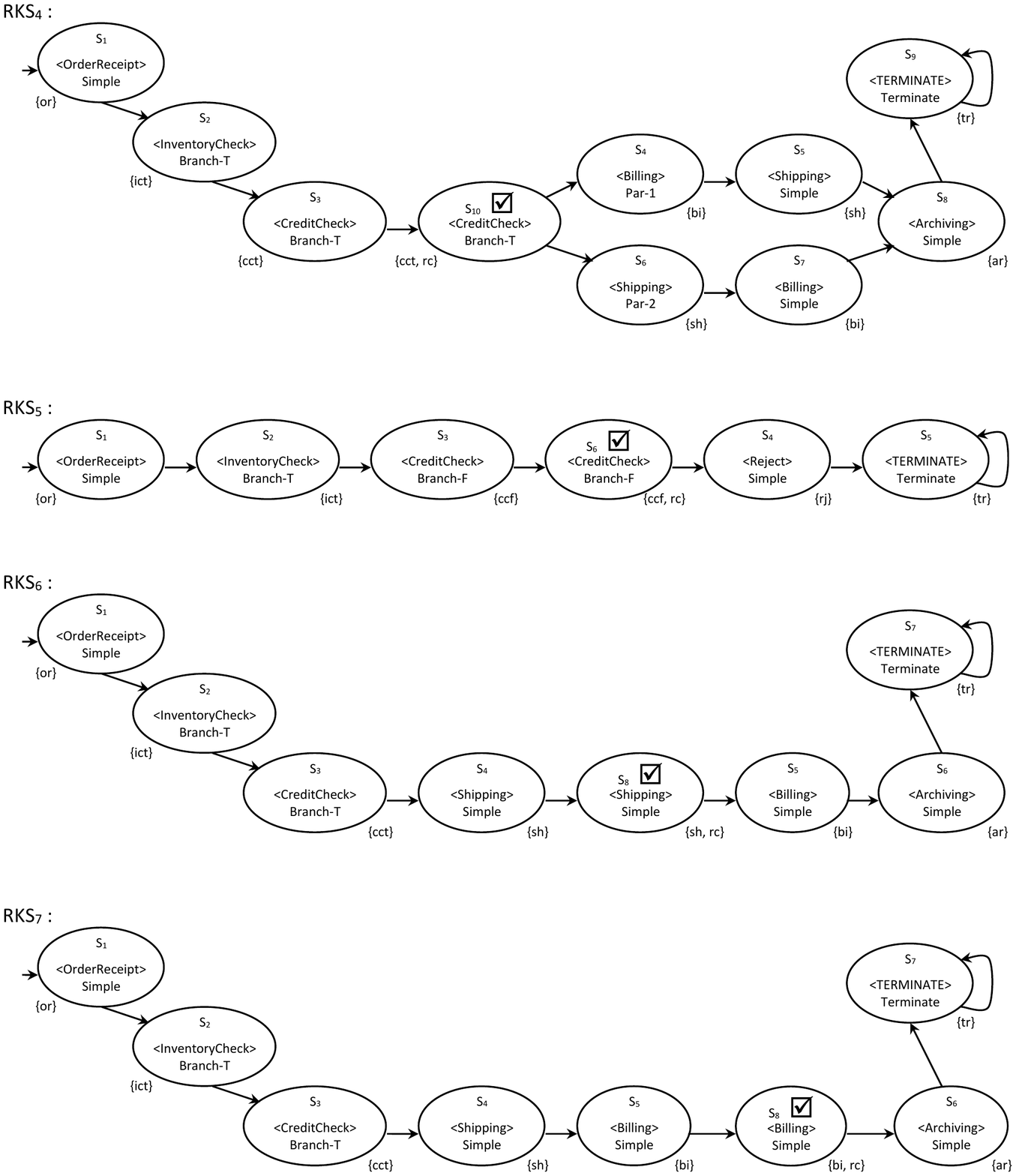} \vspace{-4.7cm}
\caption{Kripke structures of action traces of workflows reconfigured immediately after execution of
$\mathit{CreditCheck}$ or $\mathit{Shipping}$ or $\mathit{Billing}$
\label{fig:KSconfig12b}}
\end{center}
\end{figure*}

The above seven action traces can be mapped to the seven Kripke structures shown in Figure \ref{fig:KSconfig2},
where $AT_{2, 1}$, $AT_{2, 2}$, $AT_{2, 3}$, $AT_{2, 4}$, $AT_{2, 5}$, $AT_{2, 6}$, and $AT_{2, 7}$ correspond to
$KS_{2, 1}$, $KS_{2, 2}$, $KS_{2, 3}$, $KS_{2, 4}$, $KS_{2, 5}$, $KS_{2, 6}$, and $KS_{2, 7}$ respectively,
which shows clearly the mapping between action and state traces.
We combine the seven Kripke structures into an equivalent single Kripke structure
in order to represent compactly state traces of reconfigured workflows, as shown in Figures \ref{fig:KSconfig12a} and \ref{fig:KSconfig12b}.

The LTL formulae for verifying requirements on Configuration 2 given in Section \ref{sec:wfreqts2} are defined below.
Requirements C2.5.4 (the customer must not receive more than one shipment of an order -- the safety requirement)
and C2.6 (each task must be performed at most once) are met through the invariant condition on workflows defined in VDM (see Listing \ref{lst:element}),
and (therefore) are not model checked.

\(\mathit{CF_{2, 1} \triangleq
or ~\textbf{U}~ (ict ~\textbf{U}~ (cct ~\textbf{U}~ ((bi \vee sh) ~\textbf{U}~ (ar ~\textbf{U}~ \square tr)))) ~\wedge~ \square \neg rj}
\) \\
This formula verifies requirements if no \texttt{Evaluation} task fails, see Figure \ref{OfficeWorkflowConfig2}.

\(\mathit{CF_{2, 2} \triangleq
or ~\textbf{U}~ (icf ~\textbf{U}~ (scf ~\textbf{U}~ (rj ~\textbf{U}~ \square tr))) ~\wedge~ \square \neg ar}
\) \\
This formula verifies requirements if \texttt{Inventory Check} and \texttt{Supplier Check} both fail.

\(\mathit{CF_{2, 3} \triangleq
or ~\textbf{U}~ (icf ~\textbf{U}~ (sct ~\textbf{U}~ (ccf ~\textbf{U}~ (rj ~\textbf{U}~ \square tr)))) ~\wedge~ \square \neg ar}
\) \\
This formula verifies requirements if \texttt{Inventory Check} and \texttt{Credit Check} both fail.

\(\mathit{CF_{2, 4} \triangleq
or ~\textbf{U}~ (icf ~\textbf{U}~ (sct ~\textbf{U}~ (cct ~\textbf{U}~ ((bi \vee sh) ~\textbf{U}~ (ar ~\textbf{U}~ \square tr))))) ~\wedge~ \square \neg rj}
\) \\
This formula verifies requirements if \texttt{Inventory Check} fails, but the order is successfully completed.

\(\mathit{CF_{2, 5} \triangleq
or ~\textbf{U}~ (ict ~\textbf{U}~ (ccf ~\textbf{U}~ (rj ~\textbf{U}~ \square tr))) ~\wedge~ \square \neg ar}
\) \\
This formula verifies requirements if \texttt{Credit Check} fails.

$\mathit{CF_2 \triangleq CF_{2, 1} \,\vee\, CF_{2, 2} \,\vee\, CF_{2, 3} \,\vee\, CF_{2, 4} \,\vee\, CF_{2, 5}}$ \\
This formula verifies requirements on Configuration 2,
and is satisfied by the paths (starting at state $S_1$) of the Kripke structures of the action traces of \texttt{Configuration2}
(by inspection of Figure \ref{fig:KSconfig2}).

The effect of the $\mathit{Reconfigure}$ operation on the traces of actions of workflows
executing in \texttt{Configuration1} that have performed $\mathit{OrderReceipt}$ is described using the Kripke structures
in Figures \ref{fig:KSconfig12a} and \ref{fig:KSconfig12b}.
The LTL formula for reconfiguration requirement R2
(any order being processed that was received before the start of the reconfiguration must satisfy all the requirements on Configuration 2 (if possible);
otherwise, all the requirements on Configuration 1 must be satisfied) is:

$\mathit{RF \triangleq CF_2 ~\vee~ \neg CF_2 \wedge CF_1}$ \\
This formula is satisfied by the paths (starting at state $S_1$) of the Kripke structures in Figures \ref{fig:KSconfig12a} and \ref{fig:KSconfig12b}
(by inspection of Figures \ref{fig:KSconfig12a} and \ref{fig:KSconfig12b}).

\newpage
\section{Verification of Requirement R2 using Consistent Histories of CPOGs}\label{sec:appendixcpogs}

\subsection{Theorem
\(\mathit{{\bf \forall H \!\in\! \mathbb{P} \mathcal{A}
~(OrderReceipt \!\in\! H \,\wedge\, r \!\notin\! H \,\wedge\, C(H, S_\text{safe}) \neq false ~\Longrightarrow~ C(H, c_2) \neq false)}}
\)}
\label{prf:cpog_reqtR2}

Sketch proof:
$H$ is restricted to be a consistent history of $c_1$ that contains $\mathsf{OrderReceipt}$ but does not contain $\mathsf{Reject}$ or $\mathsf{Confirmation}$
to show that $H$ is a consistent history of $c_2$.

$\mathit{C(H, S_\text{safe}) \neq false}$ ~~(by hypothesis)

$\mathit{\Longleftrightarrow C(H, r + [\text{\textlnot}\mathrm{r\_done}]c_1 + [\mathrm{r\_done}]c_2 + r \rightarrow (\mathsf{Reject} + \mathsf{Confirmation})) \neq false}$ \\
(by definition of $\mathit{S_\text{safe}}$ in (\ref{eq:reconfiguration_safe}) in Section \ref{cpog-dynamic})

$\mathit{\Longrightarrow C(H, c_1) \neq false ~\wedge~ \{\mathsf{Reject}, \mathsf{Confirmation}\} \cap H = \emptyset}$ \\
(because $r \!\notin\! H$ (by hypothesis) and $r$ must precede $\mathsf{Reject}$ and $\mathsf{Confirmation}$ (by definition of $\rightarrow$))

\begin{figure}
\begin{centering}
\includegraphics[width=1\textwidth]{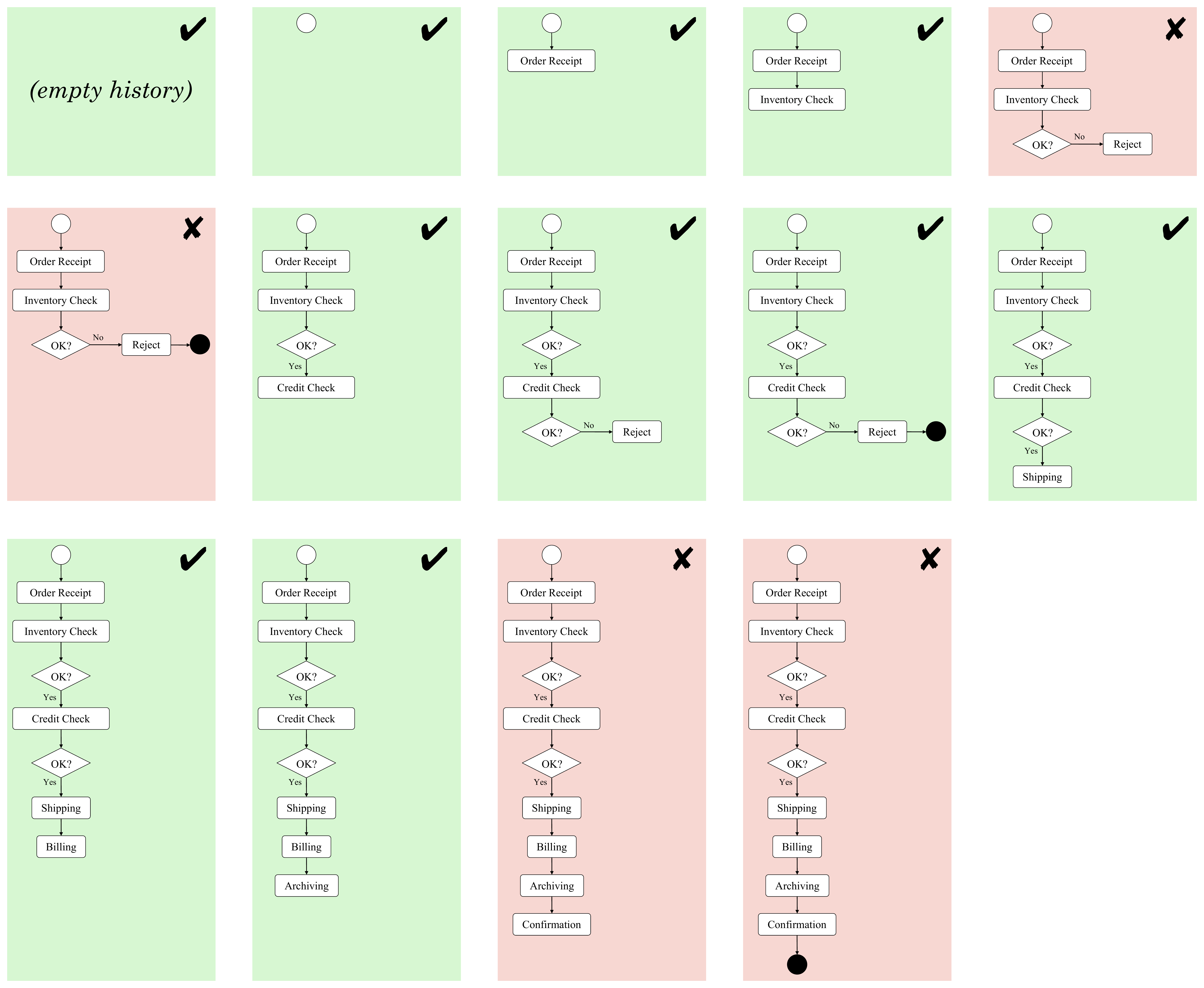}
\caption{All histories consistent with $c_1$, with histories that are also consistent with $c_2$ marked with \checkmark
\label{fig:cpog-reconfiguration-histories_1}}
\end{centering}
\end{figure}

\begin{figure}
\begin{centering}
\includegraphics[width=0.9\textwidth]{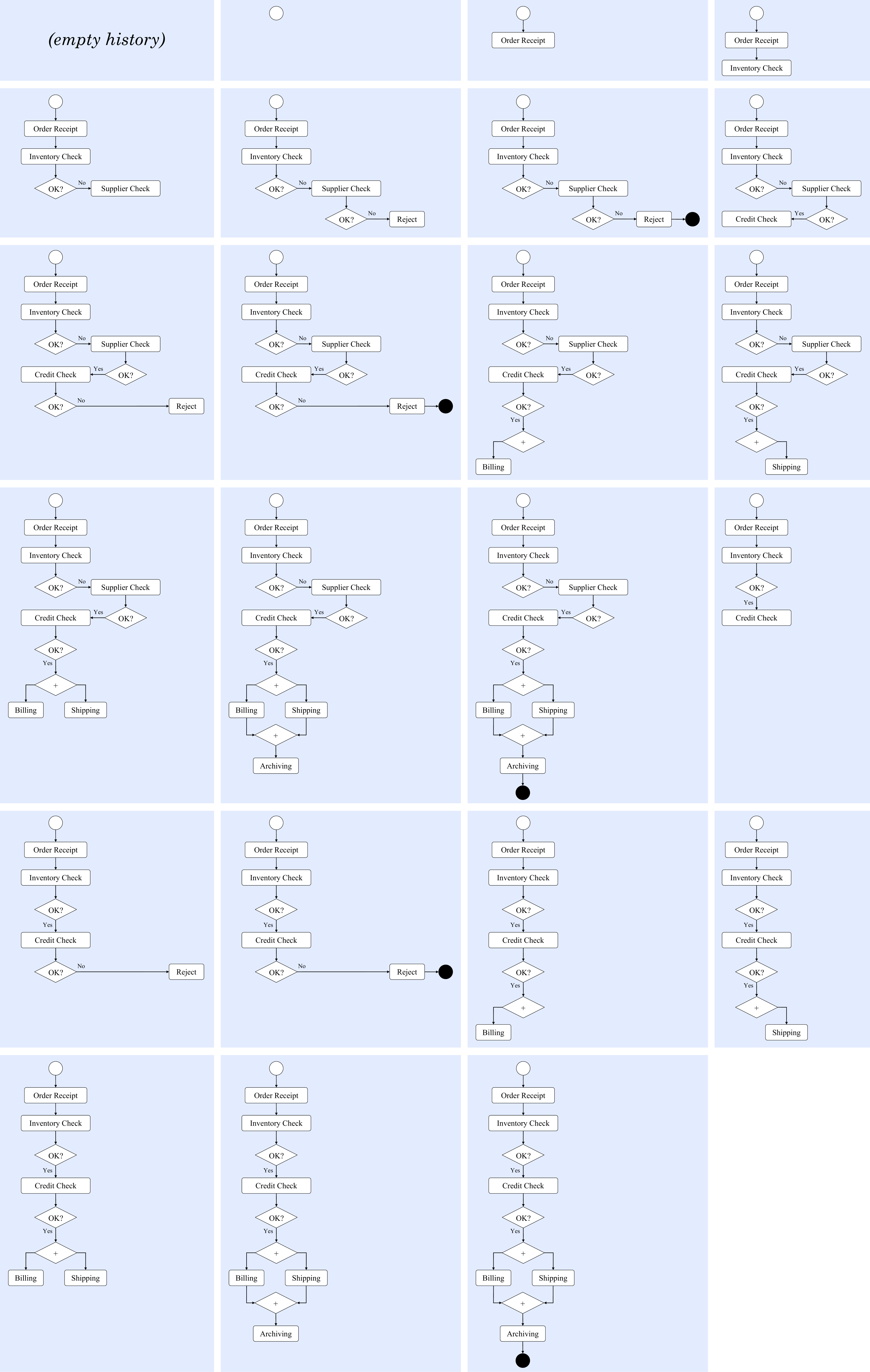}
\caption{All histories consistent with $c_2$
\label{fig:cpog-reconfiguration-histories_2}}
\end{centering}
\end{figure}

$\mathit{\Longrightarrow C(H, c_2) \neq false}$
~~(by inspection of Figures \ref{fig:cpog-reconfiguration-histories_1} and \ref{fig:cpog-reconfiguration-histories_2}, or by using a SAT solver).

%
%
%

\newpage
\section{Basic ${\bf CCS^{dp}}$}\label{sec:appendixbccsdp}

\subsection{Elaborations of
${\bf CONFIG^1 ~|~ \frac{CONFIG^2}{CONFIG^1}}$ and \\[1mm]
${\bf CONFIG^1 ~|~ \frac{ICH^2}{ICH^1} ~|~ \frac{CCH^2}{CCH^1} ~|~ \frac{SHIP^2}{SHIP^1} ~|~ \frac{BILL^2}{BILL^1} ~|~ \frac{ARC^2}{ARC^1} ~|~ \frac{0}{ARCH^1}}$}
\label{prf:reconfelabs}

$CONFIG^1 ~|~ \frac{CONFIG^2}{CONFIG^1}$ can be elaborated to a process that is weakly observationally bisimilar to $CONFIG^2$, as follows: \\[1mm]
$CONFIG^1 ~|~ \frac{CONFIG^2}{CONFIG^1} \stackrel{\tau}{\longrightarrow} 0 ~|~ CONFIG^2 \approx_o CONFIG^2$ \\
(by the Delet, Creat, and React rules, $\because CONFIG^1 \!\in\! \mathcal{P}^+ ~\wedge~ \sim_{of}$ is reflexive, and because $0$ is the identity process). \\[1mm]

$CONFIG^1 ~|~ \frac{ICH^2}{ICH^1} ~|~ \frac{CCH^2}{CCH^1} ~|~ \frac{SHIP^2}{SHIP^1} ~|~ \frac{BILL^2}{BILL^1} ~|~ \frac{ARC^2}{ARC^1} ~|~ \frac{0}{ARCH^1}$
can be elaborated to a process that is weakly observationally bisimilar to $CONFIG^2$, as follows: \\[1mm]
$CONFIG^1 ~|~ \frac{ICH^2}{ICH^1} ~|~ \frac{CCH^2}{CCH^1} ~|~ \frac{SHIP^2}{SHIP^1} ~|~ \frac{BILL^2}{BILL^1} ~|~ \frac{ARC^2}{ARC^1} ~|~ \frac{0}{ARCH^1}$ \\
\( \stackrel{\tau}{\longrightarrow}
      REC^1 ~|~ IC^1 ~|~ 0 ~|~ CC^1 ~|~ CCH^1 ~|~ SHIP^1 ~|~ BILL^1 ~|~ ARC^1 ~|~ ARCH^1 ~|~
      ICH^2 ~|~ \frac{CCH^2}{CCH^1} ~|~ \frac{SHIP^2}{SHIP^1} ~|~ \frac{BILL^2}{BILL^1} ~|~ \frac{ARC^2}{ARC^1} ~|~ \frac{0}{ARCH^1} \) \\
      (by definition of $CONFIG^1$, and by the Delet, Creat, L-Par, R-Par, and React rules, $\because ICH^1 \!\in\! \mathcal{P}^+ ~\wedge~ \sim_{of}$ is reflexive) \\
\( \stackrel{\tau}{\longrightarrow}
      REC^1 ~|~ IC^1 ~|~ 0 ~|~ CC^1 ~|~ 0 ~|~ SHIP^1 ~|~ BILL^1 ~|~ ARC^1 ~|~ ARCH^1 ~|~
      ICH^2 ~|~ CCH^2 ~|~ \frac{SHIP^2}{SHIP^1} ~|~ \frac{BILL^2}{BILL^1} ~|~ \frac{ARC^2}{ARC^1} ~|~ \frac{0}{ARCH^1} \) \\
      (by the Delet, Creat, L-Par, R-Par, and React rules, $\because CCH^1 \!\in\! \mathcal{P}^+ ~\wedge~ \sim_{of}$ is reflexive) \\
\( \stackrel{\tau}{\longrightarrow}
      REC^1 ~|~ IC^1 ~|~ 0 ~|~ CC^1 ~|~ 0 ~|~ 0 ~|~ BILL^1 ~|~ ARC^1 ~|~ ARCH^1 ~|~
      ICH^2 ~|~ CCH^2 ~|~ SHIP^2 ~|~ \frac{BILL^2}{BILL^1} ~|~ \frac{ARC^2}{ARC^1} ~|~ \frac{0}{ARCH^1} \) \\
      (by the Delet, Creat, L-Par, R-Par, and React rules, $\because SHIP^1 \!\in\! \mathcal{P}^+ ~\wedge~ \sim_{of}$ is reflexive) \\
\( \stackrel{\tau}{\longrightarrow}
      REC^1 ~|~ IC^1 ~|~ 0 ~|~ CC^1 ~|~ 0 ~|~ 0 ~|~ 0 ~|~ ARC^1 ~|~ ARCH^1 ~|~
      ICH^2 ~|~ CCH^2 ~|~ SHIP^2 ~|~ BILL^2 ~|~ \frac{ARC^2}{ARC^1} ~|~ \frac{0}{ARCH^1} \) \\
      (by the Delet, Creat, L-Par, R-Par, and React rules, $\because BILL^1 \!\in\! \mathcal{P}^+ ~\wedge~ \sim_{of}$ is reflexive) \\
\( \stackrel{\tau}{\longrightarrow}
      REC^1 ~|~ IC^1 ~|~ 0 ~|~ CC^1 ~|~ 0 ~|~ 0 ~|~ 0 ~|~ 0 ~|~ ARCH^1 ~|~
      ICH^2 ~|~ CCH^2 ~|~ SHIP^2 ~|~ BILL^2 ~|~ ARC^2 ~|~ \frac{0}{ARCH^1} \) \\
      (by the Delet, Creat, L-Par, R-Par, and React rules, $\because ARC^1 \!\in\! \mathcal{P}^+ ~\wedge~ \sim_{of}$ is reflexive) \\
\( \stackrel{\tau}{\longrightarrow}
      REC^1 ~|~ IC^1 ~|~ 0 ~|~ CC^1 ~|~ 0 ~|~ 0 ~|~ 0 ~|~ 0 ~|~ 0 ~|~
      ICH^2 ~|~ CCH^2 ~|~ SHIP^2 ~|~ BILL^2 ~|~ ARC^2 ~|~ 0 \) \\
      (by the Delet, Creat, R-Par, and React rules, $\because ARCH^1 \!\in\! \mathcal{P}^+ ~\wedge~ \sim_{of}$ is reflexive) \\
\( \sim_{of}
      REC^1 ~|~ IC^1 ~|~ ICH^2 ~|~ CC^1 ~|~ CCH^2 ~|~ SHIP^2 ~|~ BILL^2 ~|~ ARC^2 \) \\
      (by the commutativity and associativity of the parallel composition operator, and because $0$ is the identity process) \\
\( \approx_o CONFIG^2 \) ($\because ~\sim_{of}~ \subset ~\approx_o$, and by definition of $CONFIG^2$).

\subsection{Models of Designs of Configuration 1}\label{mdl:designs234}

In this section, we model Designs 2, 3, and 4 (described in Section \ref{sec:designs}) of Configuration 1;
Design 1 of the configuration is modelled in Section \ref{sec:bccsdpmodelling}.

Let $O$ be a \textbf{finite} set of possible order identifiers.

Configuration 1 of the workflow is denoted by the process $CONFIG^1$, where \\[1mm]
$CONFIG^1 \triangleq REC^1 ~|~ IC^1 ~|~ ICH^1 ~|~ CC^1 ~|~ CCH^1 ~|~ SHIP^1 ~|~ BILL^1 ~|~ ARC^1 ~|~ ARCH^1$
~This model applies to all four designs of Configuration 1.
The differences between the designs are represented by the differences in the definitions of each concurrent process of $CONFIG^1$,
particularly in the location of the recursively defined process.
In the model of Design 1, the recursively defined process is $CONFIG^1$,
which executes sequentially from $ICH^1$, $CCH^1$, and $ARCH^1$.
For Design 2, the recursively defined processes are the concurrent processes of $CONFIG^1$, each of which executes concurrently from its containing process.
Thus, the recursive definition of $CONFIG^1$ is distributed between its concurrent processes.
For Design 3, the recursively defined process is $CONFIG^1$, which executes concurrently from $REC^1$.
The model of Design 4 is a merge of the models of Designs 2 and 3 for each concurrent process of $CONFIG^1$.

\subsubsection{Model of Design 2}\label{mdl:design2}

$CONFIG^1 \triangleq REC^1 ~|~ IC^1 ~|~ ICH^1 ~|~ CC^1 ~|~ CCH^1 ~|~ SHIP^1 ~|~ BILL^1 ~|~ ARC^1 ~|~ ARCH^1$ ~where

$REC^1 \triangleq \sum_{o \in O} Receipt_o.(\overline{InventoryCheck}_o ~|~ REC^1)$
~and denotes the \texttt{Order Receipt} task.

$IC^1 \triangleq \sum_{o \in O} InventoryCheck_o.(\overline{InventoryCheckNotOK}_o + \overline{InventoryCheckOK}_o ~|~ IC^1)$
~and denotes the \texttt{Inventory Check} subtask of \texttt{Evaluation}.

$ICH^1 \triangleq \sum_{o \in O} InventoryCheckNotOK_o.(\overline{RejectIC}_o ~|~ ICH^1) + InventoryCheckOK_o.(\overline{CreditCheck}_o ~|~ ICH^1)$
~and denotes subtasks of \texttt{Evaluation} and \texttt{Rejection}.

$CC^1 \triangleq \sum_{o \in O} CreditCheck_o.(\overline{CreditCheckNotOK}_o + \overline{CreditCheckOK}_o ~|~ CC^1)$
~and denotes the \texttt{Credit Check} subtask of \texttt{Evaluation}.

$CCH^1 \triangleq \sum_{o \in O} CreditCheckNotOK_o.(\overline{RejectCC}_o ~|~ CCH^1) + CreditCheckOK_o.(\overline{Ship}_o ~|~ CCH^1)$
~and denotes subtasks of \texttt{Evaluation} and \texttt{Rejection}.

$SHIP^1 \triangleq \sum_{o \in O} Ship_o.(\overline{Bill}_o ~|~ SHIP^1)$
~and denotes the \texttt{Shipping} task,
~$BILL^1 \triangleq \sum_{o \in O} Bill_o.(\overline{Archive}_o ~|~ BILL^1)$
~and denotes the \texttt{Billing} task.

$ARC^1 \triangleq \sum_{o \in O} Archive_o.(\overline{ArchiveOK}_o ~|~ ARC^1)$
~and denotes the \texttt{Archiving} task.

$ARCH^1 \triangleq \sum_{o \in O} ArchiveOK_o.(\overline{Confirm}_o ~|~ ARCH^1)$
~and denotes the \texttt{Confirmation} task.

\subsubsection{Model of Design 3}\label{mdl:design3}

$CONFIG^1 \triangleq REC^1 ~|~ IC^1 ~|~ ICH^1 ~|~ CC^1 ~|~ CCH^1 ~|~ SHIP^1 ~|~ BILL^1 ~|~ ARC^1 ~|~ ARCH^1$ ~where

$REC^1 \triangleq \sum_{o \in O} Receipt_o.(\overline{InventoryCheck}_o ~|~ CONFIG^1)$

$IC^1 \triangleq \sum_{o \in O} InventoryCheck_o.(\overline{InventoryCheckNotOK}_o + \overline{InventoryCheckOK}_o)$

$ICH^1 \triangleq \sum_{o \in O} InventoryCheckNotOK_o.\overline{RejectIC}_o + InventoryCheckOK_o.\overline{CreditCheck}_o$

$CC^1 \triangleq \sum_{o \in O} CreditCheck_o.(\overline{CreditCheckNotOK}_o + \overline{CreditCheckOK}_o)$

$CCH^1 \triangleq \sum_{o \in O} CreditCheckNotOK_o.\overline{RejectCC}_o + CreditCheckOK_o.\overline{Ship}_o$

$SHIP^1 \triangleq \sum_{o \in O} Ship_o.\overline{Bill}_o$
\qquad\qquad\qquad ~$BILL^1 \triangleq \sum_{o \in O} Bill_o.\overline{Archive}_o$

$ARC^1 \triangleq \sum_{o \in O} Archive_o.\overline{ArchiveOK}_o$
\qquad $ARCH^1 \triangleq \sum_{o \in O} ArchiveOK_o.\overline{Confirm}_o$

\subsubsection{Model of Design 4}\label{mdl:design4}

$CONFIG^1 \triangleq REC^1 ~|~ IC^1 ~|~ ICH^1 ~|~ CC^1 ~|~ CCH^1 ~|~ SHIP^1 ~|~ BILL^1 ~|~ ARC^1 ~|~ ARCH^1$ ~where

$REC^1 \triangleq \sum_{o \in O} Receipt_o.(\overline{InventoryCheck}_o ~|~ REC^1 ~|~ CONFIG^1)$

$IC^1 \triangleq \sum_{o \in O} InventoryCheck_o.(\overline{InventoryCheckNotOK}_o + \overline{InventoryCheckOK}_o ~|~ IC^1)$

$ICH^1 \triangleq \sum_{o \in O} InventoryCheckNotOK_o.(\overline{RejectIC}_o ~|~ ICH^1) + InventoryCheckOK_o.(\overline{CreditCheck}_o ~|~ ICH^1)$

$CC^1 \triangleq \sum_{o \in O} CreditCheck_o.(\overline{CreditCheckNotOK}_o + \overline{CreditCheckOK}_o ~|~ CC^1)$

$CCH^1 \triangleq \sum_{o \in O} CreditCheckNotOK_o.(\overline{RejectCC}_o ~|~ CCH^1) + CreditCheckOK_o.(\overline{Ship}_o ~|~ CCH^1)$

$SHIP^1 \triangleq \sum_{o \in O} Ship_o.(\overline{Bill}_o ~|~ SHIP^1)$
\qquad\qquad\qquad $BILL^1 \triangleq \sum_{o \in O} Bill_o.(\overline{Archive}_o ~|~ BILL^1)$

$ARC^1 \triangleq \sum_{o \in O} Archive_o.(\overline{ArchiveOK}_o ~|~ ARC^1)$
\qquad $ARCH^1 \triangleq \sum_{o \in O} ArchiveOK_o.(\overline{Confirm}_o ~|~ ARCH^1)$

\subsection{Lemma ${\bf \forall P \!\in\! \mathcal{P} ~(CONFIG^1 ~|~ P \napprox_o CONFIG^2)}$}\label{prf:nonweakobsbisimgen}

Proof:
assumes the lemma is $false$ and constructs a sequence of transitions of $CONFIG^1 ~|~ P$ in the context $\overline{Receipt}_o \,|\, RejectIC_o$
that cannot be matched by $CONFIG^2$ in the same context to produce a contradiction.

The lemma is $true$ $~\vee~$ the lemma is $false$ (by definition of $\vee$, and because the logic used is 2-valued).

If the lemma is $false$ \\
then $\exists P \!\in\! \mathcal{P} ~(CONFIG^1 ~|~ P \approx_o CONFIG^2)$ \\
(by definition of $\neg$) \\
implies each transition sequence of $CONFIG^1 ~|~ P$ in any context
is matched by a corresponding transition sequence of $CONFIG^2$ in the same context up to $\tau$ transitions
such that the resulting process expressions are weakly observationally bisimilar in any context \\
(by definition of $\approx_o$) \\
implies each transition sequence of $CONFIG^1 ~|~ P$ in the context $\overline{Receipt}_o \,|\, RejectIC_o$
is matched by a corresponding transition sequence of $CONFIG^2$ in the same context up to $\tau$ transitions
such that the resulting process expressions are weakly observationally bisimilar in the residual context \\
(by specialisation) \\
implies the transition sequence \\[1mm]
\( CONFIG^1 ~|~ P \\
   \stackrel{Receipt_o}{\longrightarrow}
   \overline{InventoryCheck}_o ~|~ IC^1 ~|~ ICH^1 ~|~ CC^1 ~|~ CCH^1 ~|~ SHIP^1 ~|~ BILL^1 ~|~ ARC^1 ~|~ ARCH^1 ~|~ P \) \\
(by the Sum and L-Par rules) \\
\( \stackrel{\tau}{\longrightarrow}
   \overline{InventoryCheckNotOK}_o + \overline{InventoryCheckOK}_o ~|~
   ICH^1 ~|~ CC^1 ~|~ CCH^1 ~|~ SHIP^1 ~|~ BILL^1 ~|~ ARC^1 ~|~ ARCH^1 ~|~ P \) \\
(by the Sum, React, and L-Par rules) \\
\( \stackrel{\tau}{\longrightarrow}
   \overline{RejectIC}_o.CONFIG^1 ~|~ CC^1 ~|~ CCH^1 ~|~ SHIP^1 ~|~ BILL^1 ~|~ ARC^1 ~|~ ARCH^1 ~|~ P \) \\
(by the Sum, React, and L-Par rules)

is matched by one or more of the following transition sequences (by specialisation) \\[1mm]
\( CONFIG^2 \\
   \stackrel{Receipt_o}{\longrightarrow}
   \overline{InventoryCheck}_o ~|~ IC^1 ~|~ ICH^2 ~|~ CC^1 ~|~ CCH^2 ~|~ SHIP^2 ~|~ BILL^2 ~|~ ARC^2 \) \\
(by the Sum and L-Par rules) or


\( CONFIG^2 \\
   \stackrel{Receipt_o}{\longrightarrow}
   \overline{InventoryCheck}_o ~|~ IC^1 ~|~ ICH^2 ~|~ CC^1 ~|~ CCH^2 ~|~ SHIP^2 ~|~ BILL^2 ~|~ ARC^2 \\
   \stackrel{\tau}{\longrightarrow}
   \overline{InventoryCheckNotOK}_o + \overline{InventoryCheckOK}_o ~|~
   ICH^2 ~|~ CC^1 ~|~ CCH^2 ~|~ SHIP^2 ~|~ BILL^2 ~|~ ARC^2 \) \\
(by the Sum, React, and L-Par rules) or


\( CONFIG^2 \\
   \stackrel{Receipt_o}{\longrightarrow}
   \overline{InventoryCheck}_o ~|~ IC^1 ~|~ ICH^2 ~|~ CC^1 ~|~ CCH^2 ~|~ SHIP^2 ~|~ BILL^2 ~|~ ARC^2 \\
   \stackrel{\tau}{\longrightarrow}
   \overline{InventoryCheckNotOK}_o + \overline{InventoryCheckOK}_o ~|~
   ICH^2 ~|~ CC^1 ~|~ CCH^2 ~|~ SHIP^2 ~|~ BILL^2 ~|~ ARC^2 \\
   \stackrel{\tau}{\longrightarrow}
   \overline{SupplierCheck}_o ~|~ SC ~|~ CC^1 ~|~ CCH^2 ~|~ SHIP^2 ~|~ BILL^2 ~|~ ARC^2 \) \\
(by the Sum, React, and L-Par rules) or

\( CONFIG^2 \\
   \stackrel{Receipt_o}{\longrightarrow}
   \overline{InventoryCheck}_o ~|~ IC^1 ~|~ ICH^2 ~|~ CC^1 ~|~ CCH^2 ~|~ SHIP^2 ~|~ BILL^2 ~|~ ARC^2 \\
   \stackrel{\tau}{\longrightarrow}
   \overline{InventoryCheckNotOK}_o + \overline{InventoryCheckOK}_o ~|~
   ICH^2 ~|~ CC^1 ~|~ CCH^2 ~|~ SHIP^2 ~|~ BILL^2 ~|~ ARC^2 \\
   \stackrel{\tau}{\longrightarrow}
   \overline{CreditCheck}_o ~|~ CC^1 ~|~ CCH^2 ~|~ SHIP^2 ~|~ BILL^2 ~|~ ARC^2 \) \\
(by the Sum, React, and L-Par rules) or


\( CONFIG^2 \\
   \stackrel{Receipt_o}{\longrightarrow}
   \overline{InventoryCheck}_o ~|~ IC^1 ~|~ ICH^2 ~|~ CC^1 ~|~ CCH^2 ~|~ SHIP^2 ~|~ BILL^2 ~|~ ARC^2 \\
   \stackrel{\tau}{\longrightarrow}
   \overline{InventoryCheckNotOK}_o + \overline{InventoryCheckOK}_o ~|~
   ICH^2 ~|~ CC^1 ~|~ CCH^2 ~|~ SHIP^2 ~|~ BILL^2 ~|~ ARC^2 \\
   \stackrel{\tau}{\longrightarrow}
   \overline{CreditCheck}_o ~|~ CC^1 ~|~ CCH^2 ~|~ SHIP^2 ~|~ BILL^2 ~|~ ARC^2 \\
   \stackrel{\tau}{\longrightarrow}
   \overline{CreditCheckNotOK}_o + \overline{CreditCheckOK}_o ~|~ CCH^2 ~|~ SHIP^2 ~|~ BILL^2 ~|~ ARC^2 \) \\
(by the Sum, React, and L-Par rules) or


\( CONFIG^2 \\
   \stackrel{Receipt_o}{\longrightarrow}
   \overline{InventoryCheck}_o ~|~ IC^1 ~|~ ICH^2 ~|~ CC^1 ~|~ CCH^2 ~|~ SHIP^2 ~|~ BILL^2 ~|~ ARC^2 \\
   \stackrel{\tau}{\longrightarrow}
   \overline{InventoryCheckNotOK}_o + \overline{InventoryCheckOK}_o ~|~
   ICH^2 ~|~ CC^1 ~|~ CCH^2 ~|~ SHIP^2 ~|~ BILL^2 ~|~ ARC^2 \\
   \stackrel{\tau}{\longrightarrow}
   \overline{CreditCheck}_o ~|~ CC^1 ~|~ CCH^2 ~|~ SHIP^2 ~|~ BILL^2 ~|~ ARC^2 \\
   \stackrel{\tau}{\longrightarrow}
   \overline{CreditCheckNotOK}_o + \overline{CreditCheckOK}_o ~|~ CCH^2 ~|~ SHIP^2 ~|~ BILL^2 ~|~ ARC^2 \\
   \stackrel{\tau}{\longrightarrow}
   \overline{RejectCC}_o.CONFIG^2 ~|~ SHIP^2 ~|~ BILL^2 ~|~ ARC^2 \) \\
(by the Sum, React, and L-Par rules) or

\( CONFIG^2 \\
   \stackrel{Receipt_o}{\longrightarrow}
   \overline{InventoryCheck}_o ~|~ IC^1 ~|~ ICH^2 ~|~ CC^1 ~|~ CCH^2 ~|~ SHIP^2 ~|~ BILL^2 ~|~ ARC^2 \\
   \stackrel{\tau}{\longrightarrow}
   \overline{InventoryCheckNotOK}_o + \overline{InventoryCheckOK}_o ~|~
   ICH^2 ~|~ CC^1 ~|~ CCH^2 ~|~ SHIP^2 ~|~ BILL^2 ~|~ ARC^2 \\
   \stackrel{\tau}{\longrightarrow}
   \overline{CreditCheck}_o ~|~ CC^1 ~|~ CCH^2 ~|~ SHIP^2 ~|~ BILL^2 ~|~ ARC^2 \\
   \stackrel{\tau}{\longrightarrow}
   \overline{CreditCheckNotOK}_o + \overline{CreditCheckOK}_o ~|~ CCH^2 ~|~ SHIP^2 ~|~ BILL^2 ~|~ ARC^2 \\
   \stackrel{\tau}{\longrightarrow}
   \overline{Ship}_o ~|~ \overline{Bill}_o ~|~ SHIP^2 ~|~ BILL^2 ~|~ ARC^2 \) \\
(by the Sum, React, and L-Par rules) or


\( CONFIG^2 \\
   \stackrel{Receipt_o}{\longrightarrow}
   \overline{InventoryCheck}_o ~|~ IC^1 ~|~ ICH^2 ~|~ CC^1 ~|~ CCH^2 ~|~ SHIP^2 ~|~ BILL^2 ~|~ ARC^2 \\
   \stackrel{\tau}{\longrightarrow}
   \overline{InventoryCheckNotOK}_o + \overline{InventoryCheckOK}_o ~|~
   ICH^2 ~|~ CC^1 ~|~ CCH^2 ~|~ SHIP^2 ~|~ BILL^2 ~|~ ARC^2 \\
   \stackrel{\tau}{\longrightarrow}
   \overline{CreditCheck}_o ~|~ CC^1 ~|~ CCH^2 ~|~ SHIP^2 ~|~ BILL^2 ~|~ ARC^2 \\
   \stackrel{\tau}{\longrightarrow}
   \overline{CreditCheckNotOK}_o + \overline{CreditCheckOK}_o ~|~ CCH^2 ~|~ SHIP^2 ~|~ BILL^2 ~|~ ARC^2 \\
   \stackrel{\tau}{\longrightarrow}
   \overline{Ship}_o ~|~ \overline{Bill}_o ~|~ SHIP^2 ~|~ BILL^2 ~|~ ARC^2 \\
   \stackrel{\tau}{\longrightarrow}
   \overline{Bill}_o ~|~ \overline{ShipOK}_o ~|~ BILL^2 ~|~ ARC^2 \) \\
(by the Sum, L-Par, and React rules) or

\( CONFIG^2 \\
   \stackrel{Receipt_o}{\longrightarrow}
   \overline{InventoryCheck}_o ~|~ IC^1 ~|~ ICH^2 ~|~ CC^1 ~|~ CCH^2 ~|~ SHIP^2 ~|~ BILL^2 ~|~ ARC^2 \\
   \stackrel{\tau}{\longrightarrow}
   \overline{InventoryCheckNotOK}_o + \overline{InventoryCheckOK}_o ~|~
   ICH^2 ~|~ CC^1 ~|~ CCH^2 ~|~ SHIP^2 ~|~ BILL^2 ~|~ ARC^2 \\
   \stackrel{\tau}{\longrightarrow}
   \overline{CreditCheck}_o ~|~ CC^1 ~|~ CCH^2 ~|~ SHIP^2 ~|~ BILL^2 ~|~ ARC^2 \\
   \stackrel{\tau}{\longrightarrow}
   \overline{CreditCheckNotOK}_o + \overline{CreditCheckOK}_o ~|~ CCH^2 ~|~ SHIP^2 ~|~ BILL^2 ~|~ ARC^2 \\
   \stackrel{\tau}{\longrightarrow}
   \overline{Ship}_o ~|~ \overline{Bill}_o ~|~ SHIP^2 ~|~ BILL^2 ~|~ ARC^2 \\
   \stackrel{\tau}{\longrightarrow}
   \overline{Ship}_o ~|~ SHIP^2 ~|~ \overline{BillOK}_o ~|~ ARC^2 \) \\
(by the Sum, L-Par, R-Par, and React rules) or


\( CONFIG^2 \\
   \stackrel{Receipt_o}{\longrightarrow}
   \overline{InventoryCheck}_o ~|~ IC^1 ~|~ ICH^2 ~|~ CC^1 ~|~ CCH^2 ~|~ SHIP^2 ~|~ BILL^2 ~|~ ARC^2 \\
   \stackrel{\tau}{\longrightarrow}
   \overline{InventoryCheckNotOK}_o + \overline{InventoryCheckOK}_o ~|~
   ICH^2 ~|~ CC^1 ~|~ CCH^2 ~|~ SHIP^2 ~|~ BILL^2 ~|~ ARC^2 \\
   \stackrel{\tau}{\longrightarrow}
   \overline{CreditCheck}_o ~|~ CC^1 ~|~ CCH^2 ~|~ SHIP^2 ~|~ BILL^2 ~|~ ARC^2 \\
   \stackrel{\tau}{\longrightarrow}
   \overline{CreditCheckNotOK}_o + \overline{CreditCheckOK}_o ~|~ CCH^2 ~|~ SHIP^2 ~|~ BILL^2 ~|~ ARC^2 \\
   \stackrel{\tau}{\longrightarrow}
   \overline{Ship}_o ~|~ \overline{Bill}_o ~|~ SHIP^2 ~|~ BILL^2 ~|~ ARC^2 \\
   \stackrel{\tau}{\longrightarrow}
   \overline{Bill}_o ~|~ \overline{ShipOK}_o ~|~ BILL^2 ~|~ ARC^2 \\
   \stackrel{\tau}{\longrightarrow}
   \overline{ShipOK}_o ~|~ \overline{BillOK}_o ~|~ ARC^2 \) \\
(by the Sum, L-Par, and React rules) or

\( CONFIG^2 \\
   \stackrel{Receipt_o}{\longrightarrow}
   \overline{InventoryCheck}_o ~|~ IC^1 ~|~ ICH^2 ~|~ CC^1 ~|~ CCH^2 ~|~ SHIP^2 ~|~ BILL^2 ~|~ ARC^2 \\
   \stackrel{\tau}{\longrightarrow}
   \overline{InventoryCheckNotOK}_o + \overline{InventoryCheckOK}_o ~|~
   ICH^2 ~|~ CC^1 ~|~ CCH^2 ~|~ SHIP^2 ~|~ BILL^2 ~|~ ARC^2 \\
   \stackrel{\tau}{\longrightarrow}
   \overline{CreditCheck}_o ~|~ CC^1 ~|~ CCH^2 ~|~ SHIP^2 ~|~ BILL^2 ~|~ ARC^2 \\
   \stackrel{\tau}{\longrightarrow}
   \overline{CreditCheckNotOK}_o + \overline{CreditCheckOK}_o ~|~ CCH^2 ~|~ SHIP^2 ~|~ BILL^2 ~|~ ARC^2 \\
   \stackrel{\tau}{\longrightarrow}
   \overline{Ship}_o ~|~ \overline{Bill}_o ~|~ SHIP^2 ~|~ BILL^2 ~|~ ARC^2 \\
   \stackrel{\tau}{\longrightarrow}
   \overline{Bill}_o ~|~ \overline{ShipOK}_o ~|~ BILL^2 ~|~ ARC^2 \\
   \stackrel{\tau}{\longrightarrow}
   \overline{Bill}_o ~|~ BILL^2 ~|~ BillOK_o.\overline{Archive}_o.CONFIG^2 \) \\
(by the Sum, R-Par, and React rules) or

\( CONFIG^2 \\
   \stackrel{Receipt_o}{\longrightarrow}
   \overline{InventoryCheck}_o ~|~ IC^1 ~|~ ICH^2 ~|~ CC^1 ~|~ CCH^2 ~|~ SHIP^2 ~|~ BILL^2 ~|~ ARC^2 \\
   \stackrel{\tau}{\longrightarrow}
   \overline{InventoryCheckNotOK}_o + \overline{InventoryCheckOK}_o ~|~
   ICH^2 ~|~ CC^1 ~|~ CCH^2 ~|~ SHIP^2 ~|~ BILL^2 ~|~ ARC^2 \\
   \stackrel{\tau}{\longrightarrow}
   \overline{CreditCheck}_o ~|~ CC^1 ~|~ CCH^2 ~|~ SHIP^2 ~|~ BILL^2 ~|~ ARC^2 \\
   \stackrel{\tau}{\longrightarrow}
   \overline{CreditCheckNotOK}_o + \overline{CreditCheckOK}_o ~|~ CCH^2 ~|~ SHIP^2 ~|~ BILL^2 ~|~ ARC^2 \\
   \stackrel{\tau}{\longrightarrow}
   \overline{Ship}_o ~|~ \overline{Bill}_o ~|~ SHIP^2 ~|~ BILL^2 ~|~ ARC^2 \\
   \stackrel{\tau}{\longrightarrow}
   \overline{Ship}_o ~|~ SHIP^2 ~|~ \overline{BillOK}_o ~|~ ARC^2 \\
   \stackrel{\tau}{\longrightarrow}
   \overline{ShipOK}_o ~|~ \overline{BillOK}_o ~|~ ARC^2 \) \\
(by the Sum, React, and L-Par rules) or

\( CONFIG^2 \\
   \stackrel{Receipt_o}{\longrightarrow}
   \overline{InventoryCheck}_o ~|~ IC^1 ~|~ ICH^2 ~|~ CC^1 ~|~ CCH^2 ~|~ SHIP^2 ~|~ BILL^2 ~|~ ARC^2 \\
   \stackrel{\tau}{\longrightarrow}
   \overline{InventoryCheckNotOK}_o + \overline{InventoryCheckOK}_o ~|~
   ICH^2 ~|~ CC^1 ~|~ CCH^2 ~|~ SHIP^2 ~|~ BILL^2 ~|~ ARC^2 \\
   \stackrel{\tau}{\longrightarrow}
   \overline{CreditCheck}_o ~|~ CC^1 ~|~ CCH^2 ~|~ SHIP^2 ~|~ BILL^2 ~|~ ARC^2 \\
   \stackrel{\tau}{\longrightarrow}
   \overline{CreditCheckNotOK}_o + \overline{CreditCheckOK}_o ~|~ CCH^2 ~|~ SHIP^2 ~|~ BILL^2 ~|~ ARC^2 \\
   \stackrel{\tau}{\longrightarrow}
   \overline{Ship}_o ~|~ \overline{Bill}_o ~|~ SHIP^2 ~|~ BILL^2 ~|~ ARC^2 \\
   \stackrel{\tau}{\longrightarrow}
   \overline{Ship}_o ~|~ SHIP^2 ~|~ \overline{BillOK}_o ~|~ ARC^2 \\
   \stackrel{\tau}{\longrightarrow}
   \overline{Ship}_o ~|~ SHIP^2 ~|~ ShipOK_o.\overline{Archive}_o.CONFIG^2 \) \\
(by the Sum, React, and R-Par rules) or


\( CONFIG^2 \\
   \stackrel{Receipt_o}{\longrightarrow}
   \overline{InventoryCheck}_o ~|~ IC^1 ~|~ ICH^2 ~|~ CC^1 ~|~ CCH^2 ~|~ SHIP^2 ~|~ BILL^2 ~|~ ARC^2 \\
   \stackrel{\tau}{\longrightarrow}
   \overline{InventoryCheckNotOK}_o + \overline{InventoryCheckOK}_o ~|~
   ICH^2 ~|~ CC^1 ~|~ CCH^2 ~|~ SHIP^2 ~|~ BILL^2 ~|~ ARC^2 \\
   \stackrel{\tau}{\longrightarrow}
   \overline{CreditCheck}_o ~|~ CC^1 ~|~ CCH^2 ~|~ SHIP^2 ~|~ BILL^2 ~|~ ARC^2 \\
   \stackrel{\tau}{\longrightarrow}
   \overline{CreditCheckNotOK}_o + \overline{CreditCheckOK}_o ~|~ CCH^2 ~|~ SHIP^2 ~|~ BILL^2 ~|~ ARC^2 \\
   \stackrel{\tau}{\longrightarrow}
   \overline{Ship}_o ~|~ \overline{Bill}_o ~|~ SHIP^2 ~|~ BILL^2 ~|~ ARC^2 \\
   \stackrel{\tau}{\longrightarrow}
   \overline{Bill}_o ~|~ \overline{ShipOK}_o ~|~ BILL^2 ~|~ ARC^2 \\
   \stackrel{\tau}{\longrightarrow}
   \overline{ShipOK}_o ~|~ \overline{BillOK}_o ~|~ ARC^2 \\
   \stackrel{\tau}{\longrightarrow}
   \overline{BillOK}_o ~|~ BillOK_o.\overline{Archive}_o.CONFIG^2 \) \\
(by the Sum, L-Par, and React rules) or

\( CONFIG^2 \\
   \stackrel{Receipt_o}{\longrightarrow}
   \overline{InventoryCheck}_o ~|~ IC^1 ~|~ ICH^2 ~|~ CC^1 ~|~ CCH^2 ~|~ SHIP^2 ~|~ BILL^2 ~|~ ARC^2 \\
   \stackrel{\tau}{\longrightarrow}
   \overline{InventoryCheckNotOK}_o + \overline{InventoryCheckOK}_o ~|~
   ICH^2 ~|~ CC^1 ~|~ CCH^2 ~|~ SHIP^2 ~|~ BILL^2 ~|~ ARC^2 \\
   \stackrel{\tau}{\longrightarrow}
   \overline{CreditCheck}_o ~|~ CC^1 ~|~ CCH^2 ~|~ SHIP^2 ~|~ BILL^2 ~|~ ARC^2 \\
   \stackrel{\tau}{\longrightarrow}
   \overline{CreditCheckNotOK}_o + \overline{CreditCheckOK}_o ~|~ CCH^2 ~|~ SHIP^2 ~|~ BILL^2 ~|~ ARC^2 \\
   \stackrel{\tau}{\longrightarrow}
   \overline{Ship}_o ~|~ \overline{Bill}_o ~|~ SHIP^2 ~|~ BILL^2 ~|~ ARC^2 \\
   \stackrel{\tau}{\longrightarrow}
   \overline{Bill}_o ~|~ \overline{ShipOK}_o ~|~ BILL^2 ~|~ ARC^2 \\
   \stackrel{\tau}{\longrightarrow}
   \overline{ShipOK}_o ~|~ \overline{BillOK}_o ~|~ ARC^2 \\
   \stackrel{\tau}{\longrightarrow}
   \overline{ShipOK}_o ~|~ ShipOK_o.\overline{Archive}_o.CONFIG^2 \) \\
(by the Sum, R-Par, and React rules) or

\( CONFIG^2 \\
   \stackrel{Receipt_o}{\longrightarrow}
   \overline{InventoryCheck}_o ~|~ IC^1 ~|~ ICH^2 ~|~ CC^1 ~|~ CCH^2 ~|~ SHIP^2 ~|~ BILL^2 ~|~ ARC^2 \\
   \stackrel{\tau}{\longrightarrow}
   \overline{InventoryCheckNotOK}_o + \overline{InventoryCheckOK}_o ~|~
   ICH^2 ~|~ CC^1 ~|~ CCH^2 ~|~ SHIP^2 ~|~ BILL^2 ~|~ ARC^2 \\
   \stackrel{\tau}{\longrightarrow}
   \overline{CreditCheck}_o ~|~ CC^1 ~|~ CCH^2 ~|~ SHIP^2 ~|~ BILL^2 ~|~ ARC^2 \\
   \stackrel{\tau}{\longrightarrow}
   \overline{CreditCheckNotOK}_o + \overline{CreditCheckOK}_o ~|~ CCH^2 ~|~ SHIP^2 ~|~ BILL^2 ~|~ ARC^2 \\
   \stackrel{\tau}{\longrightarrow}
   \overline{Ship}_o ~|~ \overline{Bill}_o ~|~ SHIP^2 ~|~ BILL^2 ~|~ ARC^2 \\
   \stackrel{\tau}{\longrightarrow}
   \overline{Bill}_o ~|~ \overline{ShipOK}_o ~|~ BILL^2 ~|~ ARC^2 \\
   \stackrel{\tau}{\longrightarrow}
   \overline{Bill}_o ~|~ BILL^2 ~|~ BillOK_o.\overline{Archive}_o.CONFIG^2 \\
   \stackrel{\tau}{\longrightarrow}
   \overline{BillOK}_o ~|~ BillOK_o.\overline{Archive}_o.CONFIG^2 \) \\
(by the Sum, L-Par, and React rules) or

\( CONFIG^2 \\
   \stackrel{Receipt_o}{\longrightarrow}
   \overline{InventoryCheck}_o ~|~ IC^1 ~|~ ICH^2 ~|~ CC^1 ~|~ CCH^2 ~|~ SHIP^2 ~|~ BILL^2 ~|~ ARC^2 \\
   \stackrel{\tau}{\longrightarrow}
   \overline{InventoryCheckNotOK}_o + \overline{InventoryCheckOK}_o ~|~
   ICH^2 ~|~ CC^1 ~|~ CCH^2 ~|~ SHIP^2 ~|~ BILL^2 ~|~ ARC^2 \\
   \stackrel{\tau}{\longrightarrow}
   \overline{CreditCheck}_o ~|~ CC^1 ~|~ CCH^2 ~|~ SHIP^2 ~|~ BILL^2 ~|~ ARC^2 \\
   \stackrel{\tau}{\longrightarrow}
   \overline{CreditCheckNotOK}_o + \overline{CreditCheckOK}_o ~|~ CCH^2 ~|~ SHIP^2 ~|~ BILL^2 ~|~ ARC^2 \\
   \stackrel{\tau}{\longrightarrow}
   \overline{Ship}_o ~|~ \overline{Bill}_o ~|~ SHIP^2 ~|~ BILL^2 ~|~ ARC^2 \\
   \stackrel{\tau}{\longrightarrow}
   \overline{Ship}_o ~|~ SHIP^2 ~|~ \overline{BillOK}_o ~|~ ARC^2 \\
   \stackrel{\tau}{\longrightarrow}
   \overline{ShipOK}_o ~|~ \overline{BillOK}_o ~|~ ARC^2 \\
   \stackrel{\tau}{\longrightarrow}
   \overline{BillOK}_o ~|~ BillOK_o.\overline{Archive}_o.CONFIG^2 \) \\
(by the Sum, L-Par, and React rules) or

\( CONFIG^2 \\
   \stackrel{Receipt_o}{\longrightarrow}
   \overline{InventoryCheck}_o ~|~ IC^1 ~|~ ICH^2 ~|~ CC^1 ~|~ CCH^2 ~|~ SHIP^2 ~|~ BILL^2 ~|~ ARC^2 \\
   \stackrel{\tau}{\longrightarrow}
   \overline{InventoryCheckNotOK}_o + \overline{InventoryCheckOK}_o ~|~
   ICH^2 ~|~ CC^1 ~|~ CCH^2 ~|~ SHIP^2 ~|~ BILL^2 ~|~ ARC^2 \\
   \stackrel{\tau}{\longrightarrow}
   \overline{CreditCheck}_o ~|~ CC^1 ~|~ CCH^2 ~|~ SHIP^2 ~|~ BILL^2 ~|~ ARC^2 \\
   \stackrel{\tau}{\longrightarrow}
   \overline{CreditCheckNotOK}_o + \overline{CreditCheckOK}_o ~|~ CCH^2 ~|~ SHIP^2 ~|~ BILL^2 ~|~ ARC^2 \\
   \stackrel{\tau}{\longrightarrow}
   \overline{Ship}_o ~|~ \overline{Bill}_o ~|~ SHIP^2 ~|~ BILL^2 ~|~ ARC^2 \\
   \stackrel{\tau}{\longrightarrow}
   \overline{Ship}_o ~|~ SHIP^2 ~|~ \overline{BillOK}_o ~|~ ARC^2 \\
   \stackrel{\tau}{\longrightarrow}
   \overline{ShipOK}_o ~|~ \overline{BillOK}_o ~|~ ARC^2 \\
   \stackrel{\tau}{\longrightarrow}
   \overline{ShipOK}_o ~|~ ShipOK_o.\overline{Archive}_o.CONFIG^2 \) \\
(by the Sum, R-Par, and React rules) or

\( CONFIG^2 \\
   \stackrel{Receipt_o}{\longrightarrow}
   \overline{InventoryCheck}_o ~|~ IC^1 ~|~ ICH^2 ~|~ CC^1 ~|~ CCH^2 ~|~ SHIP^2 ~|~ BILL^2 ~|~ ARC^2 \\
   \stackrel{\tau}{\longrightarrow}
   \overline{InventoryCheckNotOK}_o + \overline{InventoryCheckOK}_o ~|~
   ICH^2 ~|~ CC^1 ~|~ CCH^2 ~|~ SHIP^2 ~|~ BILL^2 ~|~ ARC^2 \\
   \stackrel{\tau}{\longrightarrow}
   \overline{CreditCheck}_o ~|~ CC^1 ~|~ CCH^2 ~|~ SHIP^2 ~|~ BILL^2 ~|~ ARC^2 \\
   \stackrel{\tau}{\longrightarrow}
   \overline{CreditCheckNotOK}_o + \overline{CreditCheckOK}_o ~|~ CCH^2 ~|~ SHIP^2 ~|~ BILL^2 ~|~ ARC^2 \\
   \stackrel{\tau}{\longrightarrow}
   \overline{Ship}_o ~|~ \overline{Bill}_o ~|~ SHIP^2 ~|~ BILL^2 ~|~ ARC^2 \\
   \stackrel{\tau}{\longrightarrow}
   \overline{Ship}_o ~|~ SHIP^2 ~|~ \overline{BillOK}_o ~|~ ARC^2 \\
   \stackrel{\tau}{\longrightarrow}
   \overline{Ship}_o ~|~ SHIP^2 ~|~ ShipOK_o.\overline{Archive}_o.CONFIG^2 \\
   \stackrel{\tau}{\longrightarrow}
   \overline{ShipOK}_o ~|~ ShipOK_o.\overline{Archive}_o.CONFIG^2 \) \\
(by the Sum, React, and L-Par rules) or


\( CONFIG^2 \\
   \stackrel{Receipt_o}{\longrightarrow}
   \overline{InventoryCheck}_o ~|~ IC^1 ~|~ ICH^2 ~|~ CC^1 ~|~ CCH^2 ~|~ SHIP^2 ~|~ BILL^2 ~|~ ARC^2 \\
   \stackrel{\tau}{\longrightarrow}
   \overline{InventoryCheckNotOK}_o + \overline{InventoryCheckOK}_o ~|~
   ICH^2 ~|~ CC^1 ~|~ CCH^2 ~|~ SHIP^2 ~|~ BILL^2 ~|~ ARC^2 \\
   \stackrel{\tau}{\longrightarrow}
   \overline{CreditCheck}_o ~|~ CC^1 ~|~ CCH^2 ~|~ SHIP^2 ~|~ BILL^2 ~|~ ARC^2 \\
   \stackrel{\tau}{\longrightarrow}
   \overline{CreditCheckNotOK}_o + \overline{CreditCheckOK}_o ~|~ CCH^2 ~|~ SHIP^2 ~|~ BILL^2 ~|~ ARC^2 \\
   \stackrel{\tau}{\longrightarrow}
   \overline{Ship}_o ~|~ \overline{Bill}_o ~|~ SHIP^2 ~|~ BILL^2 ~|~ ARC^2 \\
   \stackrel{\tau}{\longrightarrow}
   \overline{Bill}_o ~|~ \overline{ShipOK}_o ~|~ BILL^2 ~|~ ARC^2 \\
   \stackrel{\tau}{\longrightarrow}
   \overline{ShipOK}_o ~|~ \overline{BillOK}_o ~|~ ARC^2 \\
   \stackrel{\tau}{\longrightarrow}
   \overline{BillOK}_o ~|~ BillOK_o.\overline{Archive}_o.CONFIG^2 \\
   \stackrel{\tau}{\longrightarrow}
   \overline{Archive}_o.CONFIG^2 \) \\
(by the Sum and React rules) or

\( CONFIG^2 \\
   \stackrel{Receipt_o}{\longrightarrow}
   \overline{InventoryCheck}_o ~|~ IC^1 ~|~ ICH^2 ~|~ CC^1 ~|~ CCH^2 ~|~ SHIP^2 ~|~ BILL^2 ~|~ ARC^2 \\
   \stackrel{\tau}{\longrightarrow}
   \overline{InventoryCheckNotOK}_o + \overline{InventoryCheckOK}_o ~|~
   ICH^2 ~|~ CC^1 ~|~ CCH^2 ~|~ SHIP^2 ~|~ BILL^2 ~|~ ARC^2 \\
   \stackrel{\tau}{\longrightarrow}
   \overline{CreditCheck}_o ~|~ CC^1 ~|~ CCH^2 ~|~ SHIP^2 ~|~ BILL^2 ~|~ ARC^2 \\
   \stackrel{\tau}{\longrightarrow}
   \overline{CreditCheckNotOK}_o + \overline{CreditCheckOK}_o ~|~ CCH^2 ~|~ SHIP^2 ~|~ BILL^2 ~|~ ARC^2 \\
   \stackrel{\tau}{\longrightarrow}
   \overline{Ship}_o ~|~ \overline{Bill}_o ~|~ SHIP^2 ~|~ BILL^2 ~|~ ARC^2 \\
   \stackrel{\tau}{\longrightarrow}
   \overline{Bill}_o ~|~ \overline{ShipOK}_o ~|~ BILL^2 ~|~ ARC^2 \\
   \stackrel{\tau}{\longrightarrow}
   \overline{ShipOK}_o ~|~ \overline{BillOK}_o ~|~ ARC^2 \\
   \stackrel{\tau}{\longrightarrow}
   \overline{ShipOK}_o ~|~ ShipOK_o.\overline{Archive}_o.CONFIG^2 \\
   \stackrel{\tau}{\longrightarrow}
   \overline{Archive}_o.CONFIG^2 \) \\
(by the Sum and React rules) or

\( CONFIG^2 \\
   \stackrel{Receipt_o}{\longrightarrow}
   \overline{InventoryCheck}_o ~|~ IC^1 ~|~ ICH^2 ~|~ CC^1 ~|~ CCH^2 ~|~ SHIP^2 ~|~ BILL^2 ~|~ ARC^2 \\
   \stackrel{\tau}{\longrightarrow}
   \overline{InventoryCheckNotOK}_o + \overline{InventoryCheckOK}_o ~|~
   ICH^2 ~|~ CC^1 ~|~ CCH^2 ~|~ SHIP^2 ~|~ BILL^2 ~|~ ARC^2 \\
   \stackrel{\tau}{\longrightarrow}
   \overline{CreditCheck}_o ~|~ CC^1 ~|~ CCH^2 ~|~ SHIP^2 ~|~ BILL^2 ~|~ ARC^2 \\
   \stackrel{\tau}{\longrightarrow}
   \overline{CreditCheckNotOK}_o + \overline{CreditCheckOK}_o ~|~ CCH^2 ~|~ SHIP^2 ~|~ BILL^2 ~|~ ARC^2 \\
   \stackrel{\tau}{\longrightarrow}
   \overline{Ship}_o ~|~ \overline{Bill}_o ~|~ SHIP^2 ~|~ BILL^2 ~|~ ARC^2 \\
   \stackrel{\tau}{\longrightarrow}
   \overline{Bill}_o ~|~ \overline{ShipOK}_o ~|~ BILL^2 ~|~ ARC^2 \\
   \stackrel{\tau}{\longrightarrow}
   \overline{Bill}_o ~|~ BILL^2 ~|~ BillOK_o.\overline{Archive}_o.CONFIG^2 \\
   \stackrel{\tau}{\longrightarrow}
   \overline{BillOK}_o ~|~ BillOK_o.\overline{Archive}_o.CONFIG^2 \\
   \stackrel{\tau}{\longrightarrow}
   \overline{Archive}_o.CONFIG^2 \) \\
(by the Sum and React rules) or

\( CONFIG^2 \\
   \stackrel{Receipt_o}{\longrightarrow}
   \overline{InventoryCheck}_o ~|~ IC^1 ~|~ ICH^2 ~|~ CC^1 ~|~ CCH^2 ~|~ SHIP^2 ~|~ BILL^2 ~|~ ARC^2 \\
   \stackrel{\tau}{\longrightarrow}
   \overline{InventoryCheckNotOK}_o + \overline{InventoryCheckOK}_o ~|~
   ICH^2 ~|~ CC^1 ~|~ CCH^2 ~|~ SHIP^2 ~|~ BILL^2 ~|~ ARC^2 \\
   \stackrel{\tau}{\longrightarrow}
   \overline{CreditCheck}_o ~|~ CC^1 ~|~ CCH^2 ~|~ SHIP^2 ~|~ BILL^2 ~|~ ARC^2 \\
   \stackrel{\tau}{\longrightarrow}
   \overline{CreditCheckNotOK}_o + \overline{CreditCheckOK}_o ~|~ CCH^2 ~|~ SHIP^2 ~|~ BILL^2 ~|~ ARC^2 \\
   \stackrel{\tau}{\longrightarrow}
   \overline{Ship}_o ~|~ \overline{Bill}_o ~|~ SHIP^2 ~|~ BILL^2 ~|~ ARC^2 \\
   \stackrel{\tau}{\longrightarrow}
   \overline{Ship}_o ~|~ SHIP^2 ~|~ \overline{BillOK}_o ~|~ ARC^2 \\
   \stackrel{\tau}{\longrightarrow}
   \overline{ShipOK}_o ~|~ \overline{BillOK}_o ~|~ ARC^2 \\
   \stackrel{\tau}{\longrightarrow}
   \overline{BillOK}_o ~|~ BillOK_o.\overline{Archive}_o.CONFIG^2 \\
   \stackrel{\tau}{\longrightarrow}
   \overline{Archive}_o.CONFIG^2 \) \\
(by the Sum and React rules) or

\( CONFIG^2 \\
   \stackrel{Receipt_o}{\longrightarrow}
   \overline{InventoryCheck}_o ~|~ IC^1 ~|~ ICH^2 ~|~ CC^1 ~|~ CCH^2 ~|~ SHIP^2 ~|~ BILL^2 ~|~ ARC^2 \\
   \stackrel{\tau}{\longrightarrow}
   \overline{InventoryCheckNotOK}_o + \overline{InventoryCheckOK}_o ~|~
   ICH^2 ~|~ CC^1 ~|~ CCH^2 ~|~ SHIP^2 ~|~ BILL^2 ~|~ ARC^2 \\
   \stackrel{\tau}{\longrightarrow}
   \overline{CreditCheck}_o ~|~ CC^1 ~|~ CCH^2 ~|~ SHIP^2 ~|~ BILL^2 ~|~ ARC^2 \\
   \stackrel{\tau}{\longrightarrow}
   \overline{CreditCheckNotOK}_o + \overline{CreditCheckOK}_o ~|~ CCH^2 ~|~ SHIP^2 ~|~ BILL^2 ~|~ ARC^2 \\
   \stackrel{\tau}{\longrightarrow}
   \overline{Ship}_o ~|~ \overline{Bill}_o ~|~ SHIP^2 ~|~ BILL^2 ~|~ ARC^2 \\
   \stackrel{\tau}{\longrightarrow}
   \overline{Ship}_o ~|~ SHIP^2 ~|~ \overline{BillOK}_o ~|~ ARC^2 \\
   \stackrel{\tau}{\longrightarrow}
   \overline{ShipOK}_o ~|~ \overline{BillOK}_o ~|~ ARC^2 \\
   \stackrel{\tau}{\longrightarrow}
   \overline{ShipOK}_o ~|~ ShipOK_o.\overline{Archive}_o.CONFIG^2 \\
   \stackrel{\tau}{\longrightarrow}
   \overline{Archive}_o.CONFIG^2 \) \\
(by the Sum and React rules) or

\( CONFIG^2 \\
   \stackrel{Receipt_o}{\longrightarrow}
   \overline{InventoryCheck}_o ~|~ IC^1 ~|~ ICH^2 ~|~ CC^1 ~|~ CCH^2 ~|~ SHIP^2 ~|~ BILL^2 ~|~ ARC^2 \\
   \stackrel{\tau}{\longrightarrow}
   \overline{InventoryCheckNotOK}_o + \overline{InventoryCheckOK}_o ~|~
   ICH^2 ~|~ CC^1 ~|~ CCH^2 ~|~ SHIP^2 ~|~ BILL^2 ~|~ ARC^2 \\
   \stackrel{\tau}{\longrightarrow}
   \overline{CreditCheck}_o ~|~ CC^1 ~|~ CCH^2 ~|~ SHIP^2 ~|~ BILL^2 ~|~ ARC^2 \\
   \stackrel{\tau}{\longrightarrow}
   \overline{CreditCheckNotOK}_o + \overline{CreditCheckOK}_o ~|~ CCH^2 ~|~ SHIP^2 ~|~ BILL^2 ~|~ ARC^2 \\
   \stackrel{\tau}{\longrightarrow}
   \overline{Ship}_o ~|~ \overline{Bill}_o ~|~ SHIP^2 ~|~ BILL^2 ~|~ ARC^2 \\
   \stackrel{\tau}{\longrightarrow}
   \overline{Ship}_o ~|~ SHIP^2 ~|~ \overline{BillOK}_o ~|~ ARC^2 \\
   \stackrel{\tau}{\longrightarrow}
   \overline{Ship}_o ~|~ SHIP^2 ~|~ ShipOK_o.\overline{Archive}_o.CONFIG^2 \\
   \stackrel{\tau}{\longrightarrow}
   \overline{ShipOK}_o ~|~ ShipOK_o.\overline{Archive}_o.CONFIG^2 \\
   \stackrel{\tau}{\longrightarrow}
   \overline{Archive}_o.CONFIG^2 \) \\
(by the Sum and React rules).


That is, \\
\( CONFIG^1 ~|~ P
   \stackrel{Receipt_o}{\Rightarrow}
   \overline{RejectIC}_o.CONFIG^1 ~|~ CC^1 ~|~ CCH^1 ~|~ SHIP^1 ~|~ BILL^1 ~|~ ARC^1 ~|~ ARCH^1 ~|~ P \) \\
   in the context $\overline{Receipt}_o \,|\, RejectIC_o$
   (by the LTS rules) and \\
\( CONFIG^2 \stackrel{Receipt_o}{\Rightarrow} Q \) in the context $\overline{Receipt}_o \,|\, RejectIC_o$ (by the LTS rules) \\
where $Q$ represents one of the processes elaborated above from $CONFIG^2$ such that \\
$\overline{RejectIC}_o.CONFIG^1 ~|~ CC^1 ~|~ CCH^1 ~|~ SHIP^1 ~|~ BILL^1 ~|~ ARC^1 ~|~ ARCH^1 ~|~ P \approx_o Q$ \\
in the residual context $RejectIC_o$
(by definition of $\approx_o$, $\because CONFIG^1 ~|~ P \approx_o CONFIG^2$) \\[-1mm]
implies $\overline{RejectIC}_o.CONFIG^1 ~|~ CC^1 ~|~ CCH^1 ~|~ SHIP^1 ~|~ BILL^1 ~|~ ARC^1 ~|~ ARCH^1 ~|~ P$ and $Q$ both perform
$\stackrel{\overline{RejectIC}_o}{\Rightarrow}$ \\
in the context $RejectIC_o$
(by the Sum and L-Par rules and by definition of $\approx_o$). \\[-1mm]
None of the processes elaborated above from $CONFIG^2$ performs $\stackrel{\overline{RejectIC}_o}{\Rightarrow}$ (by the LTS rules) \\[-0.5mm]
implies $Q$ does not perform $\stackrel{\overline{RejectIC}_o}{\Rightarrow}$ in the context $RejectIC_o$
(by specialisation; which is a contradiction).

Therefore, the lemma is not \textit{false} \\
implies the lemma is \textit{true} ($\because$ the lemma is $true$ $~\vee~$ the lemma is $false$). ~Q.E.D.

\subsection{Corollary
${\bf CONFIG^1 ~|~ \frac{CONFIG^2}{CONFIG^1} \napprox_o CONFIG^2 ~~\wedge~~}$ \\[1mm]
${\bf CONFIG^1 ~|~ \frac{ICH^2}{ICH^1} ~|~}$
             ${\bf \frac{CCH^2}{CCH^1} ~|~ \frac{SHIP^2}{SHIP^1} ~|~ \frac{BILL^2}{BILL^1} ~|~ \frac{ARC^2}{ARC^1} ~|~ \frac{0}{ARCH^1} \napprox_o CONFIG^2}$
\label{prf:nonweakobsbisimspec}}

Proof:
\( \frac{CONFIG^2}{CONFIG^1} \!\in\! \mathcal{P} ~\wedge~
   \frac{ICH^2}{ICH^1} ~|~ \frac{CCH^2}{CCH^1} ~|~ \frac{SHIP^2}{SHIP^1} ~|~ \frac{BILL^2}{BILL^1} ~|~ \frac{ARC^2}{ARC^1} ~|~ \frac{0}{ARCH^1} \!\in\! \mathcal{P} \) \\[1mm]
(by the syntax of $\mathcal{P}$) \\
implies \( CONFIG^1 ~|~ \frac{CONFIG^2}{CONFIG^1} \napprox_o CONFIG^2 ~\wedge~
           CONFIG^1 ~|~ \frac{ICH^2}{ICH^1} ~|~
                        \frac{CCH^2}{CCH^1} ~|~ \frac{SHIP^2}{SHIP^1} ~|~ \frac{BILL^2}{BILL^1} ~|~ \frac{ARC^2}{ARC^1} ~|~ \frac{0}{ARCH^1} \napprox_o CONFIG^2 \) \\[1mm]
(by Lemma \ref{prf:nonweakobsbisimgen}). ~Q.E.D.

\newpage 

\end{document}